\documentclass[preprint,
showpacs,preprintnumbers,amsmath,amssymb]{revtex4}

\usepackage{graphicx}
\usepackage{dcolumn}
\usepackage{bm,morefloats,color}

\usepackage{amssymb,bm}

\def\Eq#1{\begin{equation} #1 \end{equation}}
\def\Eqr#1{\begin{eqnarray} #1 \end{eqnarray}}
\def\Eqrsubl#1#2{\begin{subequations}\label{#1}\Eqr{#2}\end{subequations}}

\newcommand{\nn}{\nonumber}
\newcommand{\pd}{\partial}
\newcommand{\vect}[1]{\!\!\!\mbox{ \boldmath $#1$}}
\newcommand{\bea}{\begin{eqnarray}}
\newcommand{\eea}{\end{eqnarray}}

\def\Xsp{{\rm X}}

\def\Ysp{{\rm Y}}
\def\Zsp{{\rm Z}}
\def\X5sp{{\rm X}_5}
\def\Y3sp{{\rm Y}_3}
\def\Z3sp{{\rm Z}_3}

\def\Msp{{\rm M}}
\def\Nsp{{\rm N}}
\def\Ssp{{\rm S}}
\def\Wsp{{\rm W}}
\def\lap{{\triangle}}
\def\e{{\rm e}}

\begin{document}

\title{ 
Colliding $p$-branes in the dynamical intersecting brane system
}

\author{Kunihito Uzawa}
\affiliation{
Department of Physics, School of Science and Technology,
Kwansei Gakuin University, Sanda, Hyogo 669-1337, Japan
}

\date{\today}

\begin{abstract}
We discuss the dynamics of intersecting $p$-branes with cosmological 
constants in the higher-dimensional gravity theories. For the delocalized 
brane case, these solutions describe an asymptotically de Sitter or 
power-law expanding universe, while for the partially localized 
intersecting branes, they describe homogeneous and isotropic universes
at each position of the overall transverse space. We then apply 
these time-dependent branes to the study on the collision of two 0-branes 
and show that the 0$-$8-brane system or the smeared $0-p_{I}$-brane 
system can provide an example of colliding branes if they have the 
same brane charges and only one overall transverse space. 
Finally, we argue some
applications of the solutions in supergravity models. 
\end{abstract}


\maketitle


\section{Introduction}
\label{sec:introduction}
In many recent developments involving cosmological models and brane 
collisions in higher-dimensional gravity theories 
\cite{Lu:1996jk, Behrndt:2003cx, Gibbons:2005rt, Chen:2005jp, Kodama:2005fz, 
Kodama:2005cz, Kodama:2006ay, Arroja:2006zz, Koyama:2006ni, Arroja:2007ss, 
Binetruy:2007tu, Binetruy:2008ev, Arroja:2008ma, Maeda:2009tq, 
Maeda:2009zi,Gibbons:2009dr, Maeda:2009ds, Uzawa:2010zza, Minamitsuji:2010fp, 
Maeda:2010ja, Maeda:2010aj, Minamitsuji:2010kb, Nozawa:2010zg, 
Minamitsuji:2010uz, Minamitsuji:2011jt, Uzawa:2010zz, Maeda:2011sh, 
Maeda:2012xb, Minamitsuji:2012if, Blaback:2012mu, 
Uzawa:2013koa, Uzawa:2013msa, Uzawa:2014kka}, 
the dynamical $p$-branes carrying charges have played 
important roles. In the classical solution of a single $p$-brane, the 
coupling of the dilaton to the field strength includes the parameter $N$. 
Since these brane solutions with $N=4$ are related to well-known 
D-branes and M-branes in supergravity theories, they certainly exhibit 
many attractive properties in the higher-dimensional spacetime. 
Some static solutions with $N\ne 4$ also have supersymmetry after 
dimensional reductions to lower-dimensional theory 
\cite{Lu:1995cs, Cvetic:2000dm}. The time-dependent 
generalizations of these solutions are thus important examples 
of higher-dimensional gravity theories.  
The dynamical brane solution with the cosmological constant can be obtained 
by choosing the coupling constant appropriately 
\cite{Minamitsuji:2010fp, Maeda:2010aj, Minamitsuji:2010uz}. 
For a single 2-form field strength and a nontrivial dilaton, we have 
found that the dynamical single 0-brane solution describes the 
Milne universe \cite{Maeda:2010aj, Minamitsuji:2010uz}. 
The field equations give an asymptotically de Sitter solution if 
the scalar field is trivial \cite{Minamitsuji:2010uz},
which is a generalization of the Kastor-Traschen 
solution in the four-dimensional Einstein-Maxwell theory 
\cite{Kastor:1992nn, Brill:1993tm}. 
The construction of intersecting branes with a cosmological constant 
is a natural generalization of the single cosmological brane solutions. 
The time-dependent intersecting branes we have mainly discussed 
are localized only along the relative or overall transverse directions 
in a higher-dimensional background, which are delocalized intersecting brane 
systems. However, in the higher-dimensional gravity theories, 
one of the branes is localized along the relative transverse directions 
but delocalized along the overall transverse directions, which are 
partially localized branes solutions. If the background has the 
cosmological constant, there is little known about the dynamics of the 
intersecting brane system for not only the delocalized case  
but also the partially localized one. 

In the present paper,
we will explore the possible generalization of these solutions
to the case of the intersecting brane systems
with cosmological constants,  although 
similar single brane solutions have been 
analyzed in Ref.~\cite{Minamitsuji:2010uz}. 
We recall these 
arguments for constructions of the solution and  
modify the ansatz of the fields. 
A brane configuration has to satisfy an 
intersection rule which is 
an algebraic equation that relates the coupling of the dilaton to 
the dimensionality of the branes. 
The intersection rule implies that only the 0-brane can depend on time 
 and the dynamical 0-brane commutes with the static $p$-branes. 
We will study the dynamical intersecting brane solutions for 
not only the delocalized case but also the partially localized one.  

The paper is constructed as follows:
In Secs. \ref{sec:pl} and \ref{sec:cn},
we derive the dynamical intersecting brane solutions 
with cosmological constants in a $D$-dimensional theory 
following the approach developed in Ref.~\cite{Minamitsuji:2010uz}.
We then illustrate how the dynamical solution
 of two or $n$ intersecting branes arise 
under the condition of $N\ne 4$
in the $D$-dimensional theory. 
The spacetime starts with the structure of 
the combined 0-branes. 
If they do not have the same charges, 
a singularity hypersurface appears before they meet 
as the time decreases for $D>4$. 
We then discuss the dynamics of two 0-branes with static $p$-branes 
(or the dynamics of two black holes) in Sec. \ref{sec:c}. 
If there exists one uncompactified extra dimension [0$-$8-brane
system or 0$-(D-1)$-brane systems ($p\le 7$)] and two brane systems have 
the same brane charges, the solution describes a 
collision of two branes  (or two black holes), 
which is similar to the result in 
Refs.~\cite{Gibbons:2005rt, Maeda:2010aj, Maeda:2012xb}. 
In Sec.~\ref{sec:sug}, applications of these solutions to five- or 
six-dimensional 
supergravity models are discussed. We consider in detail the construction 
yielding the dynamical 0- or 1-brane in the Nishino-Salam-Sezgin model. 
We also provide brief discussions for a time-dependent brane system in
Romans' supergravity model. We describe how our Universe could be 
represented in the present formulation via an appropriate compactification 
and give the application to cosmology. 
We show that there exists no accelerating expansion of our Universe, 
although the conventional power-law expansion of the Universe is possible. 
We then discuss the dynamics of two 0- or 1-branes with smeared branes. 
If two brane systems have the same brane charges 
with smearing some dimensions, 
the solution describes a collision of two brane backgrounds.

There is a curvature singularity in the dynamical brane background if we 
set a particular value for the constant parameters. Then the solution implies 
that the presence of the singularities is signaling
possible instabilities, making the solutions sick or unphysical. 
We study the classical stability of the solutions in Sec.~\ref{sec:in}. 
Our preliminary analysis will present that the energy
of Klein-Gordon scalar fields in the dynamical brane background grows 
with time for inertial observers approaching the singularity. 
In terms of using the preliminary analysis performed in 
Refs.~\cite{oai1, Quevedo:2002tm, oai2, Cornalba:2003ze},  
the Klein-Gordon modes will be studied, arriving at the preliminary 
conclusion of instability. 
Section \ref{sec:cd} 
will be devoted to the summary and conclusions.


\section{Dynamical partially localized intersecting brane backgrounds 
with cosmological constants}
\label{sec:pl}

In this section, 
we will construct the partially localized time-dependent brane systems 
in $D$ dimensions with cosmological constants.

We consider a $D$-dimensional theory 
composed of the metric $g_{MN}$, the scalar field $\phi$, 
cosmological constants $\Lambda_I~(I=r,~s)$, 
and two antisymmetric tensor field strengths 
of rank $(p_r+2)$ and $(p_s+2)$. 
The action in $D$ dimensions is 
given by 
\Eqr{
S&=&\frac{1}{2\kappa^2}\int \left[\left(R-2\e^{\alpha_r\phi}
\Lambda_r-2\e^{\alpha_s\phi}\Lambda_s\right)\ast{\bf 1}_D
 -\frac{1}{2}\ast d\phi \wedge d\phi\right.\nn\\
 &&\left.\hspace{-0.5cm}
 -\frac{1}{2}\frac{1}{\left(p_r+2\right)!}
 \e^{\epsilon_rc_r\phi}\ast F_{(p_r+2)}\,\wedge\, 
 F_{\left(p_r+2\right)}
 -\frac{1}{2}\frac{1}{\left(p_s+2\right)!}
 \e^{\epsilon_sc_s\phi}\ast F_{(p_s+2)}\,\wedge\, F_{(p_s+2)}
 \right],
\label{pl:action:Eq}
}
where $R$ denotes the Ricci scalar constructed from the $D$-dimensional 
metric $g_{MN}$\,, $\alpha_I~(I=r,~s)$ are constants, 
$\kappa^2$ denotes the $D$-dimensional gravitational constant, 
$\ast$ is the Hodge operator in the $D$-dimensional spacetime, and 
$F_{\left(p_r+2\right)}$ and $F_{\left(p_s+2\right)}$ 
are $\left(p_r+2\right)$ and 
$\left(p_s+2\right)$-form field strengths, respectively.
The constant parameters $c_I$
and 
$\epsilon_I~(I=r,~s)$ are defined by 
\Eqrsubl{pl:parameters:Eq}{
c_I^2&=&N_I-\frac{2(p_I+1)(D-p_I-3)}{D-2},
   \label{pl:c:Eq}\\
\epsilon_I&=&\left\{
\begin{array}{cc}
 +&{\rm if~the}~p_I-{\rm brane~is~electric}\\
 -&~~~{\rm if~the}~p_I-{\rm brane~is~magnetic}\,,
\end{array} \right.
 \label{pl:epsilon:Eq}
   }
respectively.
Here $N_I$ is constant. 
The $(p_r+2)$-form, $(p_s+2)$-form 
field strengths $F_{(p_r+2)}$, $F_{(p_s+2)}$ 
are given by the 
$\left(p_r+1\right)$-form, $\left(p_s+1\right)$-form gauge 
potentials $A_{\left(p_r+1\right)}$, 
$A_{\left(p_s+1\right)}$, respectively:
\Eq{
F_{(p_r+2)}=dA_{(p_r+1)}\,,~~~~~F_{(p_s+2)}=dA_{(p_s+1)}\,.
}
For the $D$-dimensional action (\ref{pl:action:Eq}), 
the field equations read
\Eqrsubl{pl:equations:Eq}{
&&\hspace{-0.5cm}R_{MN}=\frac{2}{D-2}\left(\e^{\alpha_r\phi}\Lambda_r 
+\e^{\alpha_s\phi}\Lambda_s\right)g_{MN}
+\frac{1}{2}\pd_M\phi \pd_N \phi\nn\\
&&+\frac{1}{2}\frac{\e^{\epsilon_rc_r\phi}}
{\left(p_r+2\right)!}
\left[\left(p_r+2\right)
F_{MA_2\cdots A_{\left(p_r+2\right)}} 
{F_N}^{A_2\cdots A_{\left(p_r+2\right)}}
-\frac{p_r+1}{D-2}\,g_{MN}\,F^2_{\left(p_r+2\right)}\right]\nn\\
&&+\frac{1}{2}\frac{\e^{\epsilon_sc_s\phi}}
   {\left(p_s+2\right)!}
\left[\left(p_s+2\right)
F_{MA_2\cdots A_{\left(p_s+2\right)}} 
{F_N}^{A_2\cdots A_{\left(p_s+2\right)}}
-\frac{p_s+1}{D-2}\,g_{MN}\,F_{\left(p_s+2\right)}^2\right],
   \label{pl:Einstein:Eq}\\
&&\hspace{-0.5cm}\lap\phi-\frac{1}{2}\frac{\epsilon_rc_r}
{\left(p_r+2\right)!}
\e^{\epsilon_rc_r\phi}F^2_{\left(p_r+2\right)}
-\frac{1}{2}\frac{\epsilon_sc_s}{\left(p_s+2\right)!}
\e^{\epsilon_sc_s\phi}F^2_{\left(p_s+2\right)}
-2\alpha_r\e^{\alpha_r\phi}\Lambda_r
-2\alpha_s\e^{\alpha_s\phi}\Lambda_s=0\,,
   \label{pl:scalar:Eq}\\
&&\hspace{-0.5cm}
  d\left[\e^{\epsilon_rc_r\phi}\ast F_{\left(p_r+2\right)}\right]=0\,,
   \label{pl:gauge-r:Eq}\\
&&\hspace{-0.5cm}
d\left[\e^{\epsilon_sc_s\phi}\ast F_{\left(p_s+2\right)}\right]=0\,,
   \label{pl:gauge-s:Eq}
}
where $\lap$ denotes the Laplace operator with respect to the 
$D$-dimensional metric $g_{MN}$\,.

The $D$-dimensional metric involving the intersecting branes with cosmological 
constant can be put in the general form
\Eqr{
&&\hspace{-0.7cm}
ds^2=h^{a_r}_r(x, y, z)h_s^{a_s}(x, v, z)q_{\mu\nu}
(\Xsp)dx^{\mu}dx^{\nu}+h^{b_r}_r(x, y, z)h_s^{a_s}(x, v, z)\gamma_{ij}
(\Ysp_1)dy^idy^j\nn\\
&&\hspace{-0.5cm}
+h^{a_r}_r(x, y, z)h_s^{b_s}(x, v, z)w_{mn}(\Ysp_2)dv^{m}dv^{n}
+h^{b_r}_r(x, y, z)h_s^{b_s}(x, v, z)u_{ab}(\Zsp)dz^adz^b, 
 \label{pl:metric:Eq}
}
where $q_{\mu\nu}$ is the $(p+1)$-dimensional metric depending 
only on the $(p+1)$-dimensional coordinates $x^{\mu}$, 
$\gamma_{ij}$ is the $(p_s-p)$-dimensional metric depending 
only on the $(p_s-p)$-dimensional coordinates $y^i$, 
$w_{mn}$ is the $(p_r-p)$-dimensional metric depending 
only on the $(p_r-p)$-dimensional coordinates $v^m$ and 
$u_{ab}$ is the $(D+p-p_r-p_s-1)$-dimensional metric depending 
only on the $(D+p-p_r-p_s-1)$-dimensional coordinates $z^a$. 
Here we assume that the parameters $a_I~(I=r,~s)$ and 
$b_I~(I=r,~s)$ in the metric (\ref{pl:metric:Eq}) are given by 
\Eq{
a_I=-\frac{4\left(D-p_I-3\right)}{N_I\left(D-2\right)},~~~~~~~~~
b_I=\frac{4\left(p_I+1\right)}{N_I\left(D-2\right)}.
 \label{pl:parameter:Eq}
}
The brane configuration is illustrated 
in Table \ref{table_A}.
\begin{table}[h]
\caption{\baselineskip 14pt
Intersections of $p_r-p_s$-branes in the metric (\ref{pl:metric:Eq}), 
where $p'=p_s+p_r-p$.
}
\label{DpDp}
{\scriptsize
\begin{center}
\begin{tabular}{|c||c|c|c|c|c|c|c|c|c|c|c|c|c|c|}
\hline
Case&&0&1& $\cdots$ & $p$ & $p+1$ & $\cdots$ & $p_s$ & $p_s+1$ & 
$\cdots$ & $p'$ & $p'+1$ & $\cdots$ & $D-1$
\\
\hline
 &$p_r$ & $\circ$ & $\circ$ & $\circ$ & $\circ$ &&&& $\circ$ & $\circ$ &
  $\circ$ &&& 
\\
\cline{3-15}
$p_r-p_s$ & $p_s$ & $\circ$ & $\circ$ & $\circ$ & $\circ$ & $\circ$ &
 $\circ$ & $\circ$ && & & & &
\\
\cline{3-15}
&$x^N$ & $t$ & $x^1$ & $\cdots$ & $x^p$ & $y^1$ & $\cdots$ & $y^{p_s-p}$ 
& $v^1$ & $\cdots$ & $v^{p_r-p}$ & $z^1$ & $\cdots$ & $z^{D-p'-1}$
\\
\hline
\end{tabular}
\end{center}
}
\label{table_A}
\end{table}

The dynamical brane solutions are characterized by 
two warp factors, $h_r$ and $h_s$, depending 
on the $(D+p-p_r-p_s-1)$-dimensional coordinates transverse to the 
corresponding brane as well as 
the $(p+1)$-dimensional world-volume coordinate.
In the case of intersection involving two branes, 
the powers of warp factors have to obey the intersection rule, 
and then split the coordinates in four parts 
\cite{Behrndt:1996pm, Bergshoeff:1996rn, Bergshoeff:1997tt}. 
One is coordinates of the world-volume spacetime, $x^{\mu}$, 
which are common to the $p_r$-, $p_s$-branes. 
The others are coordinates of the overall transverse space, 
$z^a$, and coordinates of the relative transverse space, $y^i$ and $v^m$, 
which are transverse to only one of the $p_r$-, $p_s$-branes.  
In this section, we consider the intersections of a $p_r$- and a $p_s$-brane 
with the following conditions in $D$-dimension.  
We assume that the functions $h_r$ and $h_s$ depend not only on 
overall transverse coordinates but also on the corresponding 
relative coordinates and world-volume coordinates. 
We therefore may write
$h_r=h_r(x, y, z),\, h_s=h_s(x, v, z)$.

We give the expression for the field strengths $F_{\left(p_r+2\right)}$, 
$F_{\left(p_s+2\right)}$ and scalar field $\phi$ of a 
$p_r$-brane intersecting with a $p_s$-brane over a $p$-brane configuration:
\Eqrsubl{pl:ansatz:Eq}{
\e^{\phi}&=&h_r^{2\epsilon_rc_r/N_r}\,
h_s^{2\epsilon_sc_s/N_s},
  \label{pl:dilaton:Eq}\\
F_{\left(p_r+2\right)}&=&\frac{2}{\sqrt{N_r}}
d\left[h^{-1}_r(x, y, z)\right]\wedge\Omega(\Xsp)\wedge\Omega(\Ysp_2),
  \label{pl:strength-r:Eq}\\
F_{\left(p_s+2\right)}&=&\frac{2}{\sqrt{N_s}}
d\left[h^{-1}_s(x, v, z)\right]\wedge\Omega(\Xsp)\wedge\Omega(\Ysp_1),
  \label{pl:strength-s:Eq}
}
where $\Omega(\Xsp)$, $\Omega({\Ysp}_1)$, and $\Omega({\Ysp}_2)$ 
are the volume $\left(p+1\right)$-form, $\left(p_s-p\right)$-form, 
and $\left(p_r-p\right)$-form, respectively
\Eqrsubl{pl:volume:Eq}{
\Omega(\Xsp)&=&\sqrt{-q}\,dx^0\,\wedge\,dx^1\wedge\,\cdots\,\wedge\,dx^p,\\
\Omega(\Ysp_1)&=&\sqrt{\gamma}\,dy^1\,\wedge\,dy^2\,\wedge\,\cdots\,\wedge 
\,dy^{p_s-p},\\
\Omega(\Ysp_2)&=&\sqrt{w}\,dv^1\,\wedge\,dv^2\,\wedge\,\cdots\,\wedge 
\,dv^{p_r-p}.
}
Here, $q$, $\gamma$, and $w$ denote 
the determinants of the metrics $q_{\mu\nu}$, 
$\gamma_{ij}$, and $w_{mn}$, respectively.

\subsection{Power-law expanding universe}

In this subsection, we consider the field Eqs.~(\ref{pl:ansatz:Eq}) with 
$c_I~(I=r,~s)\neq 0$. 
The parameters $\alpha_I~(I=r,~s)$ are assumed to be 
\Eqr{
\alpha_r=-\epsilon_r c_r\,,~~~~~\alpha_s=-\epsilon_s c_s.
    \label{pl:alpha:Eq}
}

Let us first consider the gauge field Eqs.~(\ref{pl:gauge-r:Eq}), 
(\ref{pl:gauge-s:Eq}).
Using the assumptions (\ref{pl:metric:Eq})
and (\ref{pl:ansatz:Eq}), we have  
\Eqrsubl{pl:gauge2:Eq}{
&&\hspace{-0.3cm}
d\left[h_s^{4(\chi+1)/N_s}\pd_i h_r\left(\ast_{\Ysp_1}dy^i\right)
\wedge\Omega(\Zsp)+h_s^{4\chi/N_s}\pd_a h_r\left(\ast_{\Zsp}dz^a\right)
\wedge\Omega(\Ysp_1)\right]=0,
  \label{pl:gauge2-r:Eq}\\
&&\hspace{-0.3cm}
d\left[h_r^{4(\chi+1)/N_r}\pd_m h_s\left(\ast_{\Ysp_2}dv^m\right)
\wedge\Omega(\Zsp)+h_r^{4\chi/N_r}\pd_a h_s\left(\ast_{\Zsp}dz^a\right)
\wedge\Omega(\Ysp_2)\right]=0,
  \label{pl:gauge2-s:Eq}
 }
where $\ast_{\Ysp_1}$, $\ast_{\Ysp_2}$, and $\ast_{{\rm Z}}$ 
denote the Hodge operator on $\Ysp_1$, $\Ysp_2$, and Z, respectively, 
and $\chi$ is given by
\Eq{
\chi=
p+1-\frac{\left(p_r+1\right)\left(p_s+1\right)}{D-2}
+\frac{1}{2}\epsilon_r\epsilon_sc_rc_s.
   \label{pl:chi:Eq}
}
In the following, we discuss the case of $\chi=0$ because the 
relation $\chi=0$ is consistent with the intersection rule 
which has been found in 
\cite{Binetruy:2007tu, Maeda:2009zi, Minamitsuji:2010kb, Minamitsuji:2010uz, 
Strominger:1995ac, Townsend:1995af, Douglas:1995bn, 
Papadopoulos:1996uq, Tseytlin:1996bh, Gauntlett:1996pb, 
Cvetic:1996gq, Papadopoulos:1996ca, 
Russo:1996if, Argurio:1997gt, Tseytlin:1997cs, Ohta:1997gw, 
Youm:1997hw, Argurio:1998cp, Miao:2004bn, 
Chen:2005uw}.

Setting $\chi=0$\,, the Eq.~(\ref{pl:gauge2-r:Eq}) gives
\Eq{
h_s\lap_{\Ysp_1}h_r+\lap_{\Zsp}h_r=0,
~~~\pd_{\mu}\pd_i h_r+\frac{4}{N_s}\pd_{\mu}\ln h_s\pd_ih_r=0,
~~~\pd_{\mu}\pd_a h_r=0,
  \label{pl:gauge3:Eq}
}
where $\lap_{\Ysp_1}$, and $\lap_{\Zsp}$ are 
the Laplace operators on the space of $\Ysp_1$, and 
Z, respectively.

On the other hand, Eq.~(\ref{pl:gauge2-s:Eq}) leads to
\Eq{
h_r\lap_{\Ysp_2}h_s+\lap_{\Zsp}h_s=0,
~~~\pd_{\mu}\pd_m h_s+\frac{4}{N_r}\pd_{\mu}\ln h_r\pd_mh_s=0,
~~~\pd_{\mu}\pd_a h_s=0\,,
   \label{pl:gauge4:Eq}
}
where we used Eq.~(\ref{pl:chi:Eq}) and $\triangle_{\Ysp_2}$ is 
the Laplace operators on the space of $\Ysp_2$,

Now we consider the Einstein Eq.~(\ref{pl:Einstein:Eq}). 
Using the ansatz (\ref{pl:metric:Eq}), (\ref{pl:ansatz:Eq}), 
and intersection rule $\chi=0$, the Einstein equations become
\Eqrsubl{pl:cEinstein:Eq}{
&&\hspace{-0.5cm}R_{\mu\nu}(\Xsp)
-\frac{4}{N_r}h_r^{-1}D_{\mu}D_{\nu}h_r
-\frac{4}{N_s}h_s^{-1}D_{\mu}D_{\nu}h_s
+\frac{2}{N_r}\pd_{\mu}\ln h_r\left[\left(1-\frac{4}{N_r}\right)
\pd_{\nu}\ln h_r-\frac{4}{N_s}\pd_{\nu}\ln h_s\right]\nn\\
&&~~~~+\frac{2}{N_s}\pd_{\mu}\ln h_s\left[\left(1-\frac{4}{N_s}\right)
\pd_{\nu}\ln h_s-\frac{4}{N_r}\pd_{\nu}\ln h_r\right]\nn\\
&&~~~~-\frac{2}{D-2}\left(\Lambda_rh_r^{-2+a_rp_r}
h_s^{a_s-2\epsilon_r\epsilon_sc_rc_s/N_s}
+\Lambda_sh_r^{a_r-2\epsilon_r\epsilon_sc_rc_s/N_r}
h_s^{-2+a_sp_s}\right)q_{\mu\nu}\nn\\
&&~~~~-\frac{1}{2}q_{\mu\nu}h_r^{-4/N_r}h_s^{-4/N_s}
\left[a_rh_r^{-1}\left(h_s^{4/N_s}\lap_{\Ysp_1}h_r+\lap_{\Zsp}h_r\right)
+a_sh_s^{-1}\left(h_r^{4/N_r}\lap_{\Ysp_2}h_s+\lap_{\Zsp}h_s\right)\right]\nn\\
&&~~~~-\frac{1}{2}q_{\mu\nu}\left[a_r
h_r^{-1}\lap_{\Xsp}h_r-a_rq^{\rho\sigma}\pd_{\rho}\ln h_r
\left\{\left(1-\frac{4}{N_r}\right)
\pd_{\sigma}\ln h_r-\frac{4}{N_s}
\pd_{\sigma}\ln h_s\right\}\right.\nn\\
&&\left. ~~~~+a_sh_s^{-1}\triangle_{\Xsp}h_s
-a_sq^{\rho\sigma}\pd_{\rho}\ln h_s\left\{\left(1-\frac{4}{N_s}\right)
\pd_{\sigma}\ln h_s-\frac{4}{N_r}\pd_{\sigma}\ln h_r\right\}\right]=0,
 \label{pl:cEinstein-mu:Eq}\\
&&\hspace{-0.5cm}\frac{2}{N_r}h_r^{-1}\left(\pd_{\mu}\pd_i h_r
+\frac{4}{N_s}\pd_{\mu}\ln h_s\pd_ih_r\right)=0,~~~~
\frac{2}{N_s}h_s^{-1}\left(\pd_{\mu}\pd_m h_s
+\frac{4}{N_r}\pd_{\mu}\ln h_r\pd_mh_s\right)=0,
 \label{pl:cEinstein-mm:Eq}\\
&&\hspace{-0.5cm}\frac{2}{N_r}h_r^{-1}\pd_{\mu}\pd_a h_r
+\frac{2}{N_s}h_s^{-1}\pd_{\mu}\pd_a h_s=0,
 \label{pl:cEinstein-ma:Eq}\\
&&\hspace{-0.5cm}
R_{ij}(\Ysp_1)-\frac{1}{2}h_r^{4/N_r}\gamma_{ij}\left[b_r
h_r^{-1}\lap_{\Xsp}h_r-b_rq^{\rho\sigma}\pd_{\rho}\ln h_r
\left\{\left(1-\frac{4}{N_r}\right)
\pd_{\sigma}\ln h_r-\frac{4}{N_s}
\pd_{\sigma}\ln h_s\right\}\right.\nn\\
&&\left. ~~+a_sh_s^{-1}\triangle_{\Xsp}h_s
-a_sq^{\rho\sigma}\pd_{\rho}\ln h_s\left\{\left(1-\frac{4}{N_s}\right)
\pd_{\sigma}\ln h_s-\frac{4}{N_r}\pd_{\sigma}\ln h_r\right\}\right]\nn\\
&&~~-\frac{1}{2}\gamma_{ij}h_s^{-4/N_s}\left\{b_rh_r^{-1}
\left(h_s^{4/N_s}\lap_{\Ysp_1}h_r+\lap_{\Zsp}h_r\right)
+a_sh_s^{-1}\left(h_r^{4/N_r}\lap_{\Ysp_2}h_s+\lap_{\Zsp}h_s\right)
\right\}\nn\\
&&~~-\frac{2}{D-2}\left[\Lambda_rh_r^{-2+a_rp_r+4/N_r}
h_s^{a_s-2\epsilon_r\epsilon_sc_rc_s/N_s}
+\Lambda_sh_r^{a_r-2(\epsilon_r\epsilon_sc_rc_s-2)/N_r}
h_s^{-2+a_sp_s}\right]\gamma_{ij}=0,
 \label{pl:cEinstein-ij:Eq}\\
&&\hspace{-0.5cm}
\frac{8}{N_rN_s(D-2)^2}\left[(p_r+1)(p_s+1)-(D-2)(p_r+p_s+2)
\right]\pd_i\ln h_r\pd_m\ln h_s=0\,,
 \label{pl:cEinstein-im:Eq}\\
&&\hspace{-0.5cm}R_{mn}(\Ysp_2)-\frac{1}{2}h_s^{4/N_s}w_{mn}\left[a_r
h_r^{-1}\lap_{\Xsp}h_r-a_rq^{\rho\sigma}\pd_{\rho}\ln h_r
\left\{\left(1-\frac{4}{N_r}\right)
\pd_{\sigma}\ln h_r-\frac{4}{N_s}
\pd_{\sigma}\ln h_s\right\}\right.\nn\\
&&\left. ~~+b_sh_s^{-1}\lap_{\Xsp}h_s
-b_sq^{\rho\sigma}\pd_{\rho}\ln h_s\left\{\left(1-\frac{4}{N_s}\right)
\pd_{\sigma}\ln h_s-\frac{4}{N_r}\pd_{\sigma}\ln h_r\right\}\right]\nn\\
&&~~-\frac{1}{2}w_{mn}h_r^{-4/N_r}\left\{a_rh_r^{-1}
\left(h_s^{4/N_s}\lap_{\Ysp_1}h_r+\lap_{\Zsp}h_r\right)
+b_sh_s^{-1}\left(h_r^{4/N_r}\lap_{\Ysp_2}h_s+\lap_{\Zsp}h_s\right)
\right\}\nn\\
&&~~-\frac{2}{D-2}\left[\Lambda_rh_r^{-2+a_rp_r}
h_s^{a_s-2(\epsilon_r\epsilon_sc_rc_s-2)/N_s}
+\Lambda_sh_r^{a_r-2\epsilon_r\epsilon_sc_rc_s/N_r}
h_s^{-2+a_sp_s+4/N_s}\right]w_{mn}=0,
 \label{pl:cEinstein-mn:Eq}\\
&&\hspace{-0.5cm}
R_{ab}(\Zsp)-\frac{1}{2}h_r^{4/N_r}h_s^{4/N_s}u_{ab}\left[b_r
h_r^{-1}\lap_{\Xsp}h_r-b_rq^{\rho\sigma}\pd_{\rho}\ln h_r
\left\{\left(1-\frac{4}{N_r}\right)
\pd_{\sigma}\ln h_r-\frac{4}{N_s}\pd_{\sigma}\ln h_s\right\}\right.\nn\\
&&\left. +b_sh_s^{-1}\lap_{\Xsp}h_s
-b_sq^{\rho\sigma}\pd_{\rho}\ln h_s\left\{\left(1-\frac{4}{N_s}\right)
\pd_{\sigma}\ln h_s-\frac{4}{N_r}\pd_{\sigma}\ln h_r\right\}\right]\nn\\
&&-\frac{1}{2}u_{ab}\left[b_rh_r^{-1}
\left(h_s^{4/N_s}\lap_{\Ysp_1}h_r+\lap_{\Zsp}h_r\right)
+b_sh_s^{-1}\left(h_r^{4/N_r}\lap_{\Ysp_2}h_s+\lap_{\Zsp}h_s\right)
\right]\nn\\
&&\hspace{-0.2cm}
-\frac{2u_{ab}}{D-2}\left[\Lambda_rh_r^{-2+a_rp_r+4/N_r}
h_s^{a_s-2(\epsilon_r\epsilon_sc_rc_s-2)/N_s}
+\Lambda_sh_r^{a_r-2(\epsilon_r\epsilon_sc_rc_s-2)/N_r}
h_s^{-2+a_sp_s+4/N_s}\right]=0,
  \label{pl:cEinstein-ab:Eq}
}
where 
$D_{\mu}$ is the covariant derivative constructed from 
the metric $q_{\mu\nu}$, $\triangle_{\Xsp}$,
$\triangle_{\Ysp_1}$,
$\triangle_{\Ysp_2}$, 
and $\triangle_{\Zsp}$
are 
the Laplace operators on
$\Xsp$, $\Ysp_1$, $\Ysp_2$, and $\Zsp$,
respectively, and 
$R_{\mu\nu}(\Xsp)$, $R_{ij}(\Ysp_1)$, $R_{mn}(\Ysp_2)$,
and $R_{ab}(\Zsp)$ are the Ricci tensors
with respect to the metrics $q_{\mu\nu}(\Xsp)$, $\gamma_{ij}(\Ysp_1)$,
$w_{mn}(\Ysp_2)$\,, and $u_{ab}(\Zsp)$, respectively.

From Eqs.
(\ref{pl:cEinstein-mm:Eq}) and (\ref{pl:cEinstein-ma:Eq})\,, 
the warp factors $h_r$ and $h_s$ can be expressed as 
\Eqrsubl{pl:warp:Eq}{
&&h_r= h_0(x)+h_1(y, z),~~~~h_s=h_s(v, z)\,,~~~~~~{\rm For}~~
\pd_{\mu}h_s=0\,,
  \label{pl:warp1:Eq}\\
&&h_r= h_r(y, z),~~~~h_s= k_0(x)+k_1(v, z)\,,~~~~~~{\rm For}~~
\pd_{\mu}h_r=0.
  \label{pl:warp2:Eq}  
}
If we require that the background satisfies  
\Eqr{
\pd_{\mu}h_s=0\,,~~~p=p_r=0\,,~~~\Lambda_s=0\,,
~~~\chi=0, 
    \label{pl:condition:Eq}
}
the Einstein Equations (\ref{pl:cEinstein:Eq}) reduce to
\Eqrsubl{pl:c2Einstein:Eq}{
&&\hspace{-0.3cm}
-\frac{2}{N_r}\left[2h_r^{-1}\frac{d^2h_0}{dt^2}
-\left(1-\frac{4}{N_r}\right)\left(\pd_t\ln h_r\right)^2\right]
+\frac{2}{D-2}\Lambda_rh_r^{-2}\nn\\
&&+\frac{1}{2}h_r^{-4/N_r}h_s^{-4/N_s}
\left[a_rh_r^{-1}\left(h_s^{4/N_s}\lap_{\Ysp_1}h_1+\lap_{\Zsp}h_1\right)
+a_sh_s^{-1}\lap_{\Zsp}h_s\right]\nn\\
&&-\frac{1}{2}a_r\left[
h_r^{-1}\frac{d^2h_0}{dt^2}-\left(1-\frac{4}{N_r}\right)
\left(\pd_t\ln h_r\right)^2\right]
=0,
 \label{pl:c2Einstein-mu:Eq}\\
&&\hspace{-0.3cm}R_{ij}(\Ysp_1)+\frac{1}{2}b_rh_r^{4/N_r}\gamma_{ij}\left[
h_r^{-1}\frac{d^2h_0}{dt^2}-\left(1-\frac{4}{N_r}\right)
\left(\pd_t\ln h_r\right)^2\right]
-\frac{2}{D-2}\Lambda_rh_r^{-2+\frac{4}{N_r}}\gamma_{ij}\nn\\
&&-\frac{1}{2}\gamma_{ij}h_s^{-4/N_s}\left[b_rh_r^{-1}
\left(h_s^{4/N_s}\lap_{\Ysp_1}h_1+\lap_{\Zsp}h_1\right)
+a_sh_s^{-1}\lap_{\Zsp}h_s\right]=0,
 \label{pl:c2Einstein-ij:Eq}\\
&&\hspace{-0.3cm}R_{ab}(\Zsp)
+\frac{1}{2}b_rh_r^{4/N_r}h_s^{4/N_s}u_{ab}\left[
h_r^{-1}\frac{d^2h_0}{dt^2}-\left(1-\frac{4}{N_r}\right)
\left(\pd_t\ln h_r\right)^2\right]
-\frac{2\Lambda_r}{D-2}h_r^{-2+\frac{4}{N_r}}h_s^{\frac{4}{N_s}}u_{ab}\nn\\
&&-\frac{1}{2}u_{ab}\left[b_rh_r^{-1}
\left(h_s^{4/N_s}\lap_{\Ysp_1}h_1+\lap_{\Zsp}h_1\right)
+b_sh_s^{-1}\lap_{\Zsp}h_s\right]=0.
  \label{pl:c2Einstein-ab:Eq}
}
Note that Eq. (\ref{pl:cEinstein-mn:Eq}) becomes trivial for $p=p_r=0$. 
By combining 
the above equations and setting $p=p_r=0$, 
the Einstein equations for $N_r\ne 4$ lead to
\Eqrsubl{pl:solution1:Eq}{
&&R_{ij}(\Ysp_1)=0,~~~~
R_{ab}(\Zsp)=0,
   \label{pl:Ricci:Eq}\\
&&h_r=h_0(t)+h_1(y, z),~~~~h_s=h_s(z)\,,
   \label{pl:h:Eq}\\
&& \left(\frac{dh_0}{dt}\right)^2
+N_r\left(1-\frac{4}{N_r}\right)^{-1}\Lambda_r=0
\,,~~~~
h_s^{4/N_s}\lap_{\Ysp_1}h_1+\triangle_{\Zsp}h_1=0,
   \label{pl:warp1-1:Eq}\\
&&\triangle_{\Zsp}h_s=0\,.
   \label{pl:warp1-2:Eq}
 }

Finally, we check the scalar field equation for the case of $p=p_r=0$.
Substituting Eqs.~(\ref{pl:ansatz:Eq}), (\ref{pl:warp:Eq}), and 
(\ref{pl:condition:Eq}) and 
the intersection rule $\chi=0$ into Eq.~(\ref{pl:scalar:Eq}),
we have
\Eqr{
&&\frac{\epsilon_rc_r}{N_r}h_r^{-b_r4/N_r}h_s^{-b_s4/N_s}\left[
-h_r^{-1}\frac{d^2h_0}{dt^2}+\left(1-\frac{4}{N_r}\right)
\left(\pd_t\ln h_r\right)^2
+N_r\Lambda_rh_r^{-2}\right]\nn\\
&&~~~~~~+\frac{\epsilon_rc_r}{N_r}h_r^{-1}\left(h_s^{4/N_s}\lap_{\Ysp_1}h_1
+\lap_{\Zsp}h_1\right)
+\frac{\epsilon_sc_s}{N_s}h_s^{-1}\lap_{\Zsp}h_s=0.
  \label{pl:scalar-e:Eq}
}
Hence the scalar field equation (\ref{pl:scalar-e:Eq}) reads 
\Eqrsubl{pl:scalar solution:Eq}{
&&\frac{d^2h_0}{dt^2}=0\,,~~~
\left(\frac{dh_0}{dt}\right)^2
+N_r\left(1-\frac{4}{N_r}\right)^{-1}\Lambda_r=0\,,
~~~h_s^{4/N_s}\lap_{\Ysp_1}h_1+\lap_{\Zsp}h_1=0,
   \label{pl:scalar solution1:Eq}\\
&&\lap_{\Zsp}h_s=0.
   \label{pl:scalar solution2:Eq}
}
These are consistent with the Einstein equations (\ref{pl:solution1:Eq})\,. 
The function $h_r$ can depend on the 
coordinate $t$ only if $N_r\ne 4$\,. For $N_r=4$, 
the scalar field equation leads to $\Lambda_r=0$\,. 

We can find the solution in which the
$p_s$-brane part depends on $x^{\mu}$. 
For $p=p_s=0$, $\Lambda_r=0$\,, and 
$\pd_th_r=0$, we have 
\Eqrsubl{pl:solution3:Eq}{
&&
R_{mn}(\Ysp_2)=0,~~~~R_{ab}(\Zsp)=0,
   \label{pl:Ricci3:Eq}\\
&&h_r=h_r(z),~~~~h_s=k_0(t)+k_1(v, z)\,,
   \label{pl:h3:Eq}\\
&&\frac{d^2k_0}{dt^2}=0\,,~~~
\left(\frac{dk_0}{dt}\right)^2
+N_s\left(1-\frac{4}{N_s}\right)^{-1}\Lambda_s=0\,,
~~~h_r^{4/N_r}\lap_{\Ysp_2}k_1+\triangle_{\Zsp}k_1=0,
   \label{pl:warp3-1:Eq}\\
&&\triangle_{\Zsp}h_r=0.
   \label{pl:warp3-2:Eq}
 }
It is clear that there is a solution for $k_0(t)$
such as $\partial_th_s\ne 0$ unless $N_s=4$. For $N_r=4$, 
the field equations lead to $\Lambda_r=0$\,. 

If $F_{\left(p_r+2\right)}=0$ and $F_{\left(p_s+2\right)}=0$,
the warp factors $h_1$ and $k_1$ are trivial functions. 
Then the $D$-dimensional spacetime is no longer warped
\cite{Binetruy:2007tu}. 
Moreover, Eqs.~(\ref{pl:solution1:Eq}) and (\ref{pl:solution3:Eq}) 
 imply the two cases. First, $p_r$-, $p_s$-branes are delocalized.  
 These are localized only along the overall transverse directions. 
Second, the 0-brane is completely localized on the $p_s$- (or $p_r$-) brane 
which is localized only along the overall transverse directions, which is 
a partially localized $p_r-0$ (or $0-p_s$) brane system.

As an example, we set
\Eq{
p=p_r=0,~~~~
\gamma_{ij}=\delta_{ij}\,,~~~
u_{ab}=\delta_{ab}\,,~~~
h_s=h_s(z)\,,
 \label{pl:flat metric:Eq}
 }
where $\delta_{ij}$ and $\delta_{ab}$ are the $p_s$-  
and $(D-p_s-1)$-dimensional Euclidean metrics,
respectively. Equation (\ref{pl:warp1-1:Eq}) gives 
\Eqr{
h_0(t)=c_0t+c_1\,,~~~~
c_0=\pm\sqrt{N_r\left(\frac{4}{N_r}-1\right)^{-1}\Lambda_r}\,,
  \label{pl:h0:Eq}
} 
where $c_0$ and $c_1$ are constants.
Hence, solutions exist
for $N_r<4$ if $\Lambda_r>0$
and vice versa.

If the functions $h_1$ and $h_s$ satisfy 
the coupled partial differential equations 
\Eq{
h_s^{4/N_s}\lap_{\Ysp_1}h_1+\lap_{\Zsp}h_1=0,~~~~\lap_{\Zsp}h_s=0\,,
    \label{pl:hrhs:Eq}
} 
the harmonic function $h_s$ that satisfies the equation in 
\eqref{pl:warp1-2:Eq} takes the form 
\Eqr{
h_s(z)=1+\sum_\ell\frac{M_\ell}{|z^a-z^a_\ell|^{D-p_s-3}}, 
}
where $z_\ell^a$ is the location of the $\ell$ th $p_s$-brane and $M_\ell$ 
is constant. Since we consider the case in which the $p_s$-branes 
coincide at the same location in the overall transverse directions, 
the harmonic function $h_s$ can be written by the following form 
\cite{Youm:1999zs, Youm:1999ti, Minamitsuji:2011jt}:
\Eq{
h_s(z)=\frac{M}{|z^a- z^a_0|^{D-p_s-3}}, 
}
where $M$ is constant and the stack of $p_s$-branes is located at 
the same points $z^a_0$ along the $z$ directions. 
We can find solutions for the harmonic function 
$h_1$ in the case where each of the $p_s$-branes does not 
coincide at the same location in the overall transverse directions. 

If we set $D-p_s\neq 3$ and $2-4N_s^{-1}\left(D-p_s-3\right)\neq 0$ 
for the overall transverse space, Eq.~\eqref{pl:hrhs:Eq} 
can be solved as 
\Eqr{
h_1(y, z)=1+\sum_{\ell}\frac{M_\ell}{\left[|y^i-y^i_\ell|^2
+\frac{4M^{4/N_s}}{\{2-4N_s^{-1}\left(D-p_s-3\right)\}^2}|z^a-z^a_0|
^{2-4N_s^{-1}\left(D-p_s-3\right)}\right]^{\zeta_r}}\,,
   \label{pl:h_r solution2:Eq}
}
where $M_\ell$ is constant and $\zeta_r$ is given by
\Eq{
\zeta_r=\frac{1}{2}\left[p_s-1+
\frac{(2-4N_s^{-1})\left(D-p_s-3\right)+2}
{2-4N_s^{-1}\left(D-p_s-3\right)}\right]\,.
}
Hence, the functions $h_r$ and $h_s$ can be expressed as  
\Eqrsubl{pl:solutions1:Eq}{
\hspace{-1cm}h_r(t, y, z)&=&c_0\,t+c_1\nn\\
&&+\sum_{\ell}\frac{M_\ell}{\left[|y^i-y^i_\ell|^2
+\frac{4M^{4/N_s}}{\{2-4N_s^{-1}\left(D-p_s-3\right)\}^2}|z^a-z^a_0|
^{2-4N_s^{-1}\left(D-p_s-3\right)}\right]^{\zeta_r}},
 \label{pl:solution-r:Eq}\\
\hspace{-1cm}h_s(z)&=&\frac{M}{|z^a-z^a_0|^{D-p_s-3}},
 \label{pl:solution-s:Eq}
}
where $c_0$, $c_1$, $M_\ell$\,, and $M$ are constant parameters 
and $y^i_\ell$ and $z^a_0$ are
constants representing the positions of the branes.
The curvature singularities appear at $h_r=0$ 
in the $D$-dimensional metric (\ref{pl:metric:Eq}).
Moreover, there is also a singularity at $z^a=z^a_0$ 
unless the scalar field is trivial.

Upon setting $D-p_s=3$ and $N_s=4$, 
the solutions of Eq.~(\ref{pl:hrhs:Eq}) are 
given by  
\Eqrsubl{pl:solutions1-2:Eq}{
h_r(t, y, z)&=&c_0t+c_1
+\sum_{\ell}\frac{M_\ell}{\left[|y^i-y^i_\ell|^2
+M|z^a-z^a_0|^2\right]^
{\frac{1}{2}(p_s+1)}},
 \label{pl:solution-r-2:Eq}\\
h_s(z)&=&M\ln|z^a-z^a_0|\,.
 \label{pl:solution-s-2:Eq}
}
In the case of $D-p_s=5$ and $N_s=4$, the functions $h_r$ and $h_s$ 
can be written by 
\Eqrsubl{pl:solutions1-3:Eq}{
h_r(t, y, z)&=&c_0t+c_1
+\sum_{\ell}M_\ell\left[|y^i-y^i_\ell|^2
-p_sM\ln |z^a-z^a_0|\right]\,,
 \label{pl:solution-r-3:Eq}\\
h_s(z)&=&\frac{M}{|z^a-z^a_0|^2}\,.
 \label{pl:solution-s-3:Eq}
}
The solutions \eqref{pl:solutions1-2:Eq} and 
\eqref{pl:solutions1-3:Eq} have 
a singular hypersurface at infinity
as well as at $h_r=0$\,, because the $D$-dimensional metric depends on  
the logarithmic function of the transverse coordinates.
These solutions also give a singularity at $z^a=z^a_0$
if the dilaton is nontrivial. 

It is possible to find the solution for 
$\pd_th_r=0$ and $\pd_th_s\ne 0$ if the roles of $\Ysp_1$ and 
$\Ysp_2$ are exchanged. The solution of the field equations for 
$D-p_r\neq3$ and $D-p_r\neq5$ can be written by 
\Eqrsubl{pl:solutions2:Eq}{
\hspace{-1cm}h_s(t, v, z)&=&c_0t+c_1\nn\\
&&+\sum_{\ell}\frac{M_\ell}{\left[|v^m-v^m_\ell|^2
+\frac{4M^{4/N_r}}{\{2-4N_r^{-1}\left(D-p_r-3\right)\}^2}|z^a-z^a_0|
^{2-4N_r^{-1}\left(D-p_r-3\right)}\right]^{\zeta_s}},
 \label{pl:solution2-s:Eq}\\
\hspace{-1cm}h_r(z)&=&\frac{M}{|z^a-z^a_0|^{D-p_r-3}}\,,
 \label{pl:solution2-r:Eq}
}
where $\zeta_s$ is given by
\Eq{
\zeta_s=\frac{1}{2}\left[p_r-1+
\frac{(2-4N_r^{-1})\left(D-p_r-3\right)+2}
{2-4N_r^{-1}\left(D-p_r-3\right)}\right]\,.
}
If we set $D-p_r=3$, $D-p_r=5$\,, and $N_r=4$, the harmonic functions 
$h_r$ and $h_s$ have 
logarithmic spatial dependence like \eqref{pl:solutions1-2:Eq} and 
\eqref{pl:solutions1-3:Eq}.

Assuming $\Lambda_r>0$ and
introducing a new time coordinate $\tau$ by 
\Eq{
\frac{\tau}{\tau_0}=
\left(c_0t+c_1\right)^{\frac{(N_r-2)(D-2)+2}{N_r(D-2)}}\,,~~~~~
\tau_0=\frac{N_r(D-2)}{c_0\left[(N_r-2)(D-2)+2\right]},
}
we find the $D$-dimensional metric (\ref{pl:metric:Eq}) as
\Eqr{
ds^2&=&
\left[1+\left(\frac{\tau}{\tau_0}\right)^{-\frac{N_r(D-2)}{(N_r-2)(D-2)+2}}h_1
\right]^{-\frac{4(D-3)}{N_r(D-2)}}h_s^{a_s}
\left[-d\tau^2
\right.\nn\\
&&\left.\hspace{-1cm}
+\left\{1+\left(\frac{\tau}{\tau_0}\right)^
{-\frac{N_r(D-2)}{(N_r-2)(D-2)+2}}h_1\right\}^{\frac{4}{N_r}}
\left(\frac{\tau}{\tau_0}\right)^{\frac{4}{(N_r-2)(D-2)+2}}\left(
\gamma_{ij}dy^idy^j+h_s^{\frac{4}{N_s}}u_{ab}dz^adz^b\right)\right].\nn\\
 \label{pl:s-metric:Eq}
 }
Since $h_s$ does not approach constant in any region,
the whole spacetime cannot be homogeneous and isotropic.
But on each $z^a={\rm const}$ slice
the spacetime becomes
a homogeneous and isotropic universe. 
In the limit $\tau\rightarrow\infty$, 
the function $h_1$ can be negligible in the warp factor.
This is guaranteed by a scalar field with
the exponential potential.
The accelerating universe is obtained
on each $z^a={\rm const}$
slice if $N_r<2$,
which corresponds to the case of a positive cosmological constant.
For $\frac{2(D-3)}{D-2}<N_r<2$,
the solution provides a power-law inflationary universe,
and for $N_r>\frac{2(D-3)}{D-2}$,
the scale factor diverges at 
$\tau=\tau_{\infty}>0$,
taking the involution $\tau\to \tau_\infty-\tau$. 
Finally, for  $N_r=\frac{2(D-3)}{D-2}$,
we obtain a de Sitter universe
which will be discussed in the next subsection.

\subsection{de Sitter universe}

Next, we consider the solution with a dilaton which is 
the case of $c_I=0~(I=r~{\rm or}~s)$. 
In terms of $c_I=0$, Eq.~(\ref{pl:c:Eq}) gives 
\Eq{
N_I=\frac{2(D-p_I-3)(p_I+1)}{(D-2)}.
   \label{ds:N:Eq}
}
If we assume 
\Eqr{
c_r=0\,,~~~c_s\ne 0\,,~~~p=p_r=0\,,~~~
N_r=\frac{2(D-3)}{(D-2)}\,,~~~
\alpha_r=-\frac{N_sa_s}{2\epsilon_sc_s}\,,~~~\Lambda_s=0\,,
}
the field equations reduce to
\Eqrsubl{ds:equations:Eq}{
&&
R_{ij}(\Ysp_1)=0,~~~~~R_{ab}(\Zsp)=0,\\
&&
h_r(t, y, z)=h_0(t)+h_1(y, z),~~~~~
\left(\frac{dh_0}{dt}\right)^2
-\frac{2(D-3)^2}{(D-2)(D-1)}\Lambda_r=0,
\label{ds:warp1:Eq}\\
&&
h_s^{4/N_s}\lap_{\Ysp_1}h_1+\triangle_{\Zsp}h_1=0,~~~~~\triangle_{\Zsp}h_s=0.
   \label{ds:warp2:Eq}
   }
Then Eq.~(\ref{ds:warp1:Eq}) gives 
\Eq{
h_0=c_0t+c_1\,,
}
where $c_1$ is an integration constant and $c_0$ is given by
\Eq{
c_0=\pm(D-3)\sqrt{\frac{2}{(D-2)(D-1)}\Lambda_r}\,.
\label{ds:c0:Eq}
}
Thus there is no solution for $\Lambda_r<0$.
If the metric $u_{ab}({\rm Z})$ is assumed to be 
Eq.~(\ref{pl:flat metric:Eq}), 
the function ${h}_1$ is given by Eq.~(\ref{pl:h_r solution2:Eq}). 
Now we introduce a new time coordinate $\tau$ by 
\Eq{
c_0\tau=\ln t,
}
where we have set $c_0>0$ for simplicity. 
Then the $D$-dimensional metric 
(\ref{pl:metric:Eq}) can be expressed as
\Eqr{
ds^2&=&h_s^{a_s}\left[-\left(1+c_0^{-1}
\e^{-c_0\tau}h_1\right)^{-2}d\tau^2
+\left(1+c_0^{-1}\e^{-c_0\tau}h_1\right)^{2/(D-3)}
\left(c_0\e^{c_0\tau}\right)^{2/(D-3)}
\right.\nn\\
&&\left.\times\left\{\gamma_{ij}(\Ysp_1)dy^idy^j
+h_s^{4/N_s}u_{ab}(\Zsp)dz^adz^b\right\}\right].
   \label{ds:metric:Eq}
}
The function $h_s$ does not become constant in any region. Then, 
the $D$-dimensional spacetime cannot be de Sitter spacetime.
However, the spacetime gives a homogeneous and isotropic 
universe on each $y^i={ \rm const}$, $z^a={ \rm const}$ slice.
If we set $h_s=$const and $h_1=h_1(z)$, 
Eq.~(\ref{ds:metric:Eq}) becomes the solution which has been discussed by 
Refs.~\cite{Maki:1992tq, Maki:1994wc} (see also \cite{Ivashchuk:1996zv}). 
Furthermore, for $D=4$ and by setting $h_s=1$, 
the solution is the Kastor-Traschen one \cite{Kastor:1992nn}.


\section{The intersection involving $n$ brane backgrounds}
  \label{sec:cn}

The construction that we have analyzed in Sec.~\ref{sec:pl} 
is a special case of a more general construction of intersecting branes 
with a cosmological constant. In effect, we have been studying the special
case of intersections involving a two-brane. The time-dependent brane 
with a cosmological constant property is a 0-brane, represented by a 
2-form. To describe more general intersections on a time-dependent 
background, one simply incorporates additional branes
in a dynamical background. Without loss of the time dependence, it is
possible to also add $n$ delocalized branes. This also has one  
important further refinement. Instead of power-law expansion,
the support of a 0-brane might be accelerated expansion, 
where $D$-dimensional geometry is an asymptotically de Sitter spacetime.  
The $n$ intersection allows the time dependence of only 0-branes but
not of the $p$-branes ($p\ne 0$). 
The reason for this is that the time-dependent brane we have obtained 
can be performed in the case of $\chi=0$, where  
$\chi$ is defined by \eqref{pl:chi:Eq}. 
So the coefficient of the time dependence is simply proportional 
to the cosmological constant that we have explored in Sec~\ref{sec:pl}: 
the Einstein equations give $p=0$. 

In this section, we discuss the intersection of the 
delocalized $n$ branes 
in the higher-dimensional gravity theory with the cosmological constants. 
The general action describing the intersection involving the 
$n$ brane system is given by 
\Eqr{
S&=&
\frac{1}{2\kappa^2}\int \left[\left(R-2\sum_I\e^{\alpha_I\phi}
  \Lambda_I\right)\ast{\bf 1}_D
 -\frac{1}{2}\ast d\phi \wedge d\phi \right.\nn\\
&&\left.
 -\frac{1}{2}\sum_I \frac{\e^{\epsilon_Ic_I\phi}}{(p_I+2)!}\ast F_{(p_I+2)}
 \wedge F_{(p_I+2)}\right],
\label{cn:action:Eq}
}
where $\kappa^2$ denotes the $D$-dimensional gravitational constant, 
$R$ is the $D$-dimensional Ricci scalar constructed from 
the $D$-dimensional metric $g_{MN}$, $\phi$ is a scalar field, 
$F_{(p_I+2)}$ is the antisymmetric tensor fields of rank $(p_I+2)$, 
${\ast}$ is the Hodge dual operator in the $D$-dimensional spacetime, and 
$c_I$ and $\epsilon_I$ are constants defined by
\Eqrsubl{cn:p:Eq}{
c_I^2&=&N_I-\frac{2(p_I+1)(D-p_I-3)}{D-2}, 
\label{cn:c:Eq}\\
\epsilon_I&=&\left\{
\begin{array}{cc}
 + &~{\rm for~the~electric~brane}\,,
\\
 - &~~~{\rm for~the~magnetic~brane}\,.
\end{array} \right. 
\label{cn:epsilon:Eq}
}
Here $I$ denotes the type of the corresponding branes. 

The $D$-dimensional action (\ref{cn:action:Eq}) give the field equations
\Eqrsubl{cn:field eq:Eq}{
&&\hspace{-1cm}R_{MN}=\frac{2}{D-2}\sum_I\e^{\alpha_I\phi}\Lambda_I\,g_{MN} 
+\frac{1}{2}\pd_M\phi \pd_N \phi\nn\\
&&+\frac{1}{2}\sum_I\frac{1}{(p_I+2)!}\e^{\epsilon_Ic_I\phi}
\left[(p_I+2)F_{MA_2\cdots A_{p_I+2}} {F_N}^{A_2\cdots A_{p_I+2}}
-\frac{p_I+1}{D-2}g_{MN} F^2_{(p_I+2)}\right],
\label{cn:Einstein:Eq}\\
&&\hspace{-1cm}
\lap\phi-2\sum_I\alpha_I\e^{\alpha_I\phi}\Lambda_I
-\frac{1}{2}\sum_I\frac{\epsilon_Ic_I}{(p_I+2)!}
\e^{\epsilon_Ic_I\phi}F^2_{(p_I+2)}=0\,,
 \label{cn:scalar:Eq}\\
&&\hspace{-1cm}
d\left[\e^{\epsilon_Ic_I\phi}\ast F_{(p_I+2)}\right]=0\,, 
   \label{cn:gauge:Eq}
}
where $\lap$ denotes the Laplace operator with respect to the 
$D$-dimensional metric $g_{MN}$\,. 

We adopt the ansatz that $D$-dimensional metric can be written by 
\Eqr{
\hspace{-10mm}
ds^2=-A(t, z)dt^2
+\sum_{\alpha=1}^pB^{(\alpha)}(t, z)(dx^{\alpha})^2 
+C(t, z)u_{ab}(\Zsp) dz^a dz^b,
 \label{cn:metric:Eq}
}
where $u_{ab}(\Zsp)$ denotes the metric of the 
$(D-p-1)$-dimensional $\Zsp$ space which depends only on 
the $(D-p-1)$-dimensional coordinates $z^a$. 
Concerning the functions $A$, $B^{(\alpha)}$ and $C$, 
we assume 
\Eq{
A=\prod_I\left[h_I(t,z)\right]^{a_I},~~~
B^{(\alpha)}=\prod_I\left[h_I(t,z)\right]^{\delta^{(\alpha)}_I},~~~
C=\prod_I\left[h_I(t,z)\right]^{b_I}\,,
}
where the constants $a_I$, $b_I$ and $\delta^{(\alpha)}_I$ 
are given by
\Eq{
a_I=-\frac{4(D-p_I-3)}{N_I(D-2)},~~~~b_I=\frac{4(p_I+1)}{N_I(D-2)},~~~~
\delta^{(\alpha)}_I=\left\{
\begin{array}{cc}
a_I&~{\rm for}~~\alpha\in I\,,\\ b_I &~{\rm for}~~\alpha
\in \hspace{-.8em}/ I\,.
\end{array} \right. 
 \label{cn:parameter:Eq}
}
The function $h_I(t,z)$ is a straightforward generalization 
of the static solution associated with the brane $I$ 
in an intersecting brane system~\cite{Argurio:1997gt, Argurio:1998cp}
to the dynamical one. 

We further require that the dilaton $\phi$ and the form fields 
$F_{(p+2)}$ satisfy the following conditions
\Eq{
\e^{\phi}=\prod_Ih_I^{2\epsilon_Ic_I/N_I},~~~
F_{(p_I+2)}=\frac{2}{\sqrt{N_I}}d(h_I^{-1})\wedge\Omega(\Xsp_I)\,,
  \label{cn:fields:Eq}
}
where ${\Xsp}_I$ is the space associated with the brane $I$, 
and the volume $(p_I+1)$-form $\Omega(\Xsp_I)$ is written by 
\Eq{
\Omega(\Xsp_I)=dt\wedge dx^{p_1}\wedge \cdots \wedge
dx^{p_I}\,.
}

\subsection{Power-law expanding universe}
Firstly, we consider the Einstein equations (\ref{cn:Einstein:Eq}) with 
$c_I\neq 0~(I=0,\cdots,~n-1)$. 
We assume that the parameters $\alpha_I~(I=0,\cdots,~n-1)$ are given by 
\Eqr{
\alpha_I=-\epsilon_I c_I.
    \label{cn:alpha:Eq}
}

We impose the condition with respect to the components of 
$D$-dimensional metric~\cite{Argurio:1998cp}
\Eq{
A^{(D-p-3)}
 \prod_{\alpha=1}^p\,B^{(\alpha)} \,C=1 \,,~~~~~~~
A^{-1}\prod_{\alpha \in I}\left(B^{(\alpha)}\right)^{-1}
\e^{\epsilon_Ic_I\phi}=h^2_I\,.
   \label{cn:extremal:Eq}
}
The Einstein equations \eqref{cn:Einstein:Eq} become 
\Eqrsubl{cn:cEinstein:Eq}{
&&\hspace{-0.5cm}
\sum_{I,I'}\left(\frac{2}{N_I}\delta_{II'}-M_{II'}
\right)\pd_t\ln h_I\pd_t\ln h_{I'}
+\frac{2}{D-2}\sum_I\Lambda_Ih_I^{-2+a_Ip_I}\prod_{I'\ne I}
h_{I'}^{a_{I'}-\frac{2\varepsilon_I\varepsilon_{I'}c_Ic_{I'}}{N_{I'}}}\nn\\
&&~~~~+\frac{1}{2}\sum_{I}b_I\left[\left(1-\frac{4}{N_I}\right)\pd_t\ln h_I
-\sum_{I'\ne I}\frac{4}{N_{I'}}\pd_t\ln h_{I'}\right]\pd_t\ln h_I
\nn\\
&&~~~~-\frac{1}{2}\sum_{I}\left(\frac{4}{N_I}+b_I\right)h_I^{-1}\pd_t^2h_I
+\frac{1}{2}\prod_{I'}h_{I'}^{-4/N_{I'}}\sum_{I}a_{I}
h_{I}^{-1}\lap_{\Zsp}h_{I}=0,
   \label{cn:cEinstein-tt:Eq}\\
&&\hspace{-0.5cm}\sum_I\frac{2}{N_{I}}h^{-1}_I\pd_t\pd_ah_I+\sum_{I,I'}
\left(M_{II'}-\frac{2}{N_{I'}}\delta_{II'}
\right)\pd_t\ln h_I\pd_a\ln h_{I'}=0,
   \label{cn:cEinstein-ti:Eq}\\
&&\hspace{-0.5cm}\prod_{J'}h_{J'}^{-a_{J'}}
\sum_{\gamma}\prod_{J}h_{J}^{\delta_J^{(\gamma)}}
\sum_{I} \delta^{(\gamma)}_I\left[h_I^{-1}\pd_t^2h_I
-\left\{\left(1-\frac{4}{N_I}\right)\pd_t\ln h_I
\right.\right.\nn\\
&&\left.\left.~~~~-\sum_{I'\ne I}\frac{4}{N_{I'}}
\pd_t\ln h_{I'}\right\}\pd_t\ln h_I\right]
-\prod_{J'}h_{J'}^{-b_{J'}}
\sum_{\gamma}\prod_{J}h_{J}^{\delta_J^{(\gamma)}}\sum_I
\delta^{(\gamma)}_Ih_I^{-1}\lap_{\Zsp}h_I\nn\\
&&~~~~-\frac{4}{D-2}\sum_I\Lambda_I
h_I^{-2+\delta_I^{(\gamma)}p_I}\prod_{I'\ne I}
h_{I'}^{\delta_{I'}^{(\gamma)}
-\frac{2\varepsilon_I\varepsilon_{I'}c_Ic_{I'}}{N_{I'}}}=0,
   \label{cn:cEinstein-ab:Eq}\\
&&\hspace{-0.5cm}
R_{ab}(\Zsp)+\frac{1}{2}u_{ab}\prod_Jh_J^{4/N_J}\sum_{I}
b_I\left[ h_I^{-1}\pd_t^2h_I
-\left\{\left(1-\frac{4}{N_I}\right)\pd_t\ln h_I
-\sum_{I'\ne I}\frac{4}{N_{I'}}\pd_t\ln h_{I'}\right\}\pd_t\ln h_I\right]\nn\\
&&~~~~-\frac{1}{2}u_{ab}\sum_Ib_I h_I^{-1}\lap_{\Zsp}h_I
-\sum_{I,I'}\frac{2}{N_I}\left(M_{II'}-\frac{2}{N_{I'}}\delta_{II'}
\right)\pd_a\ln h_I\pd_b\ln h_{I'}\nn\\
&&~~~~-\frac{2}{D-2}\sum_I\Lambda_Ih_I^{-2+a_Ip_I+\frac{4}{N_I}}\prod_{I'\ne I}
h_{I'}^{a_{I'}-\frac{2(\varepsilon_I\varepsilon_{I'}c_Ic_{I'}-2)}{N_{I'}}}
u_{ab}=0,
    \label{cn:cEinstein-ab2:Eq}
}
where $R_{ab}(\Zsp)$ is the Ricci tensor with respect to  
the metric $u_{ab}(\Zsp)$, and $M_{II'}$ is defined by
\Eqr{
&&M_{II'}\equiv \frac{1}{4}\left[a_Ia_{I'}
+\sum_{\alpha}\delta^{(\alpha)}_I\delta^{(\alpha)}_{I'}
+(D-p-3)b_Ib_{I'}\right] 
+\frac{2}{N_IN_{I'}}\epsilon_I\epsilon_{I'}c_Ic_{I'}
\,.
\label{cn:M:Eq}
}
The Eq.~(\ref{cn:cEinstein-ti:Eq}) can be rewritten as
\Eq{
\sum_{I,I'}\left[M_{II'}+\frac{2}{N_{I}} \delta_{II'}
\frac{\pd_t\pd_a \ln h_I}{\pd_t \ln h_I \pd_a \ln h_I} \right]
\pd_t \ln h_I \pd_a \ln h_{I'}=0.
\label{cn:cEinstein-ti2:Eq}
}
One can find that the equation (\ref{cn:cEinstein-ti2:Eq}) 
is equivalent to satisfying that 
\Eq{
\frac{\pd_t\pd_a \ln h_I}{\pd_t \ln h_I \pd_a \ln h_I}=k_I
\,.
\label{cn:cEinstein-ti3:Eq}
}
Then we have
\Eq{
M_{II'}+\frac{2}{N_{I}}k_I \delta_{II'}=0.
\label{cn:cEinstein-ti4:Eq}
}
Eqs.~(\ref{cn:c:Eq}), (\ref{cn:parameter:Eq}) and (\ref{cn:cEinstein-ti2:Eq}) 
give 
\Eqr{
M_{II} &=&\frac{1}{4}\left[(p_I+1)a_I^2 + (p-p_I)b_I^2
+(D-p-3)b_I^2\right] +\frac{2}{N_I^2} c_I^2 \nonumber \\
&=& \frac{2}{N_I}.
  \label{cn:M2:Eq}
}
Combining the \eqref{cn:M2:Eq} with (\ref{cn:cEinstein-ti4:Eq}) 
the constant $k_I$ in~\eqref{cn:cEinstein-ti4:Eq} is $k_I=-1$, which 
implies 
\Eq{
M_{II'}= \frac{2}{N_{I'}}\delta_{II'}.
\label{cn:M3:Eq}
}
Taking account of these results, the equation (\ref{cn:cEinstein-ti:Eq}) 
yields
\Eq{
\pd_t \pd_a[h_I(t,z)]=0
\,.
}
Hence we find 
\Eq{
h_I(t, z)= K_I(t)+H_I(z)\,.
  \label{cn:warp:Eq}
}

For $I\neq I'$, ~\eqref{cn:M3:Eq} provides the intersection rule
on the dimension $\bar{p}$ of the intersection for each pair of branes
$I$ and $I'$ $(\bar{p}\leq p_I, p_{I'})$~\cite{Ohta:1997gw, Youm:1997hw}:
\Eq{
\bar{p}=\frac{(p_I+1)(p_{I'}+1)}{D-2}-1-
\frac{1}{2}\epsilon_Ic_I\epsilon_{I'}c_{I'}.
\label{cn:rule:Eq}
}

Under the assumptions given above, we next reduce the gauge field equations.
In terms of the ansatz~\eqref{cn:fields:Eq}, the Bianchi identity 
$dF_{(p_I+2)}=0$ is automatically satisfied; 
\Eq{
h_I^{-1}(2\pd_a \ln h_I \pd_b \ln h_I
+ h_I^{-1}\pd_a\pd_b h_I)
dz^a\wedge dz^b\wedge\Omega(\Xsp_I)=0.
}
Utilising \eqref{cn:fields:Eq}, the gauge field equation becomes
\Eqr{
d\left[\pd_aH_I \left(\ast_{\Zsp}dz^a\right)\wedge
\ast_{\Xsp}\Omega(\Xsp_I)\right]=0,
    \label{cn:gauge2:Eq}
 }
where we used Eqs.~(\ref{cn:extremal:Eq}), (\ref{cn:warp:Eq}), and 
$\ast_{\Xsp}$, $\ast_{\Zsp}$ are the Hodge dual
operators on $\Xsp(\equiv \cup_{I}X_I)$ and $\Zsp$, respectively.
Hence, \eqref{cn:gauge2:Eq} gives (\ref{cn:warp:Eq}) 
and we find
\Eq{
\lap_{\Zsp}H_I=0.
\label{cn:gauge3:Eq}
}
The roles of the Bianchi identity and field equations are 
interchanged for magnetic ansatz~\cite{Ohta:1997gw, Youm:1997hw, 
Argurio:1998cp}. Then the net result is the same.

In order to complete the system of equations, 
we must also consider the scalar field equation.
Substituting the ansatz for fields 
\eqref{cn:fields:Eq}, 
and the metric (\ref{cn:metric:Eq}), (\ref{cn:warp:Eq}), 
the equation of motion for the scalar field
\eqref{cn:scalar:Eq} reduces to
\Eqr{
&&-\prod_{I''}h_{I''}^{-a_{I''}}
\sum_{I}\frac{1}{N_I}\epsilon_Ic_I\left[h_I^{-1}\frac{d^2K_I}{dt^2}
-\left\{\left(1-\frac{4}{N_I}\right)\pd_t\ln h_I-
\sum_{I'\ne I}\frac{4}{N_{I'}}\pd_t\ln h_{I'}\right\}\pd_t\ln h_I
\right.\nn\\
&&\left.~~~~~-N_I\Lambda_Ih_I^{-2}\right]
+\prod_{I''}h_{I''}^{-b_{I''}}\sum_I\frac{1}{N_I}h_I^{-1}
\epsilon_Ic_I\lap_{\Zsp} H_I=0.
  \label{cn:scalar2:Eq}
}
Furthermore, \eqref{cn:scalar2:Eq} reads
\Eqrsubl{cn:enough:Eq}{
&&\hspace{-1cm}\frac{d^2K_I}{dt^2}=0,
\label{cn:KI:Eq}
\\
&&\hspace{-1cm}\triangle_{\Zsp}H_I=0,
   \label{5}\\
&&\hspace{-1cm}\sum_{I}\frac{\epsilon_Ic_I}{N_I}\left[
\left\{\left(1-\frac{4}{N_I}\right)\pd_t\ln h_I-
\sum_{I'\ne I}\frac{4}{N_{I'}}\pd_t\ln h_{I'}\right\}\pd_t\ln h_I
+N_I\Lambda_Ih_I^{-2}\right]=0\,.
\label{cn:scalar3:Eq}
}
From Eq.~(\ref{cn:KI:Eq}), we obtain
\Eq{
K_I=A_I t+B_I, 
}
where $A_I$ and $B_I$ are constants.

\subsubsection{The intersection involving the same brane}
Let us first consider the case that all cosmological constants become 
nonvanishing. If we set $\Lambda_I\ne 0$, the field equations imply 
that all functions are equal:
\Eq{
h_I(t,z)=K(t, z)=K_0(t)+K_1(z),~~~~N_I=N_{I'}\equiv N.
    \label{sa:h:Eq}
}
We can find the solutions if the function $h$ and $N$ satisfy
\Eq{
K_0(t)=A\, t+B,~~~~A=\pm\sqrt{N_I\Lambda_I/
\sum_I\left(\frac{4}{N_I}-1\right)}\,,
\label{sa:K:Eq}
}
where $B$ denotes a constant.
Then the remaining
Einstein equations~(\ref{cn:cEinstein:Eq}) are 
\Eq{
R_{ab}(\Zsp)=0.
   \label{sa:Ricci:Eq}
}
Now we set 
\Eq{
\quad u_{ab}=\delta_{ab}\,,
 \label{sa:flat:Eq}
}
where $\delta_{ab}$ is the $(D-p-1)$-dimensional Euclidean metric.
In this case, the solution for $h_I$ can be obtained explicitly as
\Eq{
K(t, z)=At+B +\sum_{k}\frac{M_{k}}{|z^a- z^a_{k}|^{D-p-3}},
  \label{sa:exact:Eq}
}
where  $M_{k}$'s are constant parameters and $z^a_{k}$
represents the positions of the branes in Z space.  
If the functions $h_I$ coincide, the locations of the  
$p_I$-brane will also coincide

In this case, all branes have the same 
total amount of charge at the same position. 

Let us consider the intersection rule in the $D$-dimensional
gravity theory. If we set $p_I=\tilde{p}$ for all $p_I$, 
the intersection rule (\ref{cn:rule:Eq}) leads to
\Eq{
\bar{p}=\tilde{p}-\frac{N}{2}.
    \label{sa:chi2:Eq}
}
Then, we find the intersection involving two $\tilde{p}$-branes:
\Eq{
\tilde{p}\cap \tilde{p}=\tilde{p}-\frac{N}{2}.
   \label{sa:int:Eq}
}
Since the number of intersections for $\tilde{p}<\frac{N}{2}$ is negative, 
there is no solution in these brane backgrounds.

If we choose $K_0=0 ~(A=B=0$), 
the metric describes the known static and extremal multi-black-hole
solution with black hole charges
$M_{k}$~\cite{Argurio:1997gt, Ohta:1997gw, Youm:1997hw, Argurio:1998cp}.

\subsubsection{A dynamical brane in the intersecting brane system}

In the following, we consider the case 
that there is only one function $h_I$ which depends on both 
$z^a$ and $t$.  We denote it with the 
subscript $\tilde{I}$, 
while other functions of $I'\neq \tilde{I}$
 are either dependent on $z^a$ or constant. 
If we assume $N_{\tilde{I}}\ne 4$\,, we have  
\Eqr{
\pd_th_{I'}=0\,,~~~p_{\tilde{I}}=0\,,~~~\Lambda_{I'}=0\,,~~~
{\rm for}~I'\ne {\tilde{I}}.
    \label{cn:condition:Eq}
}

We can find the solutions if the function $h_{\tilde{I}}$ and 
$N_{\tilde{I}}$ satisfy
\Eqr{
\hspace{-0.3cm}
h_{\tilde{I}}(t, z)= K_{\tilde{I}}(t)+H_{\tilde{I}}(z),~~~
K_{\tilde{I}}(t)=\pm\left[\left(\frac{4}{N_{\tilde{I}}}-1\right)^{-1}
N_{\tilde{I}}\Lambda_{\tilde{I}}\right]^{\frac{1}{2}}\,t+c_{\tilde{I}}\,,
~~~N_{\tilde{I}}\ne 4,
\label{cn:h:Eq}
}
where $c_{\tilde{I}}$ is constant. 
Then the remaining
Einstein equations~\eqref{cn:cEinstein:Eq} are 
\Eq{
R_{ab}(\Zsp)=0.
   \label{cn:Ricci:Eq}
}
Now we set 
\Eq{
\quad u_{ab}=\delta_{ab}\,,
 \label{cn:flat metric:Eq}
}
where $\delta_{ab}$ is the $(D-p-1)$-dimensional Euclidean metric.
In this case, the solution for $h_I$ can be written explicitly as
\Eqrsubl{cn:hI:Eq}{
h_{\tilde{I}}(t, z)&=&\pm\left[\left(\frac{4}{N_{\tilde{I}}}-1\right)^{-1}
N_{\tilde{I}}\Lambda_{\tilde{I}}\right]^{\frac{1}{2}}t
+\tilde{c}_{\tilde{I}}
+\sum_{k}\frac{M_{\tilde{I}, k}}{|z^a- z^a_{k}|^{D-p-3}}\,,\\
h_{I'}(z)&=&\tilde{c}_{I'}
+\sum_{l}\frac{M_{I',\,l}}{|z^a-z^a_{l}|^{D-p-3}}\,,
}
where $\tilde c_{\tilde{I}}$, $\tilde{c}_{\tilde{I}}$, $M_{\tilde{I}, k}$, 
and $M_{I',\,l}$ are constant parameters and 
$z^a_{k}$ and $z^a_{l}$ denote the positions of the branes in Z space.
$N_{\tilde I}<4$ leads to $\Lambda_{\tilde I}>0$
and vice versa. 
Since the functions $h_I$ coincide, the locations of the 
$p_I$-brane will also coincide
This physically means that all branes have the same 
total amount of charge at the same position. 
Here we have discussed the solution without compactification of $\Zsp$ space.
If we consider the case that $q$ dimensions of $\Zsp$ space are
smeared, we can find the different power of harmonics, i.e.
${|z^a-z^a_{k}|^{-(D-p-3-q)}}$ ($q\leq D-p-2$).

For $K_{\tilde{I}}=0 ~(A=B=0$), 
the solution describes the known static and 
extremal multi-black-hole
solution with black hole charges
$M_{\tilde{I}, k}$~\cite{Ohta:1997gw, Youm:1997hw, Argurio:1998cp}.
We can find the solution (\ref{cn:hI:Eq}) for any $N_I\ne 4$. 
If we choose ${N}_I=4$, 
the solutions have already discussed in 
Ref.~\cite{Maeda:2009zi}.

Let us consider the intersection rule in the $D$-dimensional gravity theory.
If we choose $p_{\tilde{I}}=\tilde{p}=0$ for all 
$p_{\tilde{I}}\ne p_{I'}$, 
the intersection rule (\ref{cn:rule:Eq}) leads to
\Eq{
\frac{p_{I'}+1}{D-2}-1-
\frac{1}{2}\epsilon_{\tilde{I}}c_{\tilde{I}}\epsilon_{I'}c_{I'}=0.
    \label{cn:chi2:Eq}
}

Now we discuss the application of 
the time-dependent solutions to study
the cosmology. We assume an isotropic and homogeneous three-space
in the Friedmann-Robertson-Walker (FRW) universe
after compactification.

We set the $(D-p-1)$-dimensional 
Euclidean space with $u_{ab}(\Zsp)=\delta_{ab}(\Zsp)$ and 
consider the case that there is only one function $h_I$ 
depending on both $z^a$ and $t$, which we denote it with the 
subscript $\tilde{I}$, and other functions are either dependent on
$z^a$ or constant.
If we assume $N_{\tilde{I}}\ne 4$, 
the $D$-dimensional metric can be expressed as 
\begin{eqnarray}
ds^2&=&-\prod_{I\ne \tilde{I}}h_I^{a_I}
\left[1+\left(\frac{\tau}{\tau_0}\right)^{-\frac{2}{a_{\tilde{I}}+2}}
H_{\tilde{I}}\right]^{a_{\tilde{I}}}d\tau^2\nn\\
&&
+\sum_{\alpha=1}^p\prod_{I\ne \tilde{I}}h_I^{\delta_I^{(\alpha)}}
\left[1+\left(\frac{\tau}{\tau_0}\right)
^{-\frac{2}{a_{\tilde{I}}+2}}H_{\tilde{I}}\right]^
{\delta_{\tilde{I}}^{(\alpha)}}
\left(\frac{\tau}{\tau_0}\right)^{\frac{2\delta_{\tilde{I}}^{(\alpha)}}
{a_{\tilde I}+2}}
\left(dx^{\alpha}\right)^2\nn\\
&&+\prod_{I\ne \tilde{I}}h_I^{b_I}
\left[1+\left(\frac{\tau}{\tau_0}\right)^{-\frac{2}{a_{\tilde{I}}+2}}
H_{\tilde{I}}\right]^{b_{\tilde{I}}}
\left(\frac{\tau}{\tau_0}\right)^{\frac{2b_{\tilde{I}}}{a_{\tilde{I}}+2}}
\delta_{ab}(\Zsp)dz^adz^b,
   \label{cn:metric-J:Eq}
\end{eqnarray}
where the function $H_{\tilde{I}}$ is given by 
\Eq{
H_{\tilde{I}}=\sum_{k}\frac{M_{\tilde{I},\,k}}
{|z^a- z^a_{k}|^{D-p-3}}\,, 
\label{cn:HI:Eq}
}
and the cosmic time $\tau$ defined by
\Eq{
\frac{\tau}{\tau_0}=\left(At\right)^{(a_{\tilde{I}}+2)/2},~~~~\tau_0=
\frac{2}{\left(a_{\tilde{I}}+2\right)A}.
  \label{cn:cosmic:Eq}
}
If we can regard the three-dimensional part of the overall 
transverse space Z as our Universe, 
the power of the scale factor in the fastest expanding case is 
expressed as 
\Eq{
\lambda=\frac{b_{\tilde{I}}}{a_{\tilde{I}}+2}=\left[-D+3
+\frac{N_{\tilde{I}}}{2}(D-2)\right]^{-1},
~~~~~~{\rm for}~~D>2,
  \label{cn:power:Eq}
}
where we used the $D$-dimensional metric 
(\ref{cn:metric-J:Eq}).   
Hence, we cannot find the cosmological model which
exhibits an accelerating expansion 
of our Universe. 
On the other hand, if our three-space is given by a 
three-dimensional subspace in relative transverse space, 
the power of the scale factor in the fastest expanding case is also 
given by \eqref{cn:power:Eq}. 

By taking $\tau\to \tau_\infty-\tau$, where $\tau_\infty$ is constant, 
we have accelerated expansion for $\tau_\infty>\tau$ and $\lambda<0$. 
This is equivalent to
\Eq{
N_{\tilde{I}}>2\,,~~~D>2-\frac{2}{N_{\tilde{I}}-2}\,,~~~~
{\rm or}~~~~
N_{\tilde{I}}<2\,,~~~3<D<2-\frac{2}{N_{\tilde{I}}-2}\,,
  \label{cn:condition1:Eq}
}
for $D>3$\,. 
However, the scale factor of our Universe diverges at 
$\tau=\tau_\infty$. 

On the other hand, the power of the scale factor in the fastest 
expanding case is automatically positive for 
$D=3$ and $N_{\tilde{I}}>0$\,.

Next we discuss the cosmological solution in the lower-dimensional 
effective theories. 
We compactify $d(\equiv \sum_{\alpha}d_{\alpha}+d_z)$ 
dimensions to give our Universe, 
where $d_{\alpha}$ and $d_z$ denote the compactified dimensions with
respect to the relative and overall transverse space, respectively.
The $D$-dimensional metric (\ref{cn:metric:Eq}) is written by
\Eq{
ds^2=ds^2(\Msp)+ds^2(\Nsp)
\label{cn:metric2:Eq}, 
}
where $ds^2(\Msp)$ is a $(D-d)$-dimensional metric and
$ds^2(\Nsp)$ is a metric of compactified dimensions.

In order to discuss the dynamics of the $(D-d)$-dimensional universe 
in the Einstein frame, we use the conformal transformation
\Eq{
ds^2(\Msp)=h_{\tilde{I}}^{B_{\tilde{I}}}\prod_{I\ne \tilde{I}}
h_I^{C_I}ds^2(\bar{\Msp}),
}
where $B_{\tilde{I}}$ and $C_I$ are expressed, respectively,  as 
\Eq{
B_{\tilde{I}}=-\frac{\sum_{\alpha}d_{\alpha}
{\delta_{\tilde{I}}}^{(\alpha)}+d_zb_{\tilde{I}}}{D-d-2},~~~~
C_I=-\frac{\sum_{\alpha}d_{\alpha}\delta_I^{(\alpha)}+d_zb_I}{D-d-2}.
}
The $(D-d)$-dimensional metric in the Einstein frame is thus 
given by
\begin{eqnarray}
ds^2(\bar{\Msp})&=&h_{\tilde{I}}^{-B_{\tilde{I}}}\prod_{J\ne \tilde{I}}
{h_J}^{-C_J}\left[
-h^{a_{\tilde{I}}}\prod_{I\ne \tilde{I}}h_I^{a_I}
dt^2+\sum_{{\alpha}'}h_{\tilde{I}}^{\delta_{\tilde{I}}^{({\alpha}')}}
\prod_{I\ne \tilde{I}}h_I^{\delta_I^{({\alpha}')}}
\left(dx^{{\alpha}'}\right)^2\right.\nn\\
&&\left.+h_{\tilde{I}}^{b_{\tilde{I}}}\prod_{I\ne \tilde{I}}h_I^{b_I}
\delta_{a'b'}({\Zsp}')dz^{a'}dz^{b'}\right],
  \label{cn:metric-E:Eq}
\end{eqnarray}
where $x^{{\alpha}'}$ denotes the coordinate of $(p-d_{\alpha})$-dimensional 
relative transverse space and ${\Zsp}'$ is  
$(D-p-1-d_z)$-dimensional spaces.

If we set $K_{\tilde{I}}=At$, the $(D-d)$-dimensional metric 
(\ref{cn:metric:Eq}) in the Einstein frame can be expressed as
\begin{eqnarray}
ds^2(\bar{\Msp})&=&\prod_{I\ne \tilde{I}}h_I^{-C_I}
\left[-\prod_{I\ne \tilde{I}}h_I^{a_I}
\left\{1+\left(\frac{\tau}{\tau_0}\right)^{-\frac{2}{{B'}_{\tilde{I}}+2}}
H_{\tilde{I}}\right\}^{{B'}_{\tilde{I}}}d\tau^2\right.\nn\\
&&\hspace{-0.4cm}+
\sum_{\alpha'}\prod_{I\ne \tilde{I}}h_I^{\delta^{(\alpha')}_I}
\left\{1+\left(\frac{\tau}{\tau_0}\right)
^{-\frac{2}{{B'}_{\tilde{I}}+2}}H_{\tilde{I}}\right\}^
{-B_{\tilde{I}}+\delta_{\tilde{I}}^{(\alpha')}}
\left(\frac{\tau}{\tau_0}\right)^{\frac{2\left(-B_{\tilde{I}}
+\delta_{\tilde{I}}^{(\alpha')}\right)}
{{B'}_{\tilde{I}}+2}}
\left(dx^{\alpha'}\right)^2\nn\\
&&\left.\hspace{-0.4cm}+\prod_{I\ne \tilde{I}}h_I^{b_I}
\left\{1+\left(\frac{\tau}{\tau_0}\right)^{-\frac{2}{{B'}_{\tilde{I}}+2}}
H_{\tilde{I}}\right\}^{{B'}_{\tilde{I}}+1}
\left(\frac{\tau}{\tau_0}\right)^{\frac{2\left({B'}_{\tilde{I}}+1\right)}
{{B'}_{\tilde{I}}+2}}
\delta_{a'b'}({\Zsp}')dz^{a'}dz^{b'}\right],
  \label{cn:metric-E2:Eq}
\end{eqnarray}
where ${B'}_{\tilde{I}}$ is given by 
${B'}_{\tilde{I}}=-B_{\tilde{I}}+a_{\tilde{I}}$ and we define 
the cosmic time $\tau$:  
\Eq{
\frac{\tau}{\tau_0}=\left(At\right)^{({B'}_{\tilde{I}}+2)/2},~~~~\tau_0=
\frac{2}{\left({B'}_{\tilde{I}}+2\right)A}\,.
}
Hence, in the Einstein frame, the power of the scale factor 
in the fastest expanding case is given by 
\Eq{
0<\frac{{B'}_{\tilde{I}}+1}{{B'}_{\tilde{I}}+2}<1,
~~~~~{\rm for}~~D-d-2>0
\,.
   \label{cn:power-E:Eq}
}

If the physical parameters satisfy \eqref{cn:power-E:Eq}, 
the solutions do not give 
an accelerating expansion in our Universe. 
These are the similar results with the case of the other partially 
localized and delocalized intersecting brane backgrounds. 
Although we find the exact time-dependent brane solution, 
the power exponent of the scale factor is too small. 
Furthermore, in order to 
discuss a de Sitter solution in an intersecting brane background, 
one has to consider the trivial dilaton, 
which will be discussed in the next subsection. 

\subsection{de Sitter universe}

In this subsection, we consider the Einstein 
equations (\ref{cn:Einstein:Eq}) with $c_{\tilde{I}}=0$. 
Equation (\ref{cn:c:Eq}) gives 
\Eq{
N_{\tilde{I}}=\frac{2(D-p_{\tilde{I}}-3)(p_{\tilde{I}}+1)}{(D-2)}.
   \label{ds2:N:Eq}
}

If we assume 
\Eqr{
p=p_{\tilde{I}}=0\,,~~~
N_{\tilde{I}}=\frac{2(D-3)}{(D-2)}\,,~~~
\alpha_{I'}=-\frac{N_{I'}a_{I'}}{2\epsilon_{I'}c_{I'}}\,,
~~~\Lambda_{I'}=0\,,~~~
{\rm for}~~I'\ne \tilde{I}\,,
}
the field equations reduce to
\Eqrsubl{ds2:equations:Eq}{
&&R_{ij}(\Zsp)=0\,,\\
&&h_{\tilde{I}}(t, z)=K_{\tilde{I}}(t)+H_{\tilde{I}}(z),~~~~~
\left(\frac{dK_{\tilde{I}}}{dt}\right)^2
-\frac{2(D-3)^2}{(D-2)(D-1)}\Lambda_{\tilde{I}}=0,
\label{ds2:warp1:Eq}\\
&&
\triangle_{\Zsp}H_{\tilde{I}}=0\,,~~~~\triangle_{\Zsp}h_{I'}=0.
   \label{ds2:warp2:Eq}
   }
Then Eq.~(\ref{ds2:warp1:Eq}) gives 
\Eq{
K_{\tilde{I}}(t)=c_0t+\tilde{c}\,,
   \label{ds2:exact:Eq}
}
where $\tilde{c}$ is an integration constant and 
$c_0$ is given by
\Eq{
c_0=\pm(D-3)\sqrt{\frac{2}{(D-2)(D-1)}\Lambda_{\tilde{I}}}\,.
\label{ds2:c0:Eq}
}
Thus, there is no solution for $\Lambda_{\tilde{I}}<0$.
If the metric $u_{ab}({\rm Z})$ 
is assumed to be Eq.~(\ref{cn:flat metric:Eq}), 
the function $H_{\tilde{I}}$ is given by Eq.~(\ref{cn:HI:Eq}). 
Now we introduce a new time coordinate $\tau$ by 
\Eq{
c_0\tau=\ln t\,.
  \label{ds2:cosmic:Eq}
}
The $D$-dimensional metric 
(\ref{cn:metric:Eq}) is then rewritten as
\Eqr{
ds^2&=&-\prod_{I'\ne \tilde{I}}h_{I'}^{a_{I'}}(z)
\left(1+c_0^{-1}\e^{-c_0\tau}H_{\tilde{I}}\right)^{-2}
d\tau^2+\left(1+c_0^{-1}
\e^{-c_0\tau}H_{\tilde{I}}\right)^{\frac{2}{D-3}}
\left(c_0\e^{c_0\tau}\right)^{\frac{2}{D-3}}\nn\\
&&\times\left[\sum_{\alpha=1}^p\prod_{I'\ne \tilde{I}}
\left\{h_{I'}(z)\right\}^{\delta^{(\alpha)}_{I'}}(dx^{\alpha})^2 
+\prod_{I'\ne \tilde{I}}\left\{h_{I'}(z)\right\}^{b_I}
u_{ab}(\Zsp)dz^adz^b\right].
   \label{ds2:metric:Eq}
}
The $D$-dimensional metric (\ref{ds2:metric:Eq}) implies 
that the spacetime 
describes an isotropic and homogeneous universe 
if $H_{\tilde{I}}=0$.  
In the 
region where 
the terms with 
$H_{\tilde{I}}$ are negligible
and $h_{I'}$ approaches a constant, 
which is realized in the limit $\tau\rightarrow\infty$
and for $c_0>0$, 
the $D$-dimensional spacetime becomes de Sitter universe.
If we set $h_{I'}(z)=$const and $u_{ab}=\delta_{ab}$, 
Eq.~(\ref{ds2:metric:Eq}) becomes the solution which has been discussed by 
Ref.~\cite{Maki:1992tq} (see also \cite{Ivashchuk:1996zv}). 
Furthermore, for $D=4$ 
and by setting all $h_{I'}=1$, 
the solution is 
the Kastor-Traschen one 
\cite{Kastor:1992nn}.

\subsection{The behavior of the solutions}
Now we will study the spacetime structure.
The metric has singularities at $h_{\tilde{I}}=0$ or $h_{I'}=0$\,. 
The spacetime is thus not singular when it is restricted inside the domain 
specified by the conditions 
\Eq{
h_{\tilde{I}}(t, z) = a_0+a_1\,t+K_{\tilde{I}}(z)>0\,,~~~~~
h_{I'}(z)>0\,,
}
where the function $K_{\tilde{I}}$ is defined in (\ref{cn:HI:Eq}).
The $D$-dimensional 
spacetime cannot be extended beyond this region, because 
a curvature singularity appears in the $D$-dimensional spacetime. 
The regular spacetime with branes ends up with the 
singularities. 

Since the system with $a_1>0$ has the time reversal one of $a_1<0$, 
the dynamics of the spacetime depends on the signature of $a_1$\,. 

Here we will consider the case with $a_1>0$.  
Then the function $h_{\tilde{I}}$ is positive everywhere for $t>0$ and  
the spacetime is nonsingular. 
In the limit of $t\rightarrow \infty$ and apart from a position of the branes, 
near which the geometry takes a cylindrical form of an infinite throat, 
the solution is approximately described by 
a time-dependent uniform spacetime.

Now we discuss the time evolution for $t\le 0$\,. 
The spacetime is regular everywhere and 
has a cylindrical topology near each brane at $t=0$\,. 
As time slightly decreases, a curvature singularity 
appears as $|z^a-z^a_{\alpha}|\rightarrow\infty$\,. 
The singular hypersurface cuts off more and more of the space 
as time decreases further. 
When $t$ continues to decrease, the singular hypersurface eventually splits 
and surrounds each of the $p$-brane throats individually. 
The spatial surface is finally composed of two isolated throats. 
For $t>0$, the time evolution of the $D$-dimensional spacetime 
is the time reversal of $t<0$.

For any values of fixed $z^a$ in the regular domain in the 
$D$-dimensional spacetime (\ref{cn:metric:Eq}), 
the overall transverse space tends to expand asymptotically like 
$t^{b_{\tilde{I}}}$\,. 
Thus, the solutions describe static intersecting brane systems 
composed of $p$-branes near the positions of the branes, while, 
in the far region as $|z^a-z^a_{\alpha}| \rightarrow \infty$\,, 
the solutions approach de Sitter or FRW universes with 
the power-law expansion $t^{b_{\tilde{I}}}$\,. 
The emergence of time-dependent universes is an important feature of the 
dynamical brane solutions.

\subsubsection{Asymptotic structure}
\begin{figure}[h]
 \begin{center}
\includegraphics[keepaspectratio, scale=0.28, angle=-90]{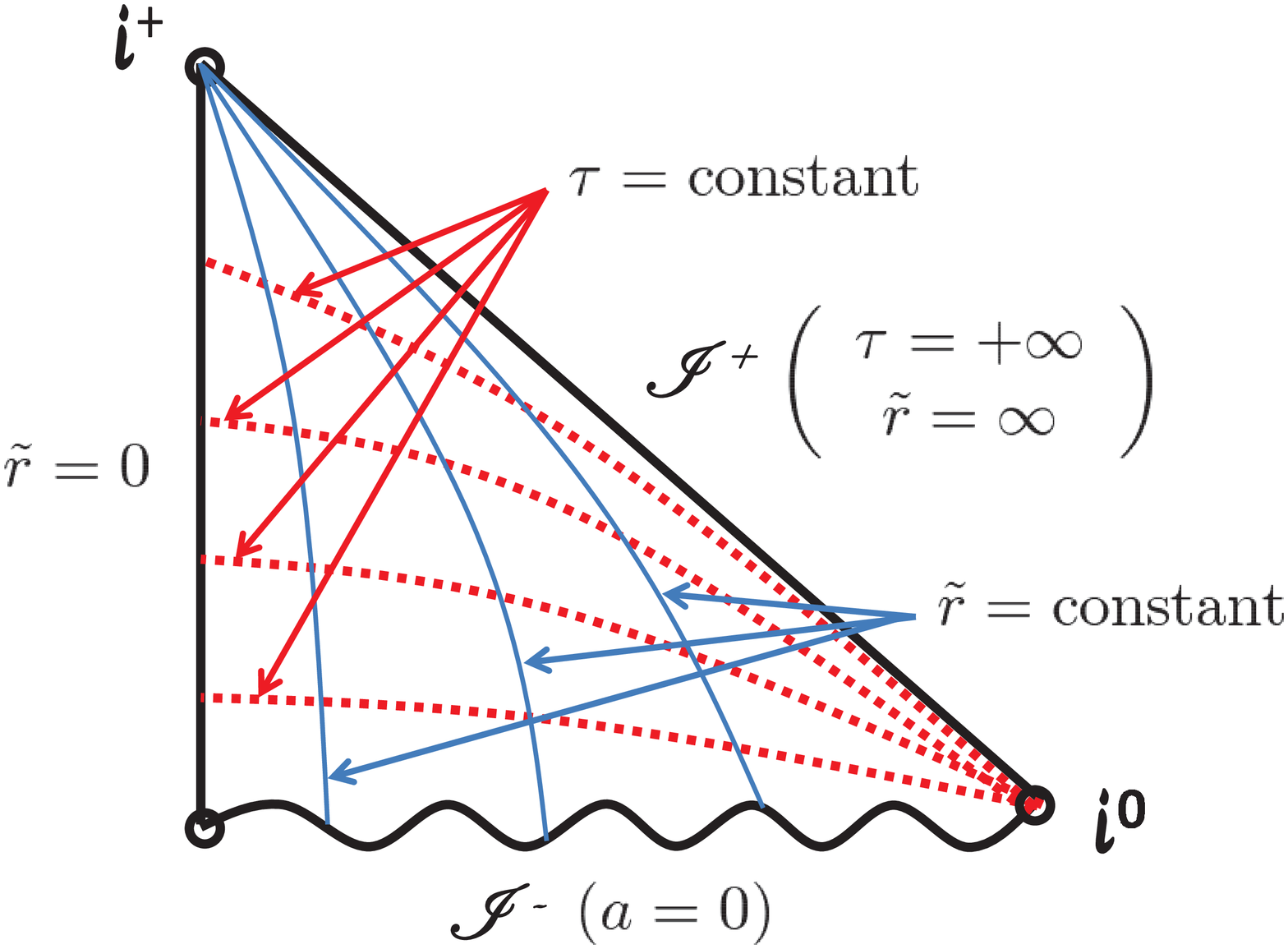}
\hskip 0.5cm
\includegraphics[keepaspectratio, scale=0.28, angle=-90]{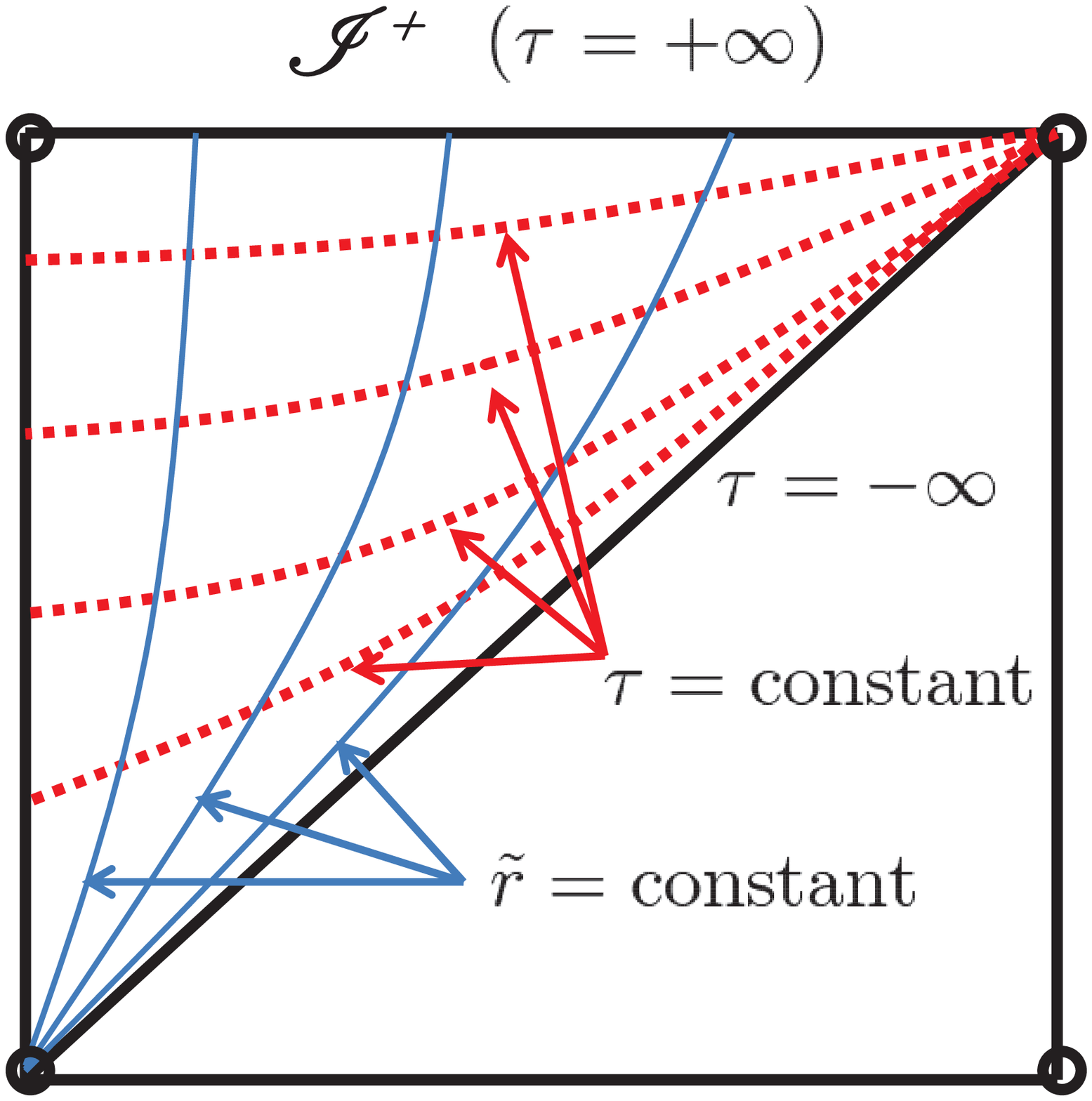}
\\
(a) \hskip 8cm (b) ~~~~~~
  \caption{\baselineskip 14pt 
Conformal diagrams of the $D$-dimensional spacetime for $p_{\tilde{I}}=0$. 
The regions corresponding to $\tilde{r}\rightarrow\infty$ give 
the original spacetime, where 
$\tilde{r}^2=\sum_\alpha(x^\alpha)^2+\delta_{ab}z^az^b$.
(a) For the case of $a_{\tilde{I}}+2\ne 0$, the metric 
(\ref{cn:metric:Eq}) approaches in the limit $\tilde{r}\rightarrow\infty$ 
to the $D$-dimensional flat FRW spacetime. 
(b) We also depict the conformal diagrams in the case of 
$a_{\tilde{I}}+2=0$. One can recognize that the asymptotic region of 
the spacetime is the de Sitter universe.}
  \label{fig:p}
 \end{center}
\end{figure}

We study the asymptotic behavior of the solutions. 
The solution describes a charged black
hole in the FRW or de Sitter universe in the limit of 
$|z^a|\rightarrow \infty$, and $H_{\tilde{I}}$ vanishes. 
First we consider the case of a power-law expanding universe. 
The function $h_{\tilde{I}}$ depends only on time $t$ in the far region 
from branes, and the resulting metric (\ref{cn:metric-J:Eq}) can be expressed 
as
\Eqr{
ds^2&=&-d\tau^2+\sum_{\alpha=1}^p
\left(\frac{\tau}{\tau_0}\right)^{\frac{2\delta_{\tilde{I}}^{(\alpha)}}
{a_{\tilde I}+2}}\left(dx^{\alpha}\right)^2
+\left(\frac{\tau}{\tau_0}\right)^{\frac{2b_{\tilde{I}}}{a_{\tilde{I}}+2}}
\delta_{ab}(\Zsp)dz^adz^b\,.
   \label{a:metric:Eq}
}
The scale factor of the relative transverse space is given by 
$a_{\rm r}(\tau)=\left(\tau/\tau_0\right)
^{b_{\tilde{I}}/a_{\tilde{I}}+2}$, while the expansion low for 
the overall transverse space is written by 
$a_{\rm t}(\tau)=\left(\tau/\tau_0\right)
^{b_{\tilde{I}}/a_{\tilde{I}}+2}$\,. 
On the other hand, for $c_{\tilde{I}}=0$ 
corresponding to de Sitter universe (\ref{ds2:metric:Eq}), 
the metric of $D$-dimensional spacetime in the far region 
from branes becomes 
\Eqr{
ds^2&=&-d\tau^2+\left(c_0\e^{c_0\tau}\right)^{\frac{2}{D-3}}
\left[\sum_{\alpha=1}^p(dx^{\alpha})^2
+u_{ab}(\Zsp)dz^adz^b\right].
   \label{a:metric2:Eq}
}
Figure \ref{fig:p} depicts the
conformal diagrams of the FRW and de Sitter universes. 

\subsubsection{Near-horizon geometry}

Next we discuss the near-horizon geometry of the solutions. We set the 
metric of $(D-p-1)$-dimensional overall transverse space:
\Eq{
\delta_{ab}(\Zsp)dz^adz^b=dr^2+r^2d\Omega^2_{(D-p-2)}\,,
}
where $\delta_{ab}$ denotes the metric of $(D-p-1)$-dimensional flat space 
and the line elements of a $(D-p-2)$-sphere (${\rm S}^{D-p-2}$) 
are given by $d\Omega^2_{(D-p-2)}$\,. 
The harmonic function $K_{\tilde{I}}$ dominates in the limit of 
$r\rightarrow 0$\,, and the time dependence 
can be ignored. Thus the system becomes static near a position of branes.
When all of the branes are located at the origin of the Z spaces, 
the solutions are rewritten as
\Eqrsubl{nh:h:Eq}{
h_{\tilde{I}}(t, r)&=&a_0+a_1\,t +\frac{M_{\tilde{I}}}{r^{D-p-3}}\,,
\label{nh:hI:Eq}\\
h_{I'}(r)&=&1+\frac{L_{I'}}{r^{D-p-3}}\,. 
  \label{nh:h':Eq}
}
Here $M_{\tilde{I}}$ and $L_{I'}$ are the mass of $p_{\tilde{I}}$-, and 
$p_{I'}$-branes, respectively. 
In the near-horizon region $r\rightarrow 0$\,, 
the dependence on $t$ in (\ref{nh:h:Eq}) is negligible. 
Then the metric is reduced to the following form:  
\Eqr{
\hspace{-1cm}
ds^2&=&r^2\left(\frac{M_{\tilde{I}}}{r^{D-p-3}}\right)^{b_{\tilde{I}}}
\prod_{I'}\left(\frac{L_{I'}}{r^{D-p-3}}\right)^{b_{I'}}
\left[-r^{-2}
\left(\frac{M_{\tilde{I}}}{r^{D-p-3}}\right)^{-\frac{4}{N_{\tilde{I}}}}
\prod_{I'}\left(\frac{L_{I'}}{r^{D-p-3}}\right)^{-\frac{4}{N_{I'}}}dt^2
\right.\nn\\
&&\left.\hspace{-1cm}+r^{-2}\sum^p_{\alpha=1}\prod_{I'}
\left(\frac{M_{\tilde{I}}}{r^{D-p-3}}\right)^{-b_{\tilde{I}}
+\delta^{(\alpha)}_{\tilde{I}}}
\left(\frac{L_{I'}}{r^{D-p-3}}\right)^{-b_{I'}
+\delta^{(\alpha)}_I}(dx^{\alpha})^2
   +\left(\frac{dr^2}{r^2}+d\Omega^2_{(D-p-2)}\right)\right].
   \label{nh:nh:Eq}
}
Thus the metric (\ref{nh:nh:Eq}) describes a warped product of 
$(p+2)$-dimensional spacetime M${}_{p+2}$ and $(D-p-2)$-dimensional 
sphere ${\rm S}^{D-p-2}$\,.

Hence, the near-brane geometry has the same metric form as the static one. 
If it has a horizon geometry, 
we can obtain a black hole solution in the time-dependent background.
In fact, some solutions, for instance, the 
M2$-$M2$-$M2, M2$-$M2$-$M5$-$M5 intersecting solution in eleven dimensions,  
give regular black hole spacetimes in the static limit 
\cite{Maeda:2009zi}. 

Our solution approaches asymptotically the dynamical universe with 
the scale factor $a(\tau)$\,, while the static solution gives a black hole. 
Then we can regard the present solution as a black hole in the expanding 
universe.

\section{Collision of 0-branes} 
\label{sec:c}

In this section, we apply our dynamical intersecting brane 
solutions found in the previous section to brane collisions. 

The functions 
$h_{\tilde{I}}$ and $h_{I'}$ are assumed to be
\Eq{
h_{\tilde{I}}(t, z)=c_0 t +\tilde c_{\tilde{I}}+H_{\tilde{I}}(z)\,,~~~~
h_{I'}=h_{I'}(z)\,.
 \label{cp:h:Eq}
}
Here $c_0$ and $\tilde c_{\tilde{I}}$ are constants, and
the function $H_{\tilde{I}}$ and $h_{I'}$ are expressed, respectively, as
\Eqr{
H_{\tilde{I}}(z)&=&\sum_{k=1}^{m}\frac{M_{\tilde{I}, k}}
{|z^a-z^a_k|^{D-p-3-d}}\,,
~~~h_{I'}(z)=\tilde{c}_{I'}
+\sum_{l=1}^{m}\frac{Q_{{I'},\,l}}{|z^a-z^a_l|^
{D-p-3-d_{I'}}}\,,
   \label{cp:h2:Eq}
}
where $\tilde{c}_{I'}$ is constant, 
$d$ and $d_{I'}$ denote the number of smeared dimension for 
$0$-brane and $p_{I'}$-brane, respectively, 
we assume $D\neq p+3+d$ and $D\neq p+3+d_{I'}$, and  
$M_{\tilde{I}, k}~ (k=1\,,\cdots\,, m)$ and 
$Q_{{I'},\,l}~ (l=1\,,\cdots\,, m)$ are mass constants of 0-brane and 
$p_{I'}$-branes located at $z^a_k$ and $z^a_l$, respectively. 
Since $h_{I'}$ is the harmonic function on the
$(D-p-1-d_{I'})$-dimensional Euclidean subspace in $\Zsp$, 
we define 
\Eqrsubl{cp:z:Eq}{
&&\hspace{-0.5cm}|z^a-z^a_k|
=\sqrt{\left(z^1-z^1_k\right)^2+\left(z^2-z^2_k\right)^2+\cdots+
\left(z^{D-p-1-d}-z^{D-p-1-d}_k\right)^2}\,,
   \label{cp:zk:Eq}\\
&&\hspace{-0.5cm}|z^a-z^a_l|
=\sqrt{\left(z^1-z^1_l\right)^2+\left(z^2-z^2_l\right)^2+\cdots+
\left(z^{D-p-1-d_{I'}}-z^{D-p-1-d_{I'}}_l\right)^2}\,.
  \label{cp:zl:Eq}
}

The metric, scalar, and gauge fields are
 given by Eqs.~(\ref{cn:metric:Eq}) and
(\ref{cn:fields:Eq}), respectively.
For $D=p+3+d$ and $D=p+3+d_{I'}$, these become
\Eqr{
H_{\tilde{I}}(z)&=&\sum_{k=1}^{m}M_{\tilde{I}, k}\, 
\ln |z^a-z^a_k|\,,
~~~h_{I'}(z)=\tilde{c}_{I'}
+\sum_{l=1}^{m}Q_{{I'},\,l}\, \ln |z^a-z^a_l|\,.
     \label{cp:h3:Eq}
}
Since the time dependence allows only for the 0-brane, 
we see that the $(D-3-d_{I'})$-brane background is critical case.
If we consider the $(D-2-d_{I'})$-brane, 
the functions $h_{\tilde{I}}$ and $h_{I'}$ are 
written by the sum of linear functions of $z$.
The possibility of brane collisions comes from the difference in 
the overall transverse dimension. 

From the solution (\ref{cp:h2:Eq}), there are curvature 
singularities at $h_{\tilde{I}}=0$ or at $h_{I'}=0$ in the 
$D$-dimensional background. 
Note that the regular $D$-dimensional spacetime
 is restricted to the region of $h_{\tilde{I}}>0$ and $h_{I'}>0$, which is 
bounded by curvature singularities.
Hence, the $D$-dimensional metric (\ref{cn:metric:Eq}) is regular 
if and only if $h_{\tilde{I}}>0$ and $h_{I'}>0$. 

The solution with $0-p_{I'}$ branes 
takes the form (\ref{cn:metric-J:Eq}), 
where we set $K_{\tilde{I}}=c_0 t$
and the function $H_{\tilde{I}}$ is given by (\ref{cn:HI:Eq}). 
We classify the behavior of the harmonic function $h_{I'}$ into 
two classes: $p_{I'}\le (D-4-d_{I'})$ and $p_{I'}= (D-2-d_{I'})$.
Since these depend on the dimensions of the $p_{I'}$-brane, 
we discuss them below separately. 
In the case of the $(D-3-d_{I'})$-brane, the harmonic function 
$h_{I'}$ diverges both at infinity and near $(D-3-d_{I'})$-branes.
Since there is no regular spacetime region near branes due to 
$h_{I'}\rightarrow -\infty$, these solutions are not physically relevant. 
In the following, we discuss the collision involving the $0-p_{I'}$ brane 
in $D$-dimensional spacetime.

\subsection{Collision of the $0-p_I$-brane in the 
asymptotically power-law expanding universe}

The harmonic function $H_{\tilde{I}}$ becomes dominant in 
the limit of $z^a\rightarrow \,z^a_k$\,, while 
the function $h_{\tilde{I}}$ depends only on time $\tau$ 
in the limit of $|z^a|\rightarrow \infty$. 
Hence, we find a static structure of the $0-p_{I'}$-brane system 
near branes. In the far region from branes, the function 
$H_{\tilde{I}}$ vanishes. 
Therefore, the metric can be written by
\Eqr{
ds^2&=&-\prod_{I'\ne \tilde{I}}h_{I'}^{a_{I'}}
\bar{h}_{\tilde{I}}^{a_{\tilde{I}}}d\tau^2+\sum_{\alpha=1}^p
\prod_{I'\ne \tilde{I}}h_{I'}^{\delta_{I'}^{(\alpha)}}
\left(\frac{\tau}{\tau_0}\right)^{\frac{2\delta_{\tilde{I}}^{(\alpha)}}
{a_{\tilde I}+2}}\bar{h}_{\tilde{I}}^{\delta_{\tilde{I}}^{(\alpha)}}
\left(dx^{\alpha}\right)^2\nn\\
&&+\prod_{I'\ne \tilde{I}}h_{I'}^{b_{I'}}
\left(\frac{\tau}{\tau_0}\right)^{\frac{2b_{\tilde{I}}}{a_{\tilde{I}}+2}}
\bar{h}_{\tilde{I}}^{b_{\tilde{I}}}\delta_{ab}(\Zsp)dz^adz^b\,,
   \label{co:sufrace2:Eq}
}
where $\bar{h}_{\tilde{I}}$ is defined by 
\Eq{
\bar{h}_{\tilde{I}}=
1+\left(\frac{\tau}{\tau_0}\right)^{-\frac{2}{a_{\tilde{I}}+2}}
H_{\tilde{I}}\,.
}

In order to analyze the brane collision, 
we consider a concrete example, in which 
two $0-p_{I'}$ branes are located at $z^a=(\pm L, 0, \ldots, 0)$. 
We will discuss the time evolution separately with respect to the signature of 
a constant $\tau_0$\,, because the behavior of spacetime strongly 
depends on it. Since the metric function is singular 
at $h_{\tilde{I}}(\tau, z)=0$ and $h_{I'}=0$, one can note that 
the regular spacetime exists inside the domain restricted by
\Eq{
h_{\tilde{I}}(\tau, z) = \left(\frac{\tau}{\tau_0}\right)^
{\frac{2}{a_{\tilde{I}}+2}}+H_{\tilde{I}}(z)>0\,,~~~~~
h_{I'}=h_{I'}(z)>0\,,
}
where the functions $H_{\tilde{I}}$ and  
$h_{I'}$ are defined in (\ref{cp:h:Eq}). 
The brane background evolves into a curvature singularity, 
because the dilaton $\phi$ diverges. 
Since the $D$-dimensional spacetime cannot be extended beyond this region, 
the regular spacetime with two 0-branes ($p+d\le 6$) ends on these singular
hypersurfaces. 
The solution with $(\tau_0)^{-2/(a_{\tilde{I}}+2)}>0$ is the time 
reversal one of $(\tau_0)^{-2/(a_{\tilde{I}}+2)}<0$, 
because the time dependence appears only in the form of 
$(\tau/\tau_0)^{2/(a_{\tilde{I}}+2)}$. 
In the following, we consider 
the case with $(\tau_0)^{-2/(a_{\tilde{I}}+2)}<0$.

For $(\tau)^{2/(a_{\tilde{I}}+2)}<0$, 
the $D$-dimensional spacetime is nonsingular, because the function 
$h_{\tilde{I}}$ is positive everywhere. 
In the limit of $(\tau)^{2/(a_{\tilde{I}}+2)}\rightarrow -\infty$, 
the $D$-dimensional spacetime becomes asymptotically 
a time-dependent uniform background, 
while the cylindrical forms of infinite 
throats exist near branes. 
  
For $\tau>0$, the spatial metric is initially  
regular everywhere. The $D$-dimensional 
spacetime has a cylindrical topology 
near each brane. 
As $\tau$ increases slightly, a singular hypersurface appears from 
the spatial infinity 
($|z^a-z^a_k|\rightarrow\infty$).  
As $\tau$ increases further, the singularity cuts the space off 
more and more. 
Since the singular hypersurface eventually splits 
and surrounds each of the brane throats, 
the spatial surface is finally composed of two isolated throats.

One notes that the transverse 
dimensions in the metric (\ref{co:sufrace2:Eq}) 
expand asymptotically as $\tau^{b_{\tilde{I}}/(a_{\tilde{I}}+2)}$ 
for fixed spatial coordinates $z^a$. 
The $D$-dimensional spacetime becomes static near branes, while the 
background approaches a FRW universe in the far region 
($|z^a-z^a_k|\rightarrow \infty$). 
Hence, the time evolution of the four-dimensional universe depends on the 
position of the observer.
For $(\tau/\tau_0)^{2/(a_{\tilde{I}}+2)}<0$, the behavior of 
$D$-dimensional spacetime is
the time reversal of the period of $(\tau/\tau_0)^
{2/(a_{\tilde{I}}+2)}>0$.

\begin{figure}[h]
 \begin{center}
     \includegraphics[keepaspectratio, scale=0.6]{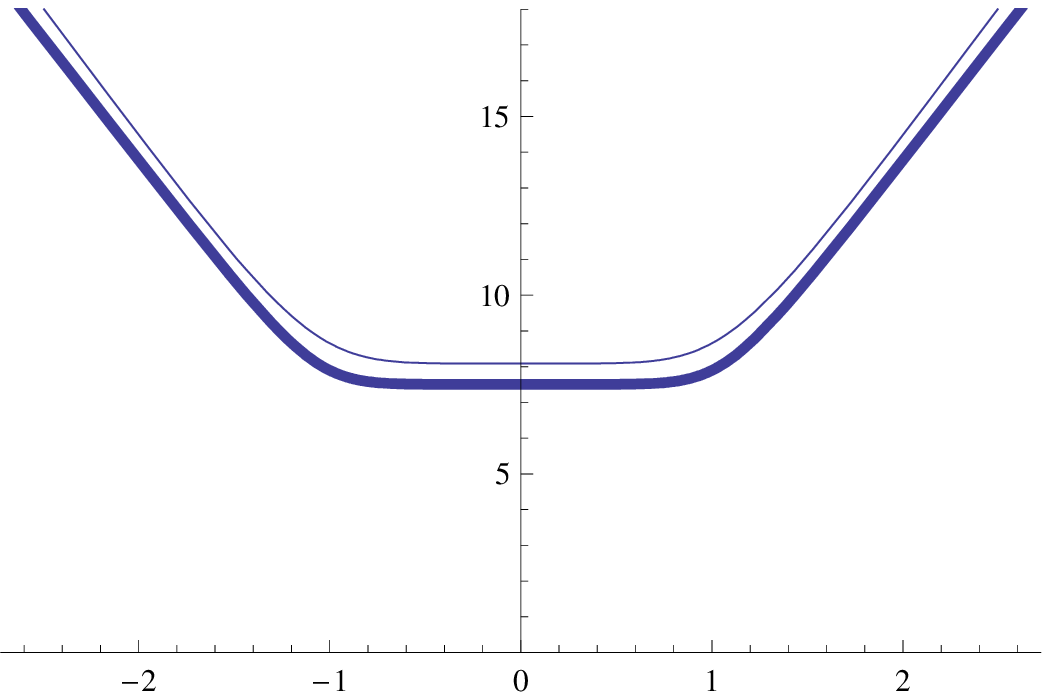}
\put(-100, 135){$d(\tau)$}
\put(10,10){$\tau$}
\hskip 2cm
\includegraphics[keepaspectratio, scale=0.6]{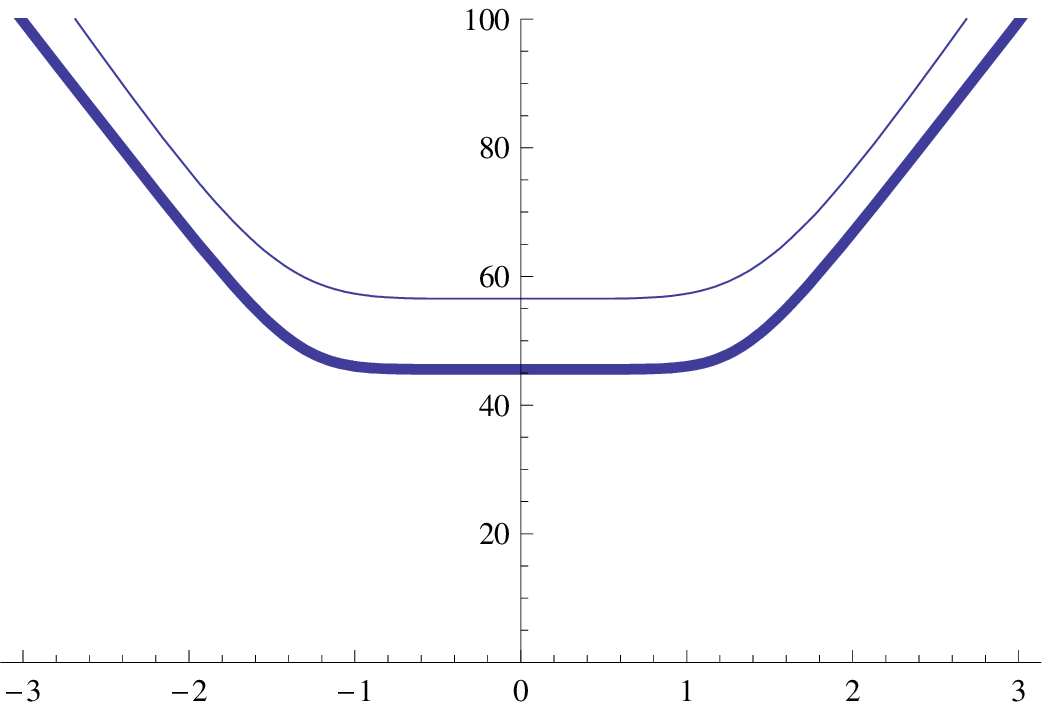}
\put(-105,135){$d(\tau)$}
\put(10,10){$\tau$}\\
(a) \hskip 7cm (b) ~~~~~~
  \caption{\baselineskip 14pt
(a) For the case of $M_1=M_2$ in the asymptotically power-law 
 expanding universe, the proper distance between two dynamical 
 0-branes given in (\ref{cp:distance:Eq}) is depicted. We fix $d=d_{I'}=0$, 
  $D=10$, $\tau_0=-1$, $M_1=1, 
M_2=1$, $N=2$\,, and $L=1$ for the $0-8$- (bold curve) and 
$0-6$- (solid curve) branes. 
The distance decreases initially ($\tau<0$)  
but turns to increase at $\tau=0$, and
then two 0-branes segregate each other. 
(b) We also show the proper distance between two dynamical 
 0-branes for $0-8$- (bold curve) and $0-6$- (solid curve) 
brane systems from the bottom in the case of $d=d_{I'}=0$, 
$M_1=10$, $M_2=1$, $N=2$, $L=1$, and $D=10$ in the 
asymptotically power-law expanding universe. 
 Although the proper distance initially decreases as $\tau(<0)$ increases, 
the distance increases as $\tau(>0)$ increases.}
  \label{fig:mil}
 \end{center}
\end{figure}

Now we define
\Eq{
z_{\perp}=\sqrt{\left(z^2\right)^2+\left(z^3\right)^2+\cdots 
+\left(z^{D-1-p-d}\right)^2}\,.
   \label{cp:zp:Eq}
}
By using the above equation, the proper distance at $z_{\perp}=0$
between two branes can be written by
\begin{eqnarray}
d(\tau)&=&\int_{-L}^L dz^1 
\left[\left(\frac{\tau}{\tau_0}\right)^{2/(a_{\tilde{I}}+2)}
+\frac{M_1}{|z^1+L|^{D-3-p_{I'}-d_{I'}}}
+\frac{M_2}{|z^1-L|^{D-3-p_{I'}-d_{I'}}}\right]^
{b_{\tilde{I}}/2}
\nn\\
&&\times
\left(1+\frac{Q_1}{|z^1+L|^{D-3-p_{I'}-d}}
+\frac{Q_2}{|z^1-L|^{D-3-p_{I'}-d}}\right)^{b_{I'}/2}
\,.
\label{cp:distance:Eq}
\end{eqnarray}
The proper distance is a monotonically increasing function of $\tau$.
We illustrate $d(\tau)$  
for the case of the $0-p_{I'}$ brane system in Fig. \ref{fig:mil}.
We consider the case of $d=d_{I'}=0$, 
$\tau_0=-1$, $Q_1=Q_2=M_1=M_2=1$, $L=1$\,, and $D=10$ or $D=8$. 
It shows that two 0-branes are initially ($\tau< 0$) approaching,
the distance $d(\tau)$ takes the minimum finite value at $\tau=0$, 
and then two 0-branes segregate each other. 
Thus they will never collide. 
Hence, we cannot discuss a brane collision in this case.

\subsection{Collision of the $0-p_{I'}$-brane 
in the asymptotically de Sitter universe}
Let us next discuss the collision in the $0-p_{I'}$-brane with a 
trivial dilaton system. We consider the case that 
the harmonic function $H_{\tilde{I}}$ and $h_{I'}$ are linear in $z$ 
 and discuss in detail the  
$0-(D-2)$-brane in $D$ dimensions as a example. 
In this case, we have one extra dimension $z$ in $\Zsp$ space.
The $D$-dimensional metric (\ref{ds2:metric:Eq}) 
can be rewritten by
\Eqr{
ds^2&=&-h_{I'}^{a_{I'}}(z)
\left(1+c_0^{-1}\e^{-c_0\tau}H_{\tilde{I}}\right)^{-2}
d\tau^2+h_{I'}^{a_{I'}}(z)
\left(1+c_0^{-1}\e^{-c_0\tau}H_{\tilde{I}}\right)^{\frac{2}{D-3}}
\left(c_0\e^{c_0\tau}\right)^{\frac{2}{D-3}}\nn\\
&&\times\left[\sum_{\alpha=1}^p(dx^{\alpha})^2 
+h_{I'}^{4/N_{I'}}(z)dz^2\right]\,,
   \label{cods:metric:Eq}
}
where the function $H_{\tilde{I}}(z)$ is written by 
\Eqr{
H_{\tilde{I}}(z)=\sum_{k=1}^{m}M_{\tilde{I}\,, k}\,|z-z_k|\,.
    \label{cods:L:Eq}
}

We consider the collision in the $0-p_{I'}$-brane system 
with charges $M_1$ and $Q_1$ at $z^1=-L$ 
and the other with charges $M_2$ and $Q_2$ at $z^1=L$.
The proper distance at $z_{\perp}=0$ between the two $0$-branes can be 
expressed as 
\Eqr{
d(\tau)&=&\int^L_{-L} dz 
\left(c_0\e^{c_0\tau}+M_1\,|\,z^1+\,L|
+M_2\,|\,z^1-\,L|\right)^{1/(D-3)}\nn\\
&&\times
\left(1+Q_1\,|\,z^1+\,L|
+Q_2\,|\,z^1-\,L|\right)^{b_{I'}/2}\,.
  \label{ds:length:Eq}
}
In the period of $c_0<0$,  
the proper distance increases as
$\tau$ increases. 
If $M_1\ne M_2$, a singular hypersurface appears at 
$\tau=\tau_s\equiv\ln\left[-(M_1|z^1+L|+M_2|z^1-L|)c_0^{-1}\right]c_0^{-1}<0$ 
when the distance is still finite.

However, in the case of the equal charges  $Q_1=Q_2=M_1=M_2=M$, the 
situation is completely different, because 
the proper distance finally vanishes at $\tau_s=
\ln\left(-2MLc_0^{-1}\right)c_0^{-1}<0$
as 
\Eqr{
d(\tau)= 2L\left(c_0\e^{c_0\tau}+2LM\right)^{1/(D-3)}
\left(1+2LM\right)^{b_{I'}/2}
\,.}
Then two branes can collide. 
A singularity is formed
at the same point and time.

Let us consider the case $p_{I'}\ne D-2$. 
The $D$-dimensional metric (\ref{ds2:metric:Eq}) 
can be written as
\Eqr{
ds^2&=&-h_{I'}^{a_{I'}}(z)
\left(1+c_0^{-1}\e^{-c_0\tau}H_{\tilde{I}}\right)^{-2}
d\tau^2+h_{I'}^{a_{I'}}(z)
\left(1+c_0^{-1}\e^{-c_0\tau}H_{\tilde{I}}\right)^{\frac{2}{D-3}}
\left(c_0\e^{c_0\tau}\right)^{\frac{2}{D-3}}\nn\\
&&\times\left[\sum_{\alpha=1}^p(dx^{\alpha})^2 
+h_{I'}^{4/N_{I'}}(z)\delta_{ab}(\Zsp)dz^adz^b\right].
   \label{cods:metric2:Eq}
}
Since the proper distance at $z_{\perp}=0$
between two branes is given by
\Eqr{
d(\tau)&=&\int^L_{-L} dz 
\left(c_0\e^{c_0\tau}+\frac{M_1}{|\,z^1+\,L|^{D-p_{I'}-3-d}}
+\frac{M_2}{|\,z^1-\,L|^{D-p_{I'}-3-d}}\right)^{1/(D-3)}
\nn\\
&&\times
\left(1+\frac{Q_1}{|z^1+L|^{D-3-p_{I'}-d}}
+\frac{Q_2}{|z^1-L|^{D-3-p_{I'}-d}}\right)^{b_{I'}/2}
\,,
\label{distance2}
}
the distance increases monotonically with respect to $\tau$.

\begin{figure}[h]
 \begin{center}
     \includegraphics[keepaspectratio, scale=0.58]{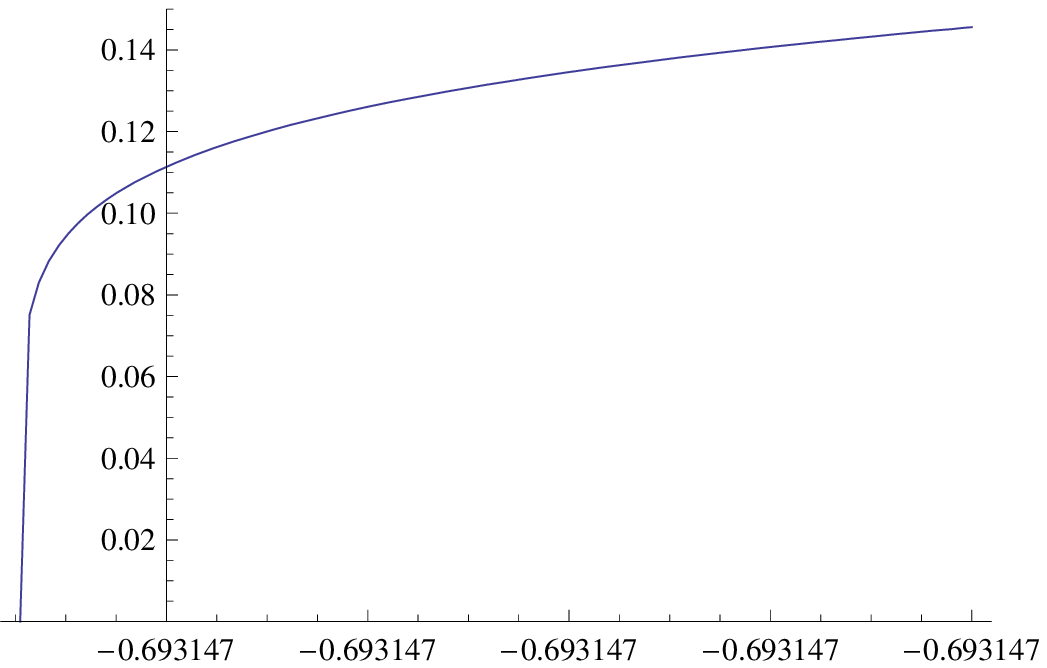}
\put(-165, 125){$d(\tau)$}
\put(10,10){$\tau$}
\hskip 2cm
\includegraphics[keepaspectratio, scale=0.58]{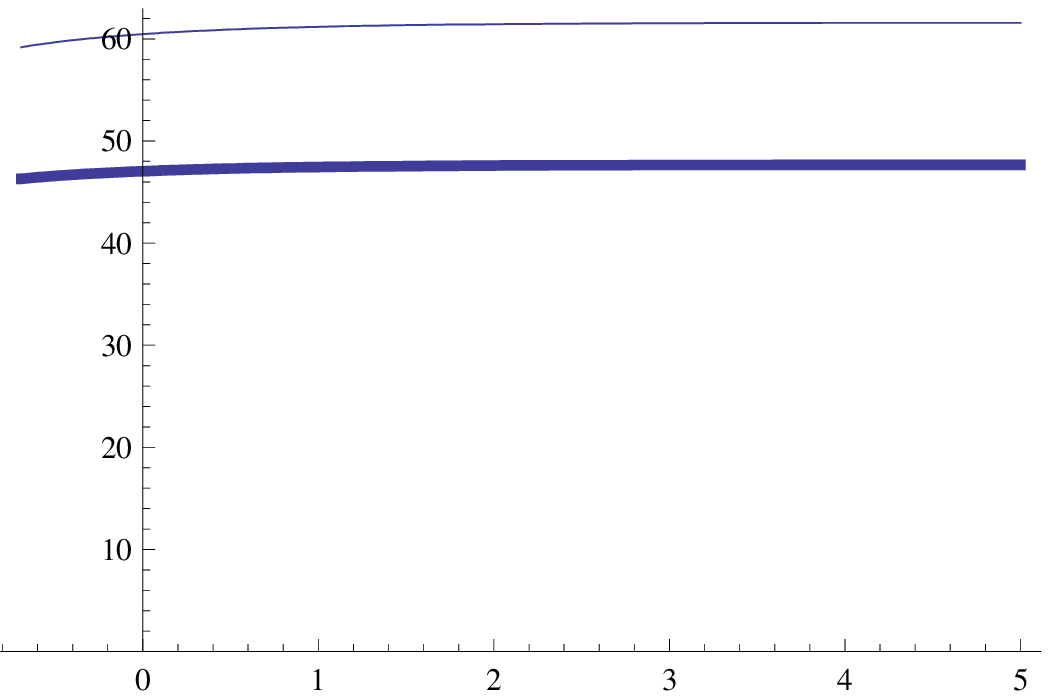}
\put(-165,125){$d(\tau)$}
\put(10,10){$\tau$}\\
(a) \hskip 7cm (b) ~~~~~~
  \caption{\baselineskip 14pt
(a) For the case of $M_1=M_2$ in the asymptotically de Sitter universe, 
we show the proper distance between two dynamical 0-branes given in 
    (\ref{ds:length:Eq}).  We set $D=10$, $c_0=-1$, $M_1=1, 
M_2=1$, $N=2$, and $L=1$ for the $0-8$-brane. The proper distance rapidly 
vanishes near where two branes collide. 
(b) We also show the proper distance between two dynamical 
 0-branes for the $0-8$- (bold curve) and 
 $0-6$- (solid curve) brane systems 
from the bottom in the case of $M_1=10$, $M_2=1$, $N=2$,  
and $D=10$ in the asymptotically de Sitter universe.  
The proper distance initially decreases as $\tau$ decreases 
and remains still finite when a singularity appears.}
  \label{fig:ds}
 \end{center}
\end{figure}

In the case of $c_0<0$, initially ($\tau=0$), $D$-dimensional space is regular
except at $|z^a-z^a_k|\rightarrow 0$\,, while this is an asymptotically 
time-dependent spacetime and has the cylindrical form of an infinite 
throat near the 0-brane.   
At $\tau=\tau_s<0$, 
a singularity appears 
from the spatial infinity ($|z^a-z^a_k|\rightarrow\infty$).
As time decreases ($\tau<0$), 
the singular hypersurface erodes the region with 
the large values of $|z^a-z^a_k|$. 
Since only the region near 0-branes remains regular, eventually 
it splits and each fragment surrounds each 0-brane
individually. 
Figure \ref{fig:ds} shows that 
this singularity appears before the proper distance $d(\tau)$ vanishes. 
Hence, the $D$-dimensional spacetime has the singularity 
before two branes collide.  
Although two 0-branes approach very slowly, 
a singularity suddenly appears
at a finite distance. Then, 
the spacetime splits into two isolated 0-brane throats.

We show $d(\tau)$ integrated numerically in Fig. \ref{fig:ds}
 for the case of $c_0<0$. 
In the future direction, the proper distance $d$ increases.
Then for $\tau>0$, each brane gradually
separates as $\tau$ increases.

\section{Applications to supergravities}
\label{sec:sug}

In the case of ten or eleven dimensions with $N=4$ and $\Lambda_I=0$, 
Eq.~(\ref{pl:action:Eq}) gives the action of supergravities. 
For instance, the bosonic part of the action of $D=11$ supergravity 
includes only 4-form field strength, while, for $D=10$, 
the constant $c$ is precisely the dilaton coupling for the 
Ramond-Ramond $\left(p+2\right)$-form in the type II supergravities. 
The dynamical solutions for the case of $N=4$
have been already discussed in Ref.~\cite{Maeda:2010aj}. In this section, 
we will discuss the time-dependent solution in six-dimensional 
Nishino-Salam-Sezgin (NSS) gauged supergravity 
and Romans' gauged supergravity models. 
The bosonic part of the six-dimensional 
NSS model~\cite{Salam:1984cj, Nishino:1984gk, Nishino:1986dc, Burgess:2003mk}
is given by the expression (\ref{pl:action:Eq}) with $\Lambda_r>0$, 
$\Lambda_s=0$, while Romans' six-dimensional 
gauged supergravity \cite{Romans:1985tw} is expressed by the action 
(\ref{pl:action:Eq}) with $\Lambda_r<0$, $\Lambda_s=0$. 

\subsection{Nishino-Salam-Sezgin gauged supergravity}
Now we consider the NSS model among the theories of $D=6$. 
The couplings of the 2-form ($p_r=0$)
and the 3-form ($p_s=1$) field strengths to the dilaton 
are given by $\epsilon_rc_r=-\frac{1}{\sqrt{2}}$ and 
$\epsilon_sc_s=-\sqrt{2}$, respectively: 
\Eqr{
S&=&\frac{1}{2\kappa^2}\int \left[\left(R-2\e^{\phi/\sqrt{2}}
\Lambda\right)\ast{\bf 1}
 -\frac{1}{2}\ast d\phi \wedge d\phi\right.\nn\\
 &&\left.\hspace{-0.5cm}
 -\frac{1}{2\cdot 2!}
 \e^{-\phi/\sqrt{2}}\ast F_{(2)}\wedge F_{(2)}
 -\frac{1}{2\cdot 3!}
 \e^{-\sqrt{2}\phi}\ast F_{(3)}\wedge F_{(3)}
 \right],
\label{nss:action:Eq}
}
where $R$ denotes the Ricci scalar constructed from the six-dimensional 
metric $g_{MN}$\,, $\kappa^2$ is the six-dimensional gravitational constant,  
$\ast$ is the Hodge operator in the six-dimensional spacetime, 
$\phi$ denotes the scalar field, $\Lambda>0$ is the cosmological 
constant, and $F_{\left(2\right)}$ and $F_{\left(3\right)}$ 
are 2-form and 3-form field strengths, respectively.
From Eq.~(\ref{pl:c:Eq}), the NSS model 
is realized by choosing $\Lambda_r=\Lambda>0$, 
$\Lambda_s=0$, $N_r=2$, and $N_s=4$.

The six-dimensional action (\ref{nss:action:Eq}) gives the field equations
\Eqrsubl{nss:equations:Eq}{
&&
R_{MN}=\frac{1}{2}\e^{\phi/\sqrt{2}}\Lambda g_{MN}
+\frac{1}{2}\pd_M\phi \pd_N \phi
+\frac{\e^{-\phi/\sqrt{2}}}{2\cdot 2!}
\left(2F_{MA}{F_N}^A
-\frac{1}{4} g_{MN} F^2_{\left(2\right)}\right)\nn\\
&&~~~~~~~~~+\frac{\e^{-\sqrt{2}\phi}}{2\cdot 3!}
\left(3F_{MAB} {F_N}^{AB}
-\frac{1}{2} g_{MN} F_{\left(3\right)}^2\right),
   \label{nss:Einstein:Eq}\\
&&
\lap\phi+
\frac{\sqrt{2}}{4\cdot 2!}\,\e^{-\phi/\sqrt{2}}\,
F^2_{\left(2\right)}+\frac{\sqrt{2}}{2\cdot 3!}\,\e^{-\sqrt{2}\phi}
F^2_{\left(3\right)}-\sqrt{2}\,\e^{\phi/\sqrt{2}}\,\Lambda=0\,,
   \label{nss:scalar:Eq}\\
&&
  d\left[\e^{-\phi/\sqrt{2}}\ast F_{\left(2\right)}\right]=0\,,
   \label{nss:gauge-r:Eq}\\
&&
d\left[\e^{-\sqrt{2}\phi}\ast F_{\left(3\right)}\right]=0\,,
   \label{nss:gauge-s:Eq}
}
where $\lap$ denotes the Laplace operator with respect to the six-dimensional 
metric $g_{MN}$\,. 

We look for solutions whose spacetime
metric has the form 
\Eqr{
ds^2&=&h^{1/2}_2(t, y, z)h_3^{1/2}(t, y, z)\left[-h^{-2}_2(t, y, z)
h^{-1}_3(t, y, z)dt^2\right.\nn\\
&&\left.+h^{-1}_3(t, y, z)dy^2
+u_{ab}(\Zsp)dz^adz^b\right], 
 \label{nss:metric:Eq}
}
where $u_{ab}(\Zsp)$ is the four-dimensional metric which
depends only on the four-dimensional coordinates $z^a$. 
The scalar field $\phi$ and field 
strengths $F_{\left(2\right)}$, and $F_{\left(3\right)}$ are written, 
respectively, by
\Eqrsubl{nss:ansatz:Eq}{
\e^{\phi}&=&\left(h_2h_3\right)^{-\sqrt{2}/2}\,,
  \label{nss:phi:Eq}\\
F_{\left(2\right)}&=&d\left[\sqrt{2}\,h^{-1}_2(t, y, z)\right]\wedge dt,
  \label{nss:fr:Eq}\\
F_{\left(3\right)}&=&d\left[h^{-1}_3(t, y, z)\right]\wedge dt\wedge dy,
  \label{nss:fs:Eq}
}

First we consider the Einstein equation (\ref{nss:Einstein:Eq}). 
By using the ansatz (\ref{nss:metric:Eq}) and (\ref{nss:ansatz:Eq}), 
the Einstein equations become
\Eqrsubl{nss:cEinstein:Eq}{
&&\frac{5}{4}h_2^{-1}\pd_t^2h_2
+\frac{3}{4}h_3^{-1}\pd_t^2h_3+
\frac{1}{4}h_2^{-2}\left(3h_2^{-1}\pd_y^2h_2+h_3^{-1}\pd_y^2h_3\right)
+\frac{1}{4}h_2^{-2}h_3^{-1}
\left(3h_2^{-1}\lap_{\Zsp}h_2+h_3^{-1}\lap_{\Zsp}h_3\right)\nn\\
&&~~~~-\frac{1}{2}h_2^{-2}h_3^{-1}\Lambda
+\frac{1}{4}\left(\pd_t\ln h_2\right)^2
+\frac{7}{4}\pd_t\ln h_2\pd_t\ln h_3
-\frac{3}{4}h_2^{-2}(1-h_3)\left(\pd_y\ln h_2\right)^2\nn\\
&&~~~~+\frac{3}{4}h_2^{-2}\pd_y\ln h_2\pd_y\ln h_3
-\frac{3}{4}h_2^{-2}h_3^{-1}(1-h_3)u^{ab}\pd_a\ln h_2\pd_b\ln h_2\nn\\
&&~~~~-\frac{1}{4}h_2^{-2}h_3^{-1}(1-h_2^2)u^{ab}\pd_a\ln h_3\pd_b\ln h_3=0,
 \label{nss:cEinstein-tt:Eq}\\
&&2h_2^{-1}\pd_t\pd_y h_2+2h_3^{-1}\pd_t\pd_y h_3
+\pd_t\ln h_2\pd_y\ln h_3+3\pd_t\ln h_3\pd_y\ln h_2=0,
 \label{nss:cEinstein-ty:Eq}\\
&&2h_2^{-1}\pd_t\pd_a h_2+h_3^{-1}\pd_t\pd_a h_3
+\pd_t\ln h_2\pd_a\ln h_3+\pd_t\ln h_3\pd_a\ln h_2=0,
 \label{nss:cEinstein-ta:Eq}\\
&&\frac{1}{4}h_2^2\left(h_2^{-1}\pd_t^2h_2-h_3^{-1}\pd_t^2h_3\right)
-\frac{1}{4}\left(h_2^{-1}\pd_y^2h_2+3h_3^{-1}\pd_y^2h_3\right)
-\frac{1}{4}h_3^{-1}
\left(h_2^{-1}\lap_{\Zsp}h_2-h_3^{-1}\lap_{\Zsp}h_3\right)\nn\\
&&~~~-\frac{1}{2}h_3^{-1}\Lambda+\frac{1}{4}\left(\pd_th_2\right)^2
-\frac{1}{4}h_2^2\pd_t\ln h_2\pd_t\ln h_3
-\frac{3}{4}(1-h_3)\left(\pd_y\ln h_2\right)^2
-\frac{5}{4}\pd_y\ln h_2\pd_y\ln h_3\nn\\
&&~~~+\frac{1}{4}h_3^{-1}(1-h_3)u^{ab}\pd_a\ln h_2\pd_b\ln h_2
-\frac{1}{4}h_3^{-1}(1-h_2^2)u^{ab}\pd_a\ln h_3\pd_b\ln h_3=0,
 \label{nss:cEinstein-yy:Eq}\\
&&h_3^{-1}\pd_y\pd_a h_3+2\pd_y\ln h_2\pd_a\ln h_2
+\pd_y\ln h_2\pd_a\ln h_3+\pd_y\ln h_3\pd_a\ln h_2=0,
 \label{nss:cEinstein-ya:Eq}\\
&&R_{ab}(\Zsp)+\frac{1}{4}h_2^2h_3u_{ab}
\left(h_2^{-1}\pd_t^2 h_2+h_3^{-1}\pd_t^2 h_3\right)
-\frac{1}{4}h_3u_{ab}
\left(h_2^{-1}\pd_y^2 h_2+h_3^{-1}\pd_y^2 h_3\right)\nn\\
&&~~~-\frac{1}{4}u_{ab}
\left(h_2^{-1}\lap_{\Zsp}h_2+h_3^{-1}\lap_{\Zsp}h_3\right)
+\frac{1}{4}h_2^2h_3u_{ab}\left[\left(\pd_t\ln h_2\right)^2
+3\pd_t\ln h_2\pd_t\ln h_3\right]\nn\\
&&~~~+\frac{1}{4}h_3\left(1-h_3\right)u_{ab}\left(\pd_y\ln h_2\right)^2
-\frac{1}{4}h_3u_{ab}\pd_y\ln h_2\pd_y\ln h_3
+\frac{1}{4}(1-h_3)u_{ab}u^{cd}\pd_c\ln h_2\pd_d\ln h_2\nn\\
&&~~~+\frac{1}{4}(1-h_2^2)u_{ab}u^{cd}\pd_c\ln h_3\pd_d\ln h_3
-(1-h_3)\pd_a\ln h_2\pd_b\ln h_2\nn\\
&&~~~-\frac{1}{2}(1-h_2^2)\pd_a\ln h_3\pd_b\ln h_3
-\frac{1}{2}\left(\pd_a\ln h_2\pd_b\ln h_3
+\pd_a\ln h_3\pd_b\ln h_2\right)-\frac{1}{2}u_{ab}\,\Lambda=0,
  \label{nss:cEinstein-ab:Eq}
}
where $\triangle_{\Zsp}$ denotes the Laplace operator on $\Zsp$ space and 
$R_{ab}(\Zsp)$ is the Ricci tensor 
constructed from the metric $u_{ab}(\Zsp)$\,.

We next consider the gauge field equations 
(\ref{nss:gauge-r:Eq}) and (\ref{nss:gauge-s:Eq}).
Under the assumption (\ref{nss:ansatz:Eq}), 
the gauge field equations are written by 
\Eqrsubl{nss:gauge2:Eq}{
&&d\left[h_3^2\,\pd_y h_2\,\Omega(\Zsp)
+h_3\,\pd_a h_2 dy\wedge\left(\ast_{\Zsp}dz^a\right)\right]=0,
  \label{nss:gauge2-r:Eq}\\
&&d\left[h_2\,\pd_a h_3\left(\ast_{\Zsp}dz^a\right)\right]=0,
  \label{nss:gauge2-s:Eq}
 }
where $\ast_{\rm Z}$ denotes the Hodge operator on Z and 
$\Omega(\Zsp)$ is the volume 4-form on Z space: 
\Eq{
\Omega(\Zsp)=\sqrt{u}\,dz^1\wedge dz^2\wedge dz^3 \wedge dz^4\,.
\label{nss:volume:Eq}
}
Here, $u$ is the determinant of the metric $u_{ab}$\,. 

Finally we consider the equation of motion for the scalar field. 
Substituting the ansatz (\ref{nss:ansatz:Eq})
 into Eq.~(\ref{nss:scalar:Eq}), we have
\Eqr{
&&h_2^2h_3\left(h_2^{-1}\pd_t^2h_2+h_3^{-1}\pd_t^2h_3\right)
+h_3\left(\pd_th_2\right)^2
+3h_2\pd_th_2\pd_th_3-h_3\left(h_2^{-1}\pd_y^2h_2
+h_3^{-1}\pd_y^2h_3\right)\nn\\
&&~~~~+h_3\left(1-h_3\right)\left(\pd_y\ln h_2\right)^2
-h_2^{-1}\pd_yh_2\pd_yh_3-h_2^{-1}\lap_{\Zsp}h_2-h_3^{-1}\lap_{\Zsp}h_3\nn\\
&&~~~~+\left(1-h_3\right)u^{ab}\pd_a\ln h_2\pd_b\ln h_2
+\left(1-h_2^2\right)u^{ab}\pd_a\ln h_3\pd_b\ln h_3-2\Lambda=0\,.
  \label{nss:scalar-e:Eq}
}

Now we consider the two cases. 
One is $\pd_th_2\ne 0$ and $\pd_th_3=0$. The other is 
$\pd_th_2= 0$ and $\pd_th_3\ne 0$. 
Upon setting $h_2=1$, the field equations reduce to
\Eqrsubl{nss:equation:Eq}{
&&R_{ab}(\Zsp)=0\,,
   \label{nss:Ricci:Eq}\\
&&h_2=1\,,~~~~h_3=k_0(t)+k_1(y)+k_2(z),
   \label{nss:h1:Eq}\\
&&\frac{d^2k_0}{dt^2}=\Lambda, ~~~~~
\frac{d^2k_0}{dt^2}=-\frac{d^2k_1}{dy^2}\,,~~~~\triangle_{\Zsp}k_2=0\,.
   \label{nss:h2:Eq}
 }
We can also choose the solution in which the
0-brane part depends on $t$. Then, we have 
\Eqrsubl{nss:solution2:Eq}{
&&R_{ab}(\Zsp)=0,
   \label{nss:Ricci2:Eq}\\
&&h_2=h_2(t, v),~~~~h_2=K_0(t)+K_1(v)\,,~~~~~h_3=1\,,
   \label{nss:h2-2:Eq}\\
&&\left(\frac{dK_0}{dt}\right)^2=2\Lambda,~~~~\triangle_{\rm W}K_1=0\,,
   \label{nss:warp2:Eq}
 }
where $\lap_{\rm W}$ denotes Laplace operator with respect to the metric 
$w_{mn}$\,: 
\Eq{
w_{mn}dv^mdv^n=dy^2+u_{ab}(\Zsp)dz^adz^b\,,~~~~
\triangle_{\rm W}K_1=\pd_y^2K_1+\triangle_{\Zsp}K_1\,.
   \label{nss:lapW:Eq}
}
Here, $w_{mn}$ is the five-dimensional metric, and $v^m$ 
denotes the five-dimensional coordinate. 

As a special example, we consider the case
\Eq{
u_{ab}=\delta_{ab}\,,~~~~~~h_2=1\,,
 \label{nss:flat:Eq}
 }
where $\delta_{ab}$ the four-dimensional Euclidean metric. 
Then, the solution for $h_3$ can be obtained explicitly as 
\cite{Minamitsuji:2010fp}
\Eqrsubl{nss:solution:Eq}{
h_2&=&1,
 \label{nss:solution-h2:Eq}\\
h_3(t, y, z)&=&\frac{\Lambda}{2}\left(t^2-y^2\right)+c_1t+c_2y
+c_3+\sum_{l=1}^N\frac{M_l}{|z^a-z^a_l|^2},
 \label{nss:solution-h3:Eq}
}
where $c_i~(i=1,~2,~3)$ and $z^a_l$ are constants and 
the parameter $M_l$ is the mass constant of 1-branes\,, 
which is located at $z^a=z^a_l$\,.  

We can obtain the solution
for $h_3=1$ and $\pd_th_2\ne 0$
if the roles of $h_2$ and $h_3$ are exchanged.
The solution of the field equations is then written as
\Eqrsubl{nss:solution3:Eq}{
h_2(t, v)&=&\epsilon\sqrt{2\Lambda}\,t+c_4+\sum_{\alpha=1}^{N'}
\frac{L_{\alpha}}
{|v^m-v^m_{\alpha}|^3}\,,
 \label{nss:solution3-r:Eq}\\
h_3&=&1\,,
 \label{nss:solution3-s:Eq}
}
where $c_4$, $v^m_\alpha$, and $L_\alpha$ are constants and 
$\epsilon=\pm 1$.
The delocalized brane solutions
in the six-dimensional NSS supergravity 
\cite{Nishino:1984gk, Salam:1984cj, Nishino:1986dc,
Gibbons:2003di,Aghababaie:2003ar} have been investigated
in Refs.~\cite{Maeda:1984gq,Maeda:1985es, Cline:2003ak, Vinet:2004bk, 
Vinet:2005dg, 
Peloso:2006cq, Tolley:2006ht, Tolley:2007et,Minamitsuji:2010fp}, 
including applications to cosmological models.
According to the intersection rule, 
the number of the intersections dimensions involving the 0-brane and 
1-brane is $-1$. Although meaningless in ordinary spacetime, 
these configurations are relevant in the Euclidean space, 
for instance, representing instantons.

In the following, we consider cosmological aspects of the solution describing 
time-dependent branes. We first study the time dependence of the 
scale factors in the 0-brane solutions after compactifying the extra 
directions, and our Universe is discussed. Next we discuss the 
dynamical 1-brane solution and apply it to the cosmology. 

\subsubsection{Cosmology in the 0-brane system}

For the solution (\ref{nss:solution3:Eq}), 
we introduce a new time coordinate $\tau$ as
\Eq{
\left(\frac{\tau}{\tau_0}\right) \equiv 
\left(\epsilon\sqrt{2\Lambda}\,t+c_4\right)^{1/4}\,,~~~~
\tau_0 \equiv \frac{4}{\epsilon\sqrt{2\Lambda}}\,.
   \label{nss:ct:Eq}
}
The six-dimensional metric is thus given by
\Eqr{
&&\hspace{-1.2cm}ds^2=
\left[1+\left(\frac{\tau}{\tau_0}\right)^{-4}\bar{h}_2(v)
\right]^{-\frac{3}{2}}
\left[-d\tau^2
+\left\{1+\left(\frac{\tau}{\tau_0}\right)^{-4}
\bar{h}_2(v)\right\}
\left(\frac{\tau}{\tau_0}\right)^2
\delta_{mn}({\rm W})dv^mdv^n
\right],
   \label{nss:metric3:Eq}
}
where $\delta_{mn}$ is the five-dimensional Euclidean metric 
and $\bar{h}_2(v)$ is defined by 
\Eq{
\bar{h}_2(v) \equiv \sum_{\alpha=1}^{N'}\frac{L_{\alpha}}
{|v^m-v^m_{\alpha}|^{3-d_s}}\,.
   \label{nss:bh:Eq}
}
Here $d_{s}$ denotes the number of smeared dimensions and 
should satisfy $0\le d_{s}\le 4$\,.

The six-dimensional spacetime implies $(\tau/\tau_0)^{-4}\,\bar{h}_2(v)=0$ 
in the limit $\tau\rightarrow\infty$\,. 
Then the scale factor of the six-dimensional space is 
proportional to $\tau$\,. 
Although the dynamical 0-brane solutions cannot give a realistic universe 
such as an accelerating expansion, a matter- or a 
radiation-dominated era, there is a possibility that appropriate 
compactification and smearing of the transverse space to the 0-brane
may lead to a realistic expansion. Now we will discuss 
this possibility. 

We consider some compactification and smearing of the extra 
directions of the solutions. 
Our Universe has to be described by the 0-brane solution
with six directions. Since 
the time direction is expressed as $t$\,,
the remaining task is to identify 
the three spatial directions from the coordinates $v^m$. 

In an approach such as the construction of the 
cosmological scenario on the basis of a dynamical brane background, 
three spatial directions are supposed to be on the overall 
transverse space to branes.  
If the spatial directions are specified with $v^m$\,, 
it also works in the present case.  
Then space is isotropic from the expression of the metric. 
Now we look for a way to realize an isotropic 
and homogeneous three-dimensional space in the 0-brane solutions.  

Since we set the coordinates $(t, v^2, v^3, v^4)$ which describes our 
Universe, it is convenient to decompose the six-dimensional metric of 
the solutions into the following form: 
\Eq{
ds^2=ds_4^2+ds^2_{\rm i}\,,
    \label{nss:c-metric:Eq}
}
where each part of the six-dimensional metric is given by 
\Eqrsubl{nss:c-metric2:Eq}{
ds^2_4&=&-h^{-3/2}_2(t, v)dt^2+h^{1/2}_2(t, v)
\delta_{\alpha\beta}dv^\alpha dv^\beta\,, \\
ds^2_{\rm i}&=&h^{1/2}_2(t, v)\delta_{ij}dv^idv^j\,.
}
Here $ds^2_4$ is the metric of the four-dimensional spacetime 
with $t,~v^{\alpha}$~$(\alpha=3, 4, 5)$\,, while $ds^2_{\rm i}$ denotes the 
metric of the internal space.   
We can obtain the compactifications of the solutions depending 
on the internal space. 

The internal space is described by the coordinates $v^i~(i=1,2)$, and  
the spatial part of our Universe $\delta_{\alpha\beta}$ is three-dimensional 
with $v^\alpha~(\alpha=3, 4, 5)$\,. 
Then $\delta_{\alpha\beta}$ and $\delta_{ij}$ are 
the three- and two-dimensional Euclidean metrics, respectively. 

Now we derive the lower-dimensional effective theory by 
compactifying the extra directions. 
In order to find a realistic universe, we compactify the 
$d$-dimensional space to be a $d$-dimensional torus, 
where $d$ is the compactified dimensions for the direction 
of internal space. The remaining noncompact space 
is the external space. 
The range of $d$ is given by $0\le d \le 1$\,, 
because the $v^1$ direction is preserved to measure the position of the 
universe in the overall transverse space. 
Hence the $v^2$ direction will be compactified, 
where the compactified direction has to be smeared out 
before the compactification. 

\begin{figure}[h]
 \begin{center}
\includegraphics[keepaspectratio, scale=0.28, angle=-90]{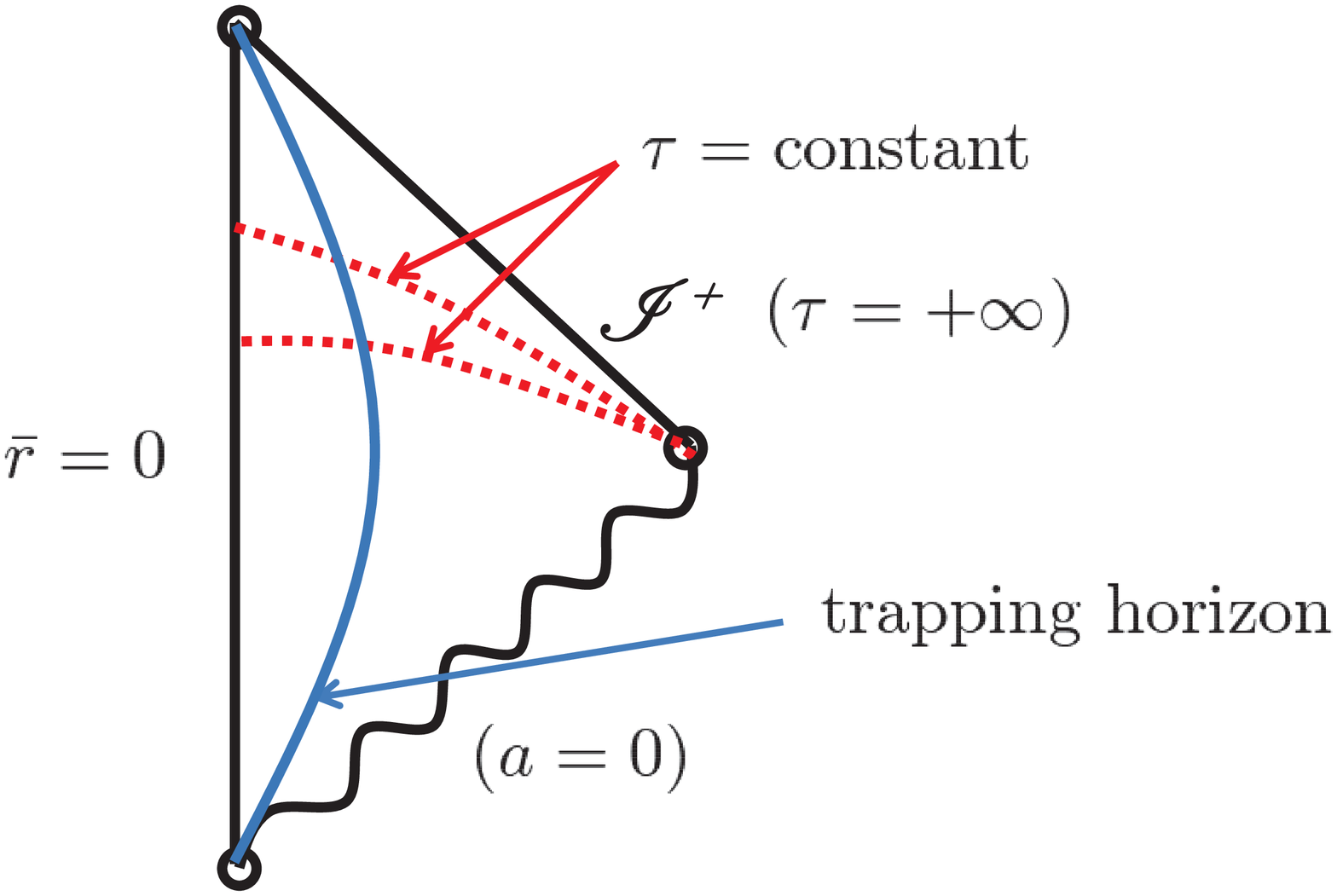}
\hskip 0.5cm
\includegraphics[keepaspectratio, scale=0.28, angle=-90]{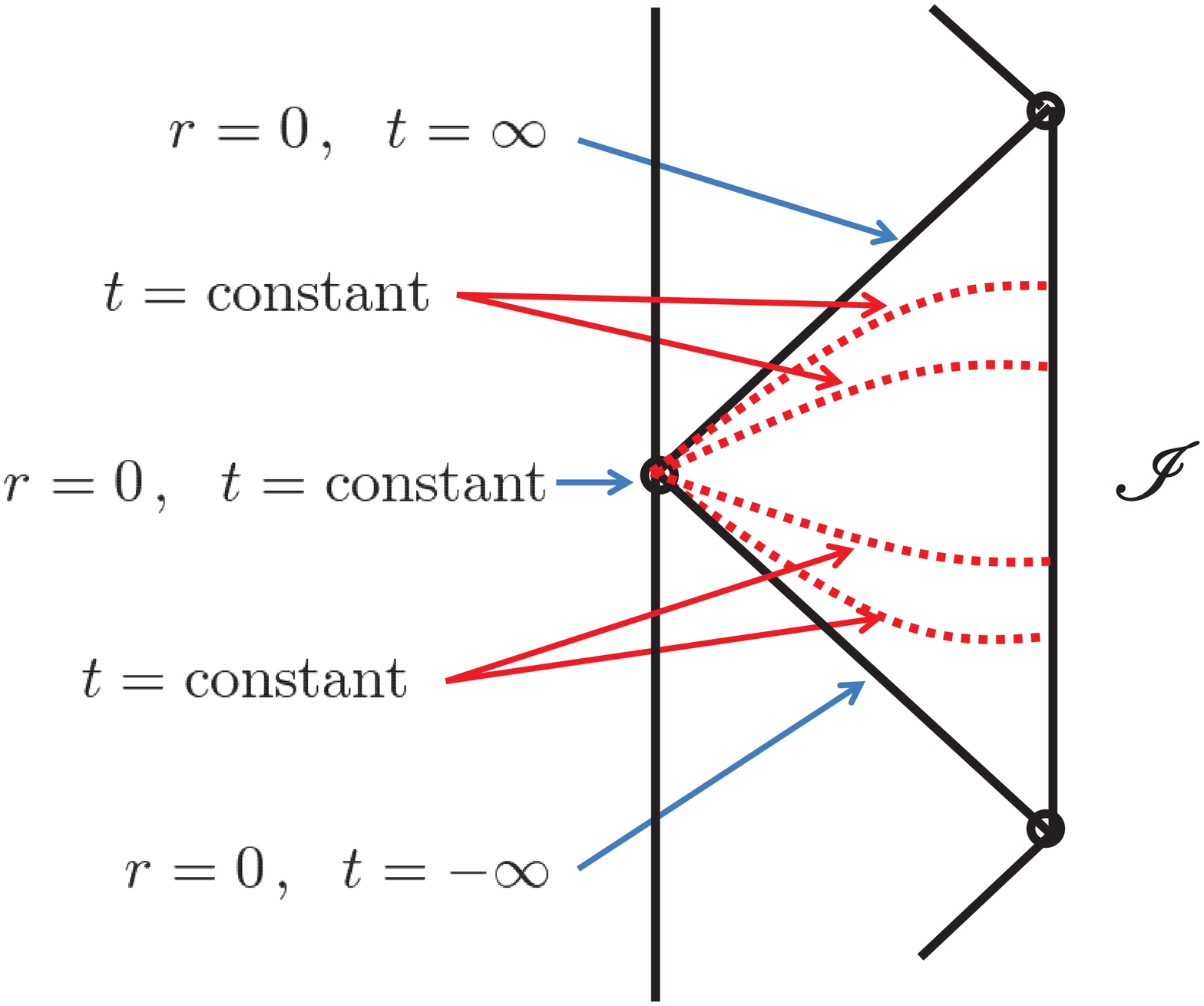}\\
(a) \hskip 7cm (b) ~~~~~~
  \caption{\baselineskip 14pt 
(a)Conformal diagrams of the Milne universe are depicted. 
The solid line denotes the trapping horizon $\bar{r}_{\rm T}\equiv 
da(\tau)/d\tau$, where $a=(\tau/\tau_0)$ \cite{Maeda:2010ja}. 
One can recognize that the asymptotic region of
the spacetime in (\ref{nss:Emetric:Eq}) corresponding to 
$\bar{r}\rightarrow\infty$ approximates the five-dimensional 
Milne universe, where $\bar{r}^2=\delta_{\alpha\beta}v^{\alpha}
v^{\beta}+\left(v^1\right)^2$. (b) The geometry of the 0-brane system 
(\ref{nss:solution3:Eq}) in the limit 
$r\rightarrow 0$ is depicted. The domain corresponding to 
$r\rightarrow 0$ with finite $t$ describes warped 
AdS${}_2\times{\rm S}^4$ spacetime \cite{Maeda:2010ja}.}
  \label{fig:milne}
 \end{center}
\end{figure}

Then the metric (\ref{nss:metric:Eq}) with $h_3=1$  
is recast into the following form: 
\Eq{
ds^2=ds^2_{\rm e}+ds^2_{\rm i}\,,
   \label{nss:c-metric3:Eq}
}
where $ds^2_{\rm e}$ is the metric of $(6-d)$-dimensional external spacetime 
and $ds^2_{\rm i}$ is the metric of compactified dimensions. 
Upon setting $d=1$, the compactified metric in the Einstein frame is
\Eqr{
d\bar{s}^2_{\rm e}&=&
\left[1+\left(\frac{\tau}{\tau_0}\right)^{-6}\bar{h}_2(v)
\right]^{-\frac{5}{3}}
\left[-d\tau^2+\left\{1+\left(\frac{\tau}{\tau_0}\right)^{-6}
\bar{h}_2(v)\right\}^2
\left(\frac{\tau}{\tau_0}\right)^{2}\right.\nn\\
&&\left.\times
\left\{\delta_{\alpha\beta}dv^{\alpha}dv^{\beta}
+\left(dv^1\right)^2\right\}\right],
   \label{nss:Emetric:Eq}
}
where $d\bar{s}^2_{\rm e}$ is the five-dimensional metric in 
the Einstein frame and the constant parameters $\tau_0$ and 
the cosmic time $\tau$ are defined, respectively, as 
\Eq{
\frac{\tau}{\tau_0} \equiv \left(\epsilon\sqrt{2\Lambda}
t\right)^{1/6},~~~~
\tau_0 \equiv \frac{6}{\epsilon\sqrt{2\Lambda}}\,.
   \label{nss:pa:Eq}
}
Since the power exponent of the scale factor is given by 1\,, 
the metric of four-dimensional spacetime 
in the Einstein frame implies that 
the solutions gives rise to a Milne universe. 
To construct a realistic cosmological model such as in the inflationary
scenario, it would be necessary to add some new ingredients in the 
background. Figure \ref{fig:milne} depicts the
conformal diagrams of the five-dimensional spacetime in the limit 
$\tau\rightarrow\infty$. Hence, the asymptotic
regions of the present spacetime (\ref{nss:Emetric:Eq}) resemble the
five-dimensional Milne universe. 

Finally, we discuss the near-horizon geometry of the 0-brane solution. 
When all of the 0-branes are located at the origin of the overall transverse
space, the solution can be expressed as  
\Eq{
h_2(t, r)=\epsilon\sqrt{2\Lambda}\,t+c_4+\frac{L}{r^3}\,,~~~~~~~~
r^2 \equiv \delta_{mn}\, v^m v^n\,,
\label{nh:h2:Eq}
}
where $L$ is the total mass of 0-branes  
\Eq{
L \equiv \sum_{\alpha=1}^{N'} L_{\alpha}\,. 
}
In the near-horizon limit $r\rightarrow 0$\,, 
the dependence on $t$ in (\ref{nh:h2:Eq}) is negligible. 
The six-dimensional metric is thus reduced to the following form:  
\Eqrsubl{nh:nh2-metric:Eq}{
ds^2&=&\left(\frac{L}{r^3}\right)^{-1/6}
\left[ds^2_{\rm AdS_2}+L^{2/3}d\Omega^2_{(4)}\right], \\
ds^2_{\rm AdS_2}&\equiv&-\left(\frac{L^{4/3}}{r^4}\right)^{-1}dt^2
+\frac{L^{2/3}}{r^2}\,dr^2\,,
}
where $\delta_{mn}\, dv^m dv^n=dr^2+r^2d\Omega^2_{(4)}$ has been performed. 
The line elements of a two-dimensional AdS space (AdS$_2$) 
and a four-sphere with the unit radius (${\rm S}^4$) 
are given by $ds^2_{\rm AdS_2}$ and $d\Omega^2_{(4)}$\,, respectively. 
Then the six-dimensional metric (\ref{nh:nh2-metric:Eq}) in 
the near-horizon limit of the 
0-brane system describes a warped product of AdS${}_2$ and ${\rm S}^4$\,. 
Figure \ref{fig:milne} shows the geometry of the
AdS${}_2$ and ${\rm S}^4$\,.

\subsubsection{Cosmology in the 1-brane system}
Now we discuss the cosmological evolution for the time-dependent 1-brane 
solution (\ref{nss:solution:Eq}).
We define the cosmic time $\tau$, which is given by
\Eq{
\left(\frac{\tau}{\tau_0}\right) \equiv 
\left(\frac{\Lambda}{2}\,t^2\right)^{1/4}\,,~~~~
\tau_0 \equiv \frac{2\sqrt{2}}{\sqrt{\Lambda}}\,,~~~~c_1=0\,.
   \label{nss:ct2:Eq}
}
The six-dimensional metric is expressed as 
\Eqr{
&&\hspace{-0.8cm}ds^2=
\left[1+\left(\frac{\tau}{\tau_0}\right)^{-4}\bar{h}_3(y, z)
\right]^{-\frac{1}{2}}
\left[-d\tau^2
+\left(\frac{\tau}{\tau_0}\right)^{-2}dy^2
\right.\nn\\
&&\left.+\left\{1+\left(\frac{\tau}{\tau_0}\right)^{-4}
\bar{h}_3(y, z)\right\}
\left(\frac{\tau}{\tau_0}\right)^2
\delta_{ab}(\Zsp)dz^adz^b\right],
   \label{nss:metric4:Eq}
}
where $\bar{h}_3(y, z)$ is defined by 
\Eq{
\bar{h}_3(y, z) \equiv -\frac{\Lambda}{2} y^2+c_2y
+c_3+\sum_{l=1}^N\frac{M_l}{|z^a-z^a_l|^{2-d_{s}}}\,.
}
Here $d_{s}$ is the number of smeared dimensions and 
should satisfy $0\le d_{s}\le 3$\,. 
In order to fix the location of our Universe in the transverse space,
let us assume that at least one direction of $z^a~(a=1,\ldots,4)$ is 
not smeared.  

Now we apply the 1-brane solution to the lower-dimensional effective theory. 
Let us consider a compactification and smearing of the extra 
directions of the 1-brane solution. 
First of all, our Universe is described by  the solutions with the 
six-dimensional coordinates $t,~y,~z^a~(a=1,\ldots,4)$\,. 
The time direction is written by $t$. 
Our choice is to take the three-dimensional 
from the overall transverse space with $z^a$. 
The four-dimensional universe is spanned by $t$, $z^2$, $z^3$, and $z^4$\,, 
for instance. The $z^1$ direction is preserved to measure the position 
of our Universe in the overall transverse space of the 1-brane. 
Since the metric depends on $z^a$ explicitly, we have to smear 
out ${z}^2$\,, ${z}^3$\,, and ${z}^4$ so as to define our Universe. 
Then the number of the smeared directions 
$d_{s}$ should satisfy the condition $d_{s}=3$\,. 

It is necessary to take that $c_2=0$ and $\Lambda=0$ in (\ref{nss:metric:Eq}) 
to compactify the $y$ direction.  
We compactify the $y$ direction to fit our Universe, where $y$ denotes the
compactified dimensions with respect to the world volume of the 1-brane. 
The metric (\ref{nss:metric:Eq}) with $h_2=1$ is then described by 
(\ref{nss:c-metric3:Eq}). 

In terms of the conformal transformation
\Eq{
ds^2_{\rm e}=h_3^{1/6}d\bar{s}^2_{\rm e}\,,
}
we can rewrite the $(6-d)$-dimensional metric in the Einstein frame. 
If we set $d=1$, the five-dimensional metric in the Einstein frame is
\Eqr{
d\bar{s}^2_{\rm e}&=&-h_3^{-2/3}(t, z)dt^2
+h_3^{1/3}(t, z)\delta_{ab}(\Zsp)dz^adz^b\,, 
 \label{nss:metric5:Eq}
}
where $d\bar{s}^2_{\rm e}$ is the metric of five-dimensional external 
spacetime in the Einstein frame.
For $h_3=c_1t+\bar{h}_3(z)$, the metric (\ref{nss:metric5:Eq}) 
is thus rewritten as
\Eqr{
d\bar{s}^2_{\rm e}&=&-
\left[1+\left(\frac{\tau}{\tau_0}\right)^{-3/2}\bar{h}_3(z)
\right]^{-2/3}d\tau^2\nn\\
&&+\left[1+\left(\frac{\tau}{\tau_0}\right)^{-3/2}\bar{h}_3(z)
\right]^{1/3}\left(\frac{\tau}{\tau_0}\right)^{1/2}
\left[\delta_{\alpha\beta}dz^{\alpha}dz^{\beta}
+\left(dz^1\right)^2\right], 
 \label{nss:metric6:Eq}
}
where the spatial part of our Universe $\delta_{\alpha\beta}$ 
is three-dimensional with $z^\alpha~(\alpha=2, 3, 4)$\,, and 
the constant parameters $\tau_0$ and 
the cosmic time $\tau$ are defined, respectively, as 
\Eq{
\frac{\tau}{\tau_0} \equiv \left(c_1t\right)^{2/3},~~~~
\tau_0 \equiv \frac{3}{2c_1}\,.
   \label{nss:pa2:Eq}
}

Unfortunately, the power exponent of the four-dimensional 
universe in the Einstein frame becomes 1/4. 
Hence, we have to conclude that, in order to obtain a realistic expansion
of the universe in this type of models, one has to include additional 
fields on the background.

Let us finally consider the case of the near-horizon limit that 
the spacetime metric and the functions $h_2$ and 
$h_3$ satisfy (\ref{nss:solution:Eq}). 
If we consider the case where $N$ 1-branes are located at the origin of
the Z space, we have
\Eq{
h_3(t, r)=\frac{\Lambda}{2} \left(t^2-y^2\right)+c_1t+c_2y
+c_3+\frac{M}{r^2}\,,~~~~~~~~
r^2 \equiv \delta_{ab}\, z^a z^b\,,
\label{nh:h3:Eq}
}
where $M$ is the total mass of 1-branes  
\Eq{
M \equiv \sum_{l=1}^{N} M_{l}\,. 
}
Since the dependence on $t$ and $y$ in (\ref{nh:h3:Eq}) is negligible 
in the near-horizon limit $r\rightarrow 0$\,, 
 the six-dimensional metric is reduced to the following form: 
\Eq{
ds^2=\left(\frac{M}{r^2}\right)^{-1/2}
\left[-dt^2+dy^2+M\left\{\frac{dr^2}{r^2}
+d\Omega^2_{(3)}\right\}\right], 
}
where $\delta_{ab}\, dz^a dz^b=dr^2+r^2d\Omega^2_{(3)}$ has been used. 
The line elements of a three-dimensional space (M${}_3$) 
and a three-sphere  
are given by $ds^2_{\rm M_3}$, $d\Omega^2_{(3)}$\,, respectively. 
Thus we see that the near-horizon limit of the 1-brane system is a 
warped product of M${}_3$ with a certain internal 3-space with a circle.

\subsection{Collision of the 0-brane in Nishino-Salam-Sezgin 
gauged supergravity}
\label{sec:ns-c}

We next study the behavior of the time-dependent 0-brane 
solution (\ref{nss:solution3:Eq})\,. 
Substituting (\ref{nss:solution3:Eq}) into the metric 
(\ref{nss:metric:Eq})\,, the six-dimensional metric is expressed as 
\Eq{
ds^2=-\left[\epsilon\sqrt{2\Lambda}\,t+c_4+\bar{h}_2(v)\right]^{-3/2}
dt^2+\left[\epsilon\sqrt{2\Lambda}\,t+c_4+\bar{h}_2(v)\right]^{1/2}
w_{mn}dv^mdv^n\,,
   \label{nss:surface:Eq}
}
where $w_{mn}$ is given by (\ref{nss:lapW:Eq}), and 
the function $\bar{h}_2(v)$ is defined by (\ref{nss:bh:Eq}).
Since the time dependence appears through the function $h_2$, 
the next task is to study the time evolution of the solutions carefully. 
Hereafter we will consider it by focusing upon collision of 
0-branes. We also discuss smearing out some of the directions 
in the transverse space to decrease the number of transverse dimensions 
to 0-brane effectively.  
  
Now we consider the case that the number of the smeared direction 
is given by $d_{s}$\,. Then the function $\bar{h}_2(v)$ 
can be expressed as
\Eq{
\bar{h}_2(v) \equiv \sum_{\alpha}\frac{L_{\alpha}}
{|v^m-v^m_{\alpha}|^{3-d_{s}}}\,,
   \label{nss:sme:Eq}
}
where $d_{s}$ is the number of smeared dimensions and 
should satisfy $0\le d_{s}\le 4$\,, and we assume that 
at least one direction of $v^m~(m=1,\ldots,5)$ is not smeared 
in order to fix the location of our universe in the transverse space. 
In the following, 
we will use the function (\ref{nss:sme:Eq}).

We will discuss the asymptotic behavior of the time-dependent solutions. 
In the limit of $v^m\rightarrow \,v^m_{\alpha}$, 
the time dependence in the function $h_2$ can be ignored because 
the harmonic function $\bar{h}_2(v)$ dominates near a position of 0-brane. 
On the other hand, 
function $\bar{h}_2(v)$ vanishes in the limit of $v^m\rightarrow \infty$. 
Then the system becomes static near 0-brane 
while $h_2$ depends only on time $t$ in the far region from 0-branes. 
Thus the six-dimensional metric in the limit of $v^m\rightarrow \infty$\,, 
is rewritten by
\Eqr{
ds^2=-\left(\epsilon\sqrt{2\Lambda}\,t+c_4\right)^{-3/2}
dt^2+\left(\epsilon\sqrt{2\Lambda}\,t+c_4\right)^{1/2}
w_{mn}dv^mdv^n\,.
   \label{nss:surface2:Eq}
}

The metric has singularity at $h_2=0$\,. Then the spacetime is regular 
if it is restricted inside the domain specified by the conditions,  
\Eq{
h_2(t, v)=\epsilon\sqrt{2\Lambda}\,t+c_4+\bar{h}_2(v)>0\,,
}
where the function $\bar{h}_2(v)$ is defined in (\ref{nss:sme:Eq}). 
Since the spacetime evolves into a curvature singularity, 
the six-dimensional spacetime cannot be extended beyond this region. 
The regular spacetime with 0-branes ends up with the singularities. 

The evolution of the spacetime highly depends on the signature of 
$\tilde{\Lambda}\left(\equiv \epsilon\sqrt{2\Lambda}\right)$\,. 
The system with $\tilde{\Lambda}>0$ has the time 
reversal one of $\tilde{\Lambda}<0$. 
Now we will discuss the case with $\tilde{\Lambda}<0$.  
For $t<0$, the spacetime is not singular because 
the function $h_2$ is positive everywhere. 
In the limit of $t\rightarrow -{\infty}$, the solution is approximately 
given by a time-dependent uniform spacetime apart 
from a position of 0-branes. In the vicinity of branes, the geometry 
takes a cylindrical form of infinite throat.   

We study the time evolution for $t>0$ and $c_4=0$\,. 
At $t=0$\,, the spacetime is regular everywhere and 
has a cylindrical topology near each 0-brane. 
As time slightly evolves, a curvature singularity 
appears as $|v^m-v^m_{\alpha}|\rightarrow\infty$\,. 
The singular hypersurface cuts off more and more of the 
space as time increases further. 
When time continues to evolve, the singular hypersurface eventually splits 
and surrounds each of the 0-brane throats individually. 
Hence the spatial surface is composed of each isolated throats. 
For $t<0$, the time evolution of the six-dimensional spacetime 
is the time reversal of $t>0$.

Since the metric (\ref{nss:surface2:Eq}) in the regular domain implies that 
the overall transverse space tends to expand asymptotically like $t^{1/4}$\,, 
for any values of fixed $v^m$, the solutions describe static 
0-branes near the positions of the branes. 
In the far region as $|v^m-v^m_{\alpha}| 
\rightarrow \infty$\,, the solutions approach 
FRW universes with the power law expansion $t^{1/4}$\,. The emergence of 
FRW universes is an important feature of the time-dependent 
0-brane solutions.

We will discuss whether two 0-branes can collide or not. 
We put the two 0-branes at $\vect{v}_1=(0, 0, \ldots, 0)$ 
and $\vect{v}_2=(\xi, 0, \ldots, 0)$\,, 
where $\xi$ is a constant. 
If we introduce the following quantity
\Eq{
\tilde{v}=\sqrt{\left(v^2\right)^2+\left(v^3\right)^2+\cdots 
+\left(v^{5-d_{s}}\right)^2}\,,
}
the proper distance at $\tilde{v}=0$ between the two 0-branes 
is given by  
\Eqr{
d(t)&=&\int^{\xi}_0 dv^1 
\left(\epsilon\sqrt{2\Lambda}\,t+c_4+\frac{L_1}{|v^1|^{3-d_{s}}}
+\frac{L_2}{|v^1-\xi|^{3-d_{s}}}\right)^{1/4}\,, 
\label{nss:distance:Eq}
}
where $L_1$ and $L_2$ are the charges of the 0-brane. 
For $\epsilon=-1$, this is a monotonically decreasing function of $t$. 
The behavior of the proper length 
is different depending on the number of the smeared directions $d_{s}$\,. 
We will discuss it for each of the values of $d_{s}$ below. 

\begin{figure}[h]
 \begin{center}
\includegraphics[keepaspectratio, scale=0.55, angle=0]{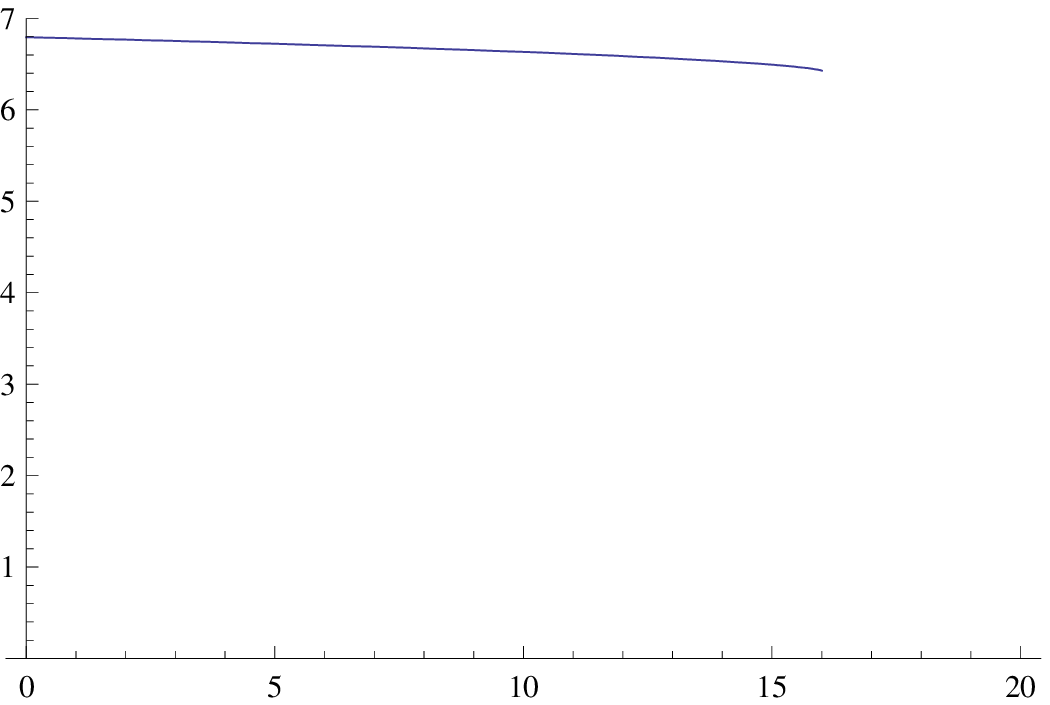}
\put(-175,120){$d(t)$}
\put(10,10){$t$}
\hskip 2.0cm
\includegraphics[keepaspectratio, scale=0.55, angle=0]{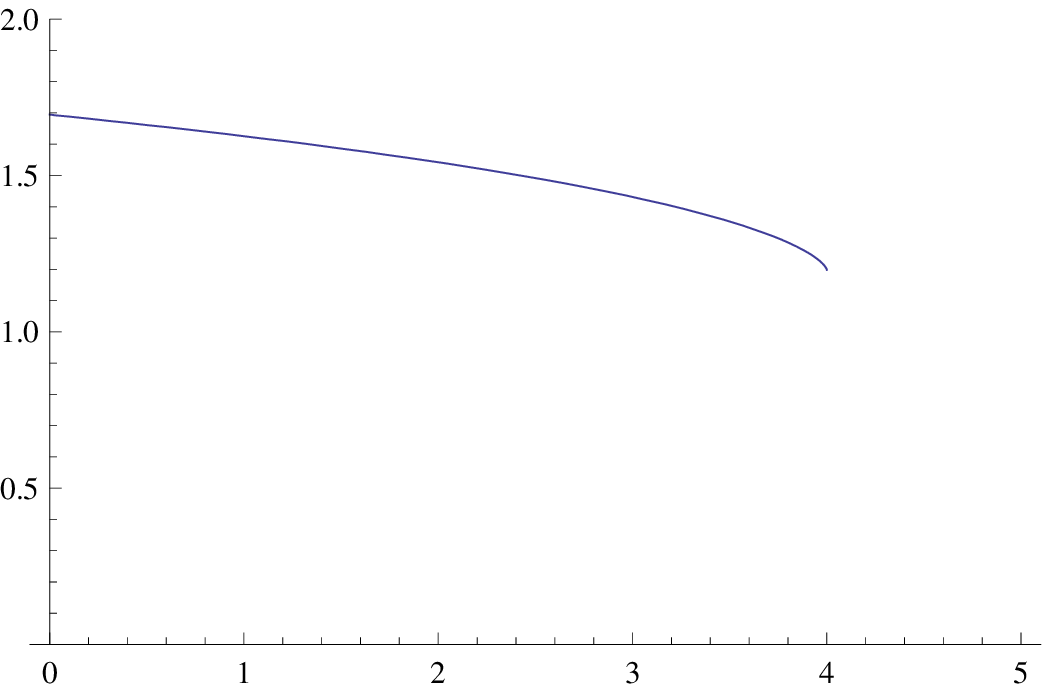}
\put(-175,120){$d(t)$}
\put(10,10){$t$}\\
(a) \hskip 7cm (b) ~~~~~~
  \caption{\baselineskip 14pt 
 The time evolution of the proper distance between two dynamical 0-branes 
 for $d_s=0$ (a) and $d_s=2$ (b) in the six-dimensional 
 Nishino-Salam-Sezgin gauged supergravity. For both cases, 
 the two 0-brane charges are identical, 
$L_1=L_2=1$ and the parameters are taken as 
$c_4=0$\,, $\Lambda=0.5$\,, $\epsilon=-1$\,, and $\xi=1$. 
The result is also the same and a singularity develops before
collision of 0-branes.}
  \label{fig:nss0}
 \end{center}
\end{figure}

First we consider the case with $d_s \le 3$. 
The proper length is plotted in Fig.~\ref{fig:nss0} for the cases with 
$d_s=0$ and $d_s=2$. Since both cases show that a singularity appears
before the proper distance becomes zero, the singularity between 
two 0-branes appears before collision. The two 0-branes approach 
very slowly, and then the singular hypersurface suddenly appears at a 
finite proper distance. The spacetime finally splits into two isolated 
0-brane throats. Therefore one cannot see collision of the 0-branes in these 
examples. For the other case with $d_s=1$, the result is the same.

Next we consider the case with $d_{s}=4$\,, and 
assume that the $v^m$ directions apart from $v^1$ are smeared. 
Since the function $\bar{h}_2$ is linear in $v$\,, 
the behavior of the proper distance is different from the previous case.  
The six-dimensional metric is now given by (\ref{nss:surface:Eq})\,. 
By choosing $v=v^1$\,, the harmonic function $\bar{h}_2$ is written by 
\Eq{
\bar{h}_2(v)=\sum_{\alpha=1}^{N'}L_{\alpha}
|\,v-\,v_{\alpha}|\,.
    \label{nss:bh2:Eq}
}

We discuss the time-dependent solutions in the case that one 0-brane 
charge $L_1$ is located at $v=0$ and the other $L_2$ at $v=\xi$\,. 
The proper length between the two 0-branes is given by
\Eqr{
d(t)&=&\int^{\xi}_0 dv \left[\epsilon\sqrt{2\Lambda}\,t+c_4
   +\left(L_1|v|+L_2|v-\xi|\right)\right]^{1/4}\,.
  \label{nss:length:Eq}
}
For $\epsilon=-1$, the proper distance decreases with time. 
If we set $L_1\ne L_2$\,, a singularity appears again at 
a certain finite time $t=t_{\rm s}$\,, 
while the proper distance is still finite, where $t_{\rm s}$ is defined as 
\Eq{
t_{\rm s}\equiv\frac{c_4+L_1|v|+L_2|v-\xi|}{\sqrt{2\Lambda}}\,. 
}
This is the same result as the case with $d_{s}\leq 3$\,. 

On the other hand, two 0-branes have the same brane charge $L_1=L_2=L$\,, 
the proper distance vanishes at a certain finite time $t=t_{\rm c}$\,, 
where $t_{\rm c}$ is defined by   
\Eq{
t_{\rm c} \equiv \frac{c_4+L\xi}{\sqrt{2\Lambda}}\,.
}
Hence two 0-branes can collide completely. 

\begin{figure}[h]
 \begin{center}
\includegraphics[keepaspectratio, scale=0.55, angle=0]{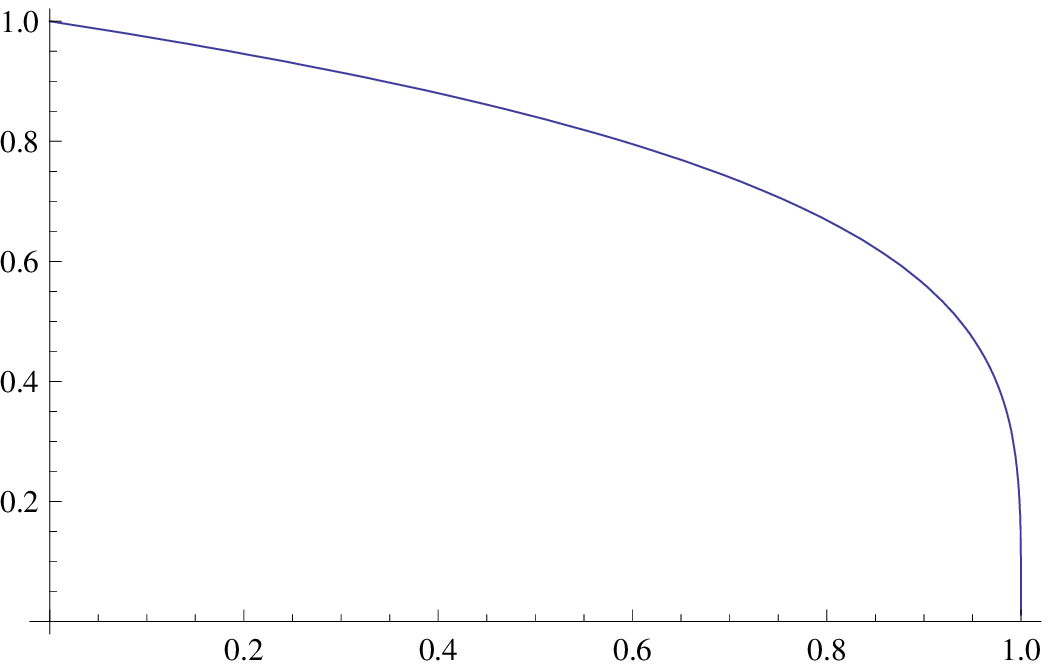}
\put(-175,115){$d(t)$}
\put(10,10){$t$}
\hskip 2.0cm
\includegraphics[keepaspectratio, scale=0.55, angle=0]{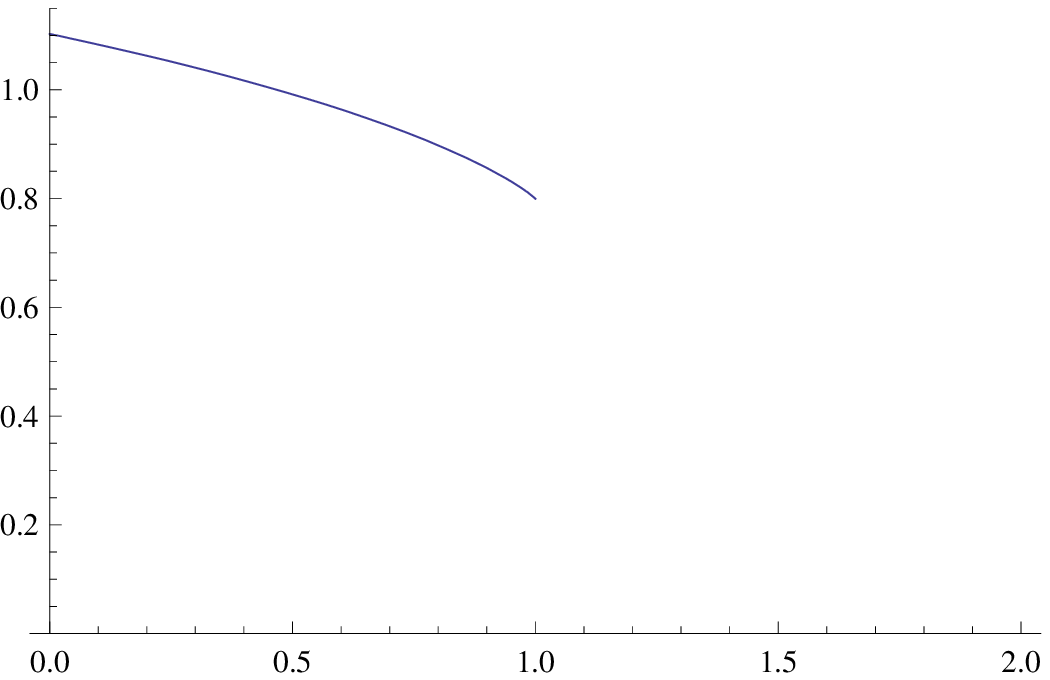}
\put(-175,115){$d(t)$}
\put(10,10){$t$}\\
(a) \hskip 7cm (b) ~~~~~~
  \caption{\baselineskip 14pt 
 The time evolution of the proper distance between two dynamical 0-branes 
 for $L_1=L_2=1$ (a) and $L_1=2$, $L_2=1$ (b) in the six-dimensional 
 Nishino-Salam-Sezgin gauged supergravity. We fix $d_s=4$\,, $c_4=0$\,, 
 $\Lambda=0.5$\,, 
$\epsilon=-1$\,, and $\xi=1$. 
The proper distance rapidly vanishes near two
0-branes collide for the case of $L_1=L_2=1$. 
While for the case of $L_1=2$, $L_2=1$, it 
is still finite when a curvature singularity appears.}
  \label{fig:nss}
 \end{center}
\end{figure}

In terms of $t_{\rm c}$\,, the proper length is expressed as 
\Eqr{
d(t)=L\left[-\sqrt{2\Lambda}(t-t_{\rm c})\right]^{1/4}\,.
} 
If we choose the values as 
$c_4=0$\,, $\Lambda=0.5$\,,  $\xi=1$ and 
$\epsilon=-1$\,, the proper distance $d(t)$ is plotted in 
Fig.~\ref{fig:nss} 
for the two cases (a) the same 0-brane charges $L_1=L_2=1$ and (b) different 
charges $L_1=2$, $L_2=1$\,. 
In the case (a) the two 0-branes can collide completely. 
However, in the case (b) a singularity appears before collision, 
as we have already discussed analytically.

\subsection{Collision of the 1-brane in Nishino-Salam-Sezgin 
gauged supergravity}

Now we apply our time-dependent solutions to 
a collision of 1-brane systems. 
In the case of $h_2=1$, the function $h_3$ is assumed to be
\Eq{
h_3(t, y, z)=\frac{\Lambda}{2}\left(t^2-y^2\right)+c_3+\tilde{h}(z)\,,
 \label{nsc:h:Eq}
}
where $c_3$ is a constant parameter, we choose $c_1=c_2=0$, and
the harmonic function $\tilde{h}$ is expressed as
\Eqrsubl{nsc:h2:Eq}{
\tilde{h}(z)&=&\sum_{l=1}^{N}\frac{M_l}{|z^a-z^a_l|^{2-d_s}}\,,
~~~~~~{\rm for}~~d_s\neq 2\,,
\label{nsc:h2-1:Eq}
\\
\tilde{h}(z)&=&\sum_{l=1}^{N}M_l\, \ln |z^a-z^a_l|\,,
~~~~{\rm for}~~ d_s=2\,.
\label{nsc:h2-2:Eq}
}
Here $M_l$ are charges of 1-branes located at $z^a=z^a_l$ and 
\Eq{
|z^a-z^a_l|
=\sqrt{\left(z^1-z^1_l\right)^2+\left(z^2-z^2_l\right)^2+\cdots+
\left(z^{4-d_s}-z^{4-d_s}_l\right)^2}\,,
}
 because the harmonic function $\tilde{h}$ is defined on the
$(4-d_s)$-dimensional Euclidean subspace in $\Zsp$.
The six-dimensional metric, scalar field, 
and gauge field of the solution are
 given by Eqs.~(\ref{nss:metric:Eq}), and
(\ref{nss:ansatz:Eq}), respectively.
We see that $d_s=2$ case is critical.
For $d_s=3$, the function $\tilde{h}$ is written by 
the sum of linear functions of $z$. 
The possibility of 1-brane collisions depends on the difference 
in the transverse dimensions because the behaviors of the gravitational 
field in the transverse space depends on the number of the transverse 
dimensions.

Although the six-dimensional metric (\ref{nss:metric:Eq}) is regular 
if and only if $h_3>0$, the spacetime 
shows curvature singularities at $h_3=0$. 
Hence the regular six-dimensional spacetime
 is restricted to the region of $h_3>0$, which is 
bounded by curvature singularities.

Let us study the time evolution for time-dependent 1-brane 
solution (\ref{nss:solution:Eq}). 
We perform the following coordinate transformation:
\begin{eqnarray}
&&t=\sqrt{\frac{2}{\Lambda}}\,
\tilde{t} \cosh \tilde{y}~~,~~~y=\sqrt{\frac{2}{\Lambda}}\,
\tilde{t} \sinh \tilde{y}
\,.
\end{eqnarray}
If we choose $c_1=c_2=0$, we find
\begin{eqnarray}
\hspace{-0.5cm}ds^2&=&\frac{2}{\Lambda}\,
\left[1+\tilde{t}^{-2}\tilde{h}(\tilde{z})\right]^{-1/2}\tilde{t}^{-1}
\left[-d\tilde{t}^2+\tilde{t}^2\left\{d\tilde{y}^2
+\left(1+\tilde{t}^{-2}\tilde{h}(\tilde{z})\right)
\delta_{ab}d\tilde{z}^ad\tilde{z}^b\right\}
\right]
\nonumber \\
\hspace{-0.5cm}&=&\frac{2}{\Lambda}\,
\left[1+16\bar{t}^{-4}\tilde{h}(\tilde{z})\right]^{-1/2}
\left[-d\bar{t}^2+\frac{1}{4}\,\bar{t}^2
\left\{d\tilde{y}^2+
\left(1+16\bar{t}^{-4}\tilde{h}(\tilde{z})\right)
\delta_{ab}d\tilde{z}^ad\tilde{z}^b\right\}\right],
\label{nsc:sufrace:Eq_m}
\end{eqnarray}
where $\tilde{h}(\tilde{z})$, $\bar{t}$, and $\tilde{z}^a$ are defined, 
respectively, by 
\begin{eqnarray}
\tilde{h}(\tilde{z})=c_3
+\frac{\Lambda}{2}\sum_{l=1}^N\frac{M_l}{|\tilde{z}^a-\tilde{z}^a_l|^2}\,,
~~~~~~\bar{t}=2\tilde{t}^{1/2},~~~~~~~
\tilde{z}^a=\sqrt{\frac{\Lambda}{2}}~z^a\,.
\end{eqnarray}
Here, $\tilde{t}$ obeys  
$\Lambda(t^2-y^2)/2=\tilde{t}^2$.
The six-dimensional metric (\ref{nsc:sufrace:Eq_m}) represents 
a homogeneous and isotropic spacetime whose scale factor evolves 
as the cosmic time 
$\bar{t}$, which is described as the Milne universe. 
Hence, we can consider
that the present solution with $\Lambda>0$ gives a system of 
1-branes in the Milne universe. The existence of the expanding Milne 
universe is guaranteed by the scalar field with the exponential potential 
in the six-dimensional action (\ref{nss:action:Eq}). 

Now let us consider the collision of 1-branes. 
The solution (\ref{nss:metric:Eq}) without 0-branes can be written 
in the form
\Eq{
ds^2=\left[\frac{\Lambda}{2}\left(t^2-y^2\right)
+c_3+\tilde{h}(z)\right]^{-\frac{1}{2}}
\left(-dt^2+dx^2\right)+\left[\frac{\Lambda}{2}\left(t^2-y^2\right)
+c_3+\tilde{h}(z)\right]^{\frac{1}{2}}
u_{ab}dz^adz^b\,,
   \label{nsc:surface:Eq}
}
where we choose $c_1=c_2=0$, $u_{ab}$ denotes the four-dimensional metric, 
and the function $\tilde{h}(z)$ is given by (\ref{nsc:h2:Eq}).
The behavior of the harmonic function $\tilde{h}(z)$ is divided into two
classes depending on the dimensions of the 1-brane, that is, 
$d_s\ne 2$ and $d_s=2$, which we will study below separately.
For $d_s=2$, the harmonic function $\tilde{h}(z)$ diverges
both at infinity and near 1-branes.
In particular, there is no regular spacetime region near 1-branes, 
because $\tilde{h}(z)\rightarrow -\infty$\,.
Hence, such a 1-brane solution is not physically relevant. 

Since the harmonic function $\tilde{h}(z)$ becomes dominant 
in the limit of $z^a\rightarrow z^a_l$ (near 1-branes), 
we find a static structure of the 1-brane system.
In the far region from 1-branes, that is,
in the limit of $|z^a-z^a_l|\rightarrow \infty$,
the function $h_3$ depends only on time $t$, because
$\tilde{h}(z)$ vanishes. The metric is thus written by
\Eqr{
ds^2=\left[\frac{\Lambda}{2}\left(t^2-y^2\right)+c_3\right]^{-\frac{1}{2}}
\left(-dt^2+dx^2\right)
+\left[\frac{\Lambda}{2}\left(t^2-y^2\right)+c_3\right]^{\frac{1}{2}}
u_{ab}dz^adz^b\,.
   \label{nsc:surface2:Eq}
}

In the following, we will analyze one concrete example, in which 
two 1-branes are located at $\,\vect{z}_1=(0, 0,\ldots,0)$ and 
$\,\vect{z}_2=(z_0, 0,\ldots,0)$\, in order to study in more detail. 
Since the metric function is singular at $h_3=0$\,, 
the regular spacetime exists inside the domain restricted by
\Eq{
h_3(t, z) = \frac{\Lambda}{2}\left(t^2-y^2\right)+c_3+\tilde{h}(z)>0\,,
}
where the function $\tilde{h}(z)$ is given by (\ref{nsc:h2:Eq}). 
The six-dimensional 
spacetime cannot be extended beyond this region, because
not only does the dilaton $\phi$ diverge but also the spacetime 
evolves into a curvature singularity.

The regular spacetime with two 1-branes ends on these singularities.
The time dependence appears in the form of $\frac{\Lambda}{2} t^2$. 
For $t^2>y^2$ and $c_3=0$, the function $h$ is positive everywhere and  
the six-dimensional spacetime is not singular. It is asymptotically 
a time-dependent uniform spacetime except for 
near branes in the limit of $z^a\rightarrow z^a_l$\,, 
where the background geometry becomes the cylindrical
forms of infinite throats

When $t\le 0$, the spatial metric is initially ($t\rightarrow-\infty$) 
regular apart from $y\rightarrow\pm\infty$\,, and 
the spacetime has a cylindrical topology near each 1-brane. 
As $t$ evolves slightly, a curvature singularity appears at 
$y\rightarrow\pm\infty$ and from a 
far region ($|z^1|\rightarrow\infty$). 
As $t$ evolves further, the singularity cuts off the 
space. As the time continues to increase, the singular hypersurface 
eventually splits and surrounds each of the 1-brane throats individually. 
Then the spatial surface is composed of two isolated throats.

The six-dimensional metric (\ref{nsc:surface2:Eq}) implies that the 
transverse dimensions expand asymptotically as $t^{1/2}$ 
for fixed spatial coordinates $y$ and $z^a$\,. 
However, this is observer dependent, because it becomes static near branes, 
and the spacetime approaches a Friedmann-Robertson-Walker 
universe in the far region ($|z^1|\rightarrow \infty$), 
which expands in the background isotropically. 

\begin{figure}[h]
 \begin{center}
\includegraphics[keepaspectratio, scale=0.5, angle=0]{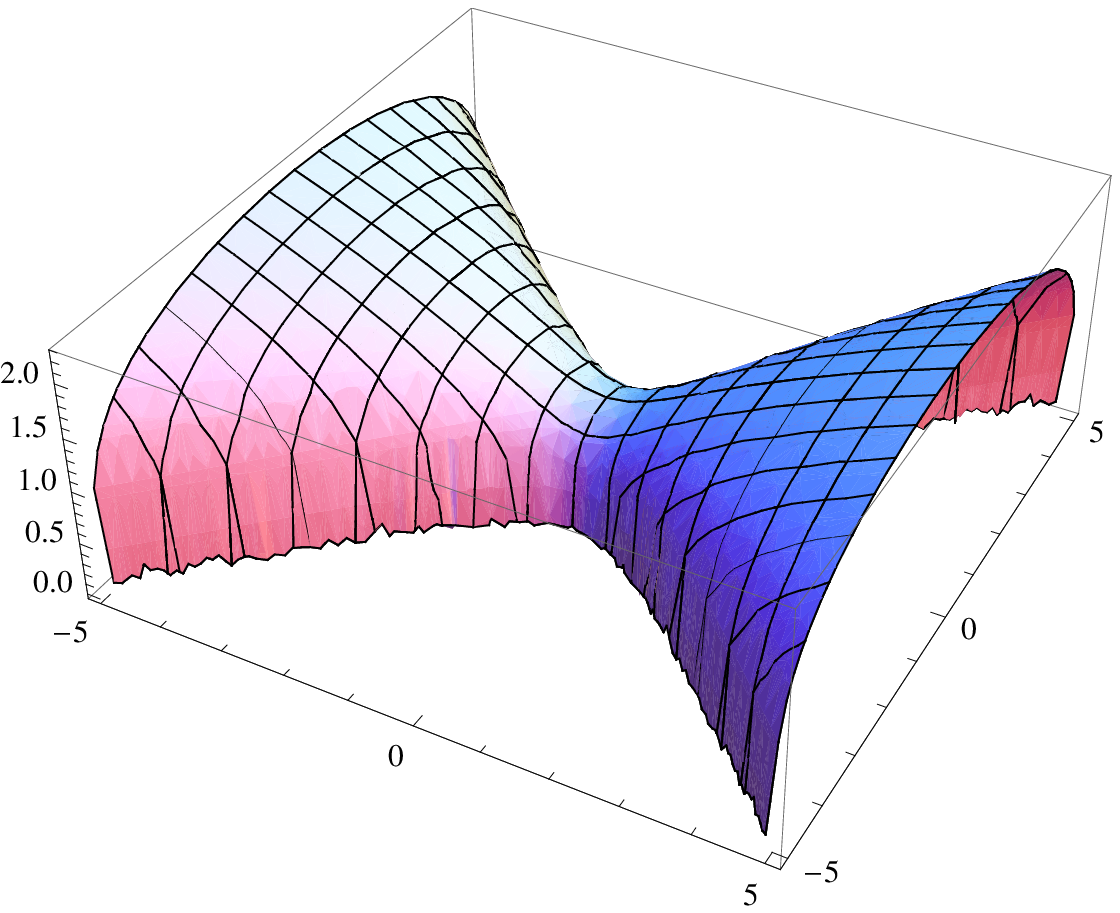}
\put(-195, 80){$d(t, y)$}
\put(-105, 0){$t$}
\put(-5,40){$y$}
\hskip 2.cm
\includegraphics[keepaspectratio, scale=0.5, angle=0]{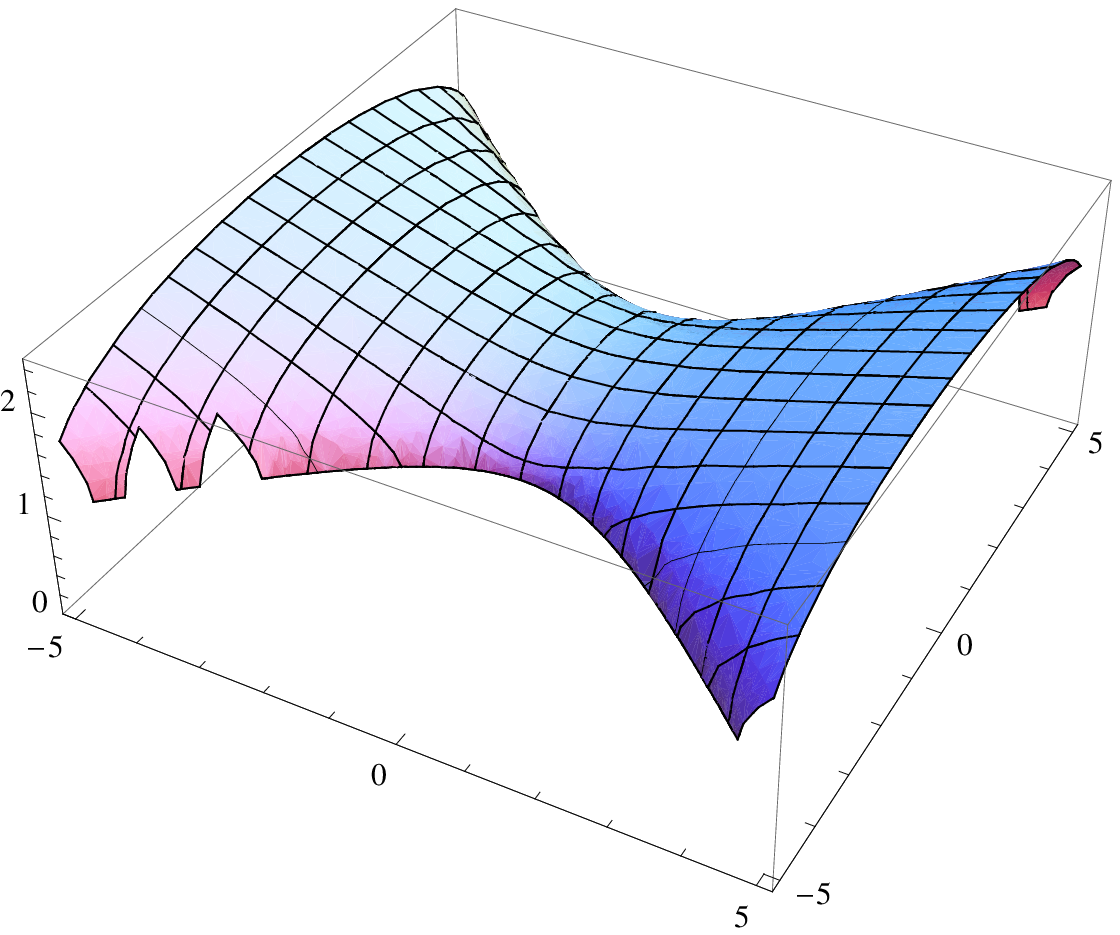}
\put(-195, 80){$d(t, y)$}
\put(-105, 0){$t$}
\put(-5,40){$y$}\\
(a) \hskip 7cm (b) ~~~~~~
  \caption{\baselineskip 14pt 
 The time evolution of the proper distance between two dynamical 
 1-branes for (a) $d_s=3$ and (b) $d_s=1$ in the six-dimensional 
 Nishino-Salam-Sezgin gauged supergravity. 
 We fix $c_3=0$\,, $M_1=M_2=1$\,, $z_0=1$\,, and $\Lambda=2$. 
The proper distance rapidly vanishes near where two
1-branes collide for the case of $d_s=3$,  
while for the case of $d_s=1$, it 
is still finite when a curvature singularity appears.}
  \label{fig:ns}
 \end{center}
\end{figure}

\begin{figure}[h]
 \begin{center}
\includegraphics[keepaspectratio, scale=0.5, angle=0]{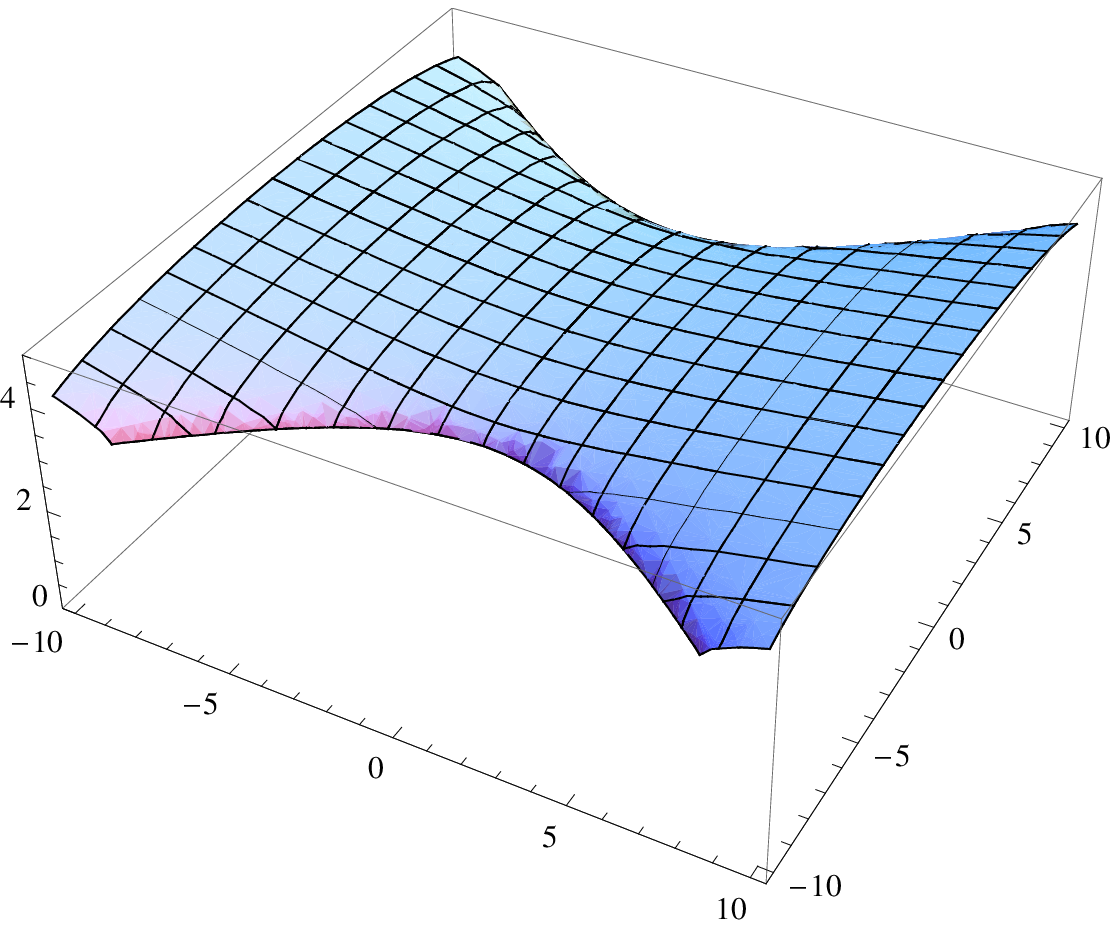}
\put(-195, 80){$d(t, y)$}
\put(-105, 0){$t$}
\put(-5,40){$y$}
\hskip 2.cm
\includegraphics[keepaspectratio, scale=0.5, angle=0]{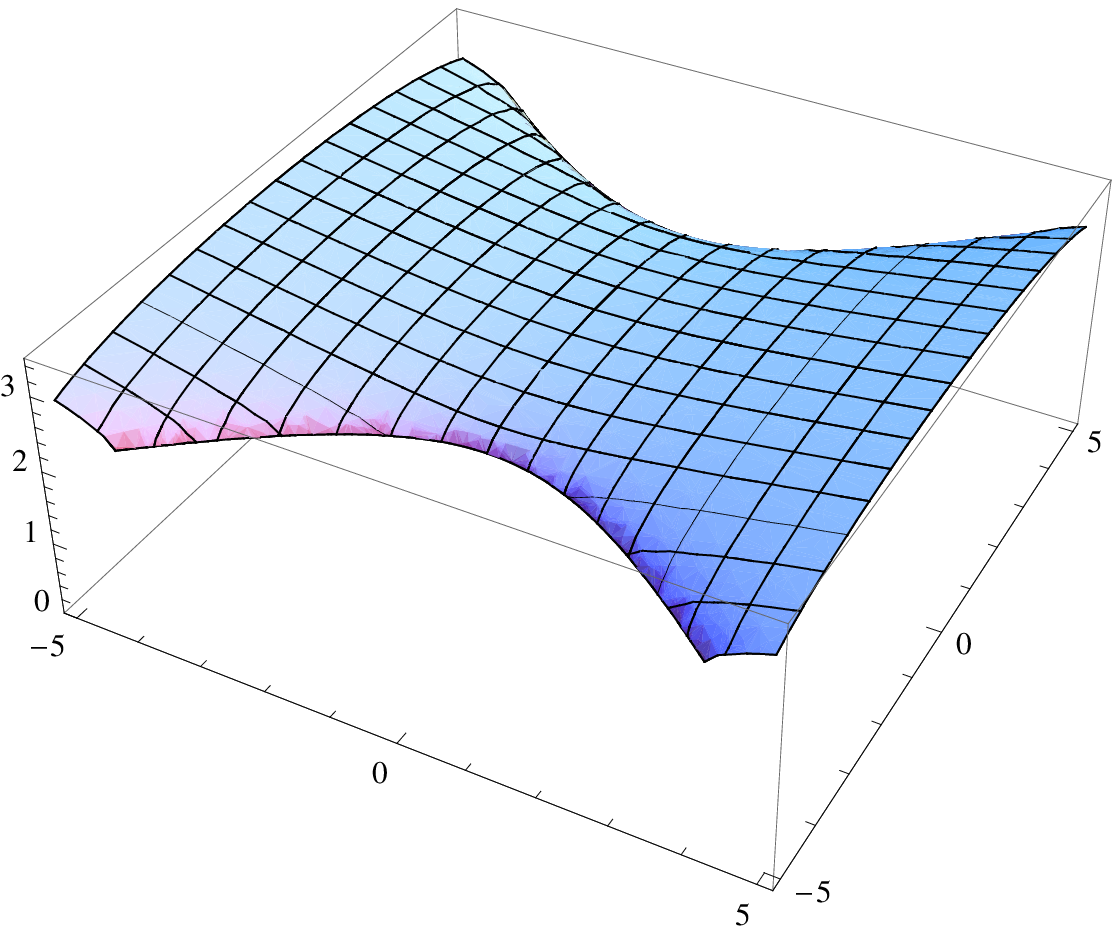}
\put(-195, 80){$d(t, y)$}
\put(-105, 0){$t$}
\put(-5,40){$y$}\\
(a) \hskip 7cm (b) ~~~~~~
  \caption{\baselineskip 14pt 
The proper distance between two dynamical 1-branes given in 
(\ref{nsc:distance:Eq}) is depicted for (a) $M_1=10$, $M_2=1$
and (b) $M_1=2$\,, $M_2=1$ in the six-dimensional 
 Nishino-Salam-Sezgin gauged supergravity. 
 We fix $c_3=0$\,, $d_s=0$\,, $z_0=1$, and $\Lambda=2$. 
In both cases, a singularity appears at $t=t_{\rm s}<0$ when the distance is
still finite.}
  \label{fig:ns1}
 \end{center}
\end{figure}

If we define  
\Eq{
\bar{z}=\sqrt{\left(z^2\right)^2+\left(z^3\right)^2+\cdots 
+\left(z^{4-d_s}\right)^2}\,,
}
the proper length at $\bar{z}=0$
between two 1-branes is written by
\begin{eqnarray}
d(t, y)&=&\int_{0}^{z_0} dz^1 
\left[\frac{\Lambda}{2}\left(t^2-y^2\right)+c_3+\frac{M_1}{|z^1|^{2-d_s}}
+\frac{M_2}{|z^1-z_0|^{2-d_s}}\right]^{\frac{1}{4}}
\,.
\label{nsc:distance:Eq}
\end{eqnarray}
This is a monotonically decreasing function of time for $t\le 0$.
In Fig.~\ref{fig:ns}, we show $d(t, y)$  
for the case of the 1-brane system.
We set $\Lambda=2$, $z_0=1$, $c_3=0$, and $M_1=M_2=1$. 
All of the six-dimensional space is initially ($t=-\infty$) regular
except at $|y|\rightarrow \infty$ and $|z^1-z_0|\rightarrow\infty$. 
Although the spacetime becomes asymptotically 
time dependent and has the cylindrical form of an infinite 
throat near the 1-brane, the singularity appears 
from a far region ($|z^1-z_0|\rightarrow\infty$) 
and $|y|\rightarrow \infty$\,. 
As time increases ($t<0$), 
the singularity erodes the region with the large $|y|$ region. 
The region of transverse space is also invaded in time. 
As a result, only the region of small $|y|$ and near 1-branes remains regular. 
When we study the evolution on the $y$ and $z^a$ plane, the singularity 
appears at infinity $|z^1|\rightarrow\infty$\,, 
$|y|\rightarrow \infty$\,, and comes to the region of two 1-branes. 
A singular hypersurface eventually surrounds each 1-brane 
individually, and then the regular regions near 1-branes split 
into two isolated throats. 
For the period of $t>0$, we find the time-reversed 
behavior of the case of $t<0$. 
Figures \ref{fig:ns} and \ref{fig:ns1} show that this singularity appears 
 before the distance $d$ vanishes.

Then a singularity between two branes forms before their collision 
except for $d_s=3$.
Two 1-branes approach very slowly, a singularity suddenly appears
at a finite distance, and the six-dimensional spacetime splits
into two isolated 1-brane throats. 

\begin{figure}[h]
 \begin{center}
\includegraphics[keepaspectratio, scale=0.5, angle=0]{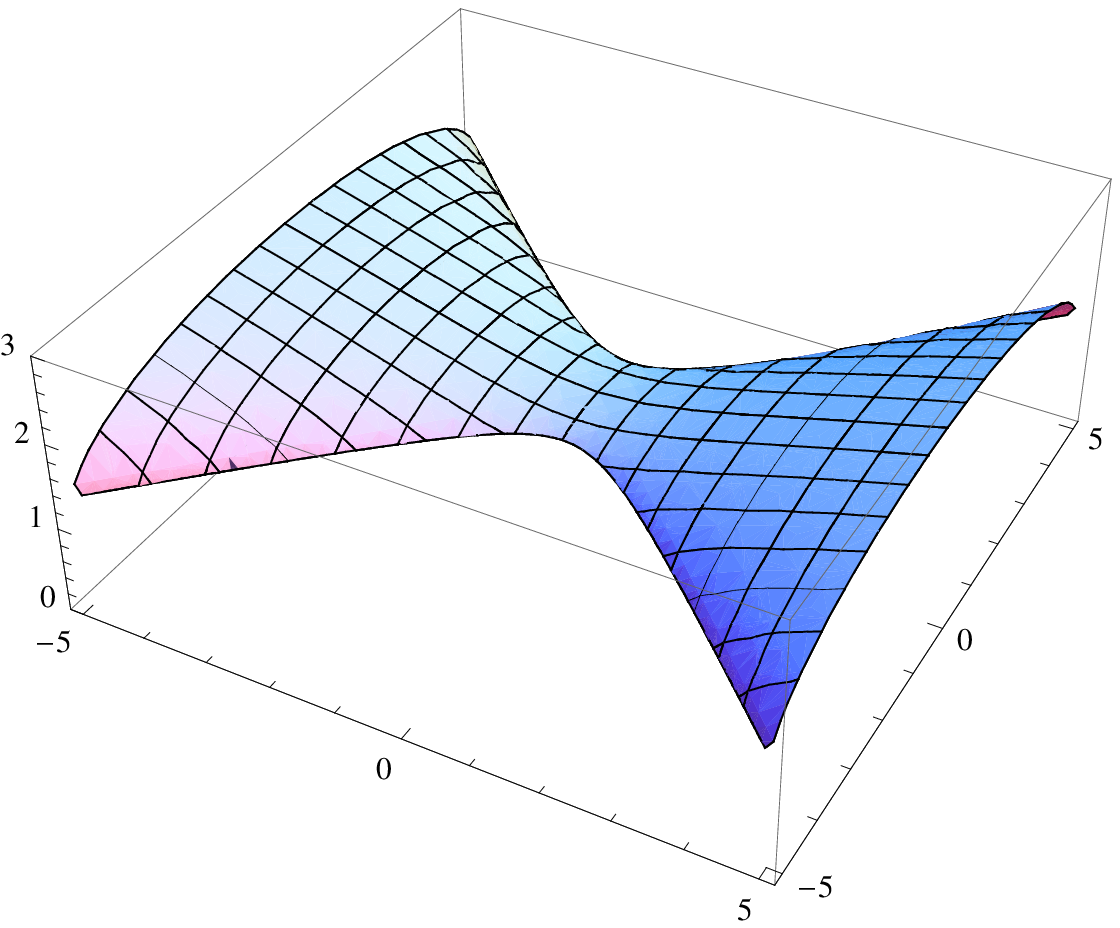}
\put(-195, 80){$d(t, y)$}
\put(-105, 0){$t$}
\put(-5,40){$y$}
\hskip 2.cm
\includegraphics[keepaspectratio, scale=0.5, angle=0]{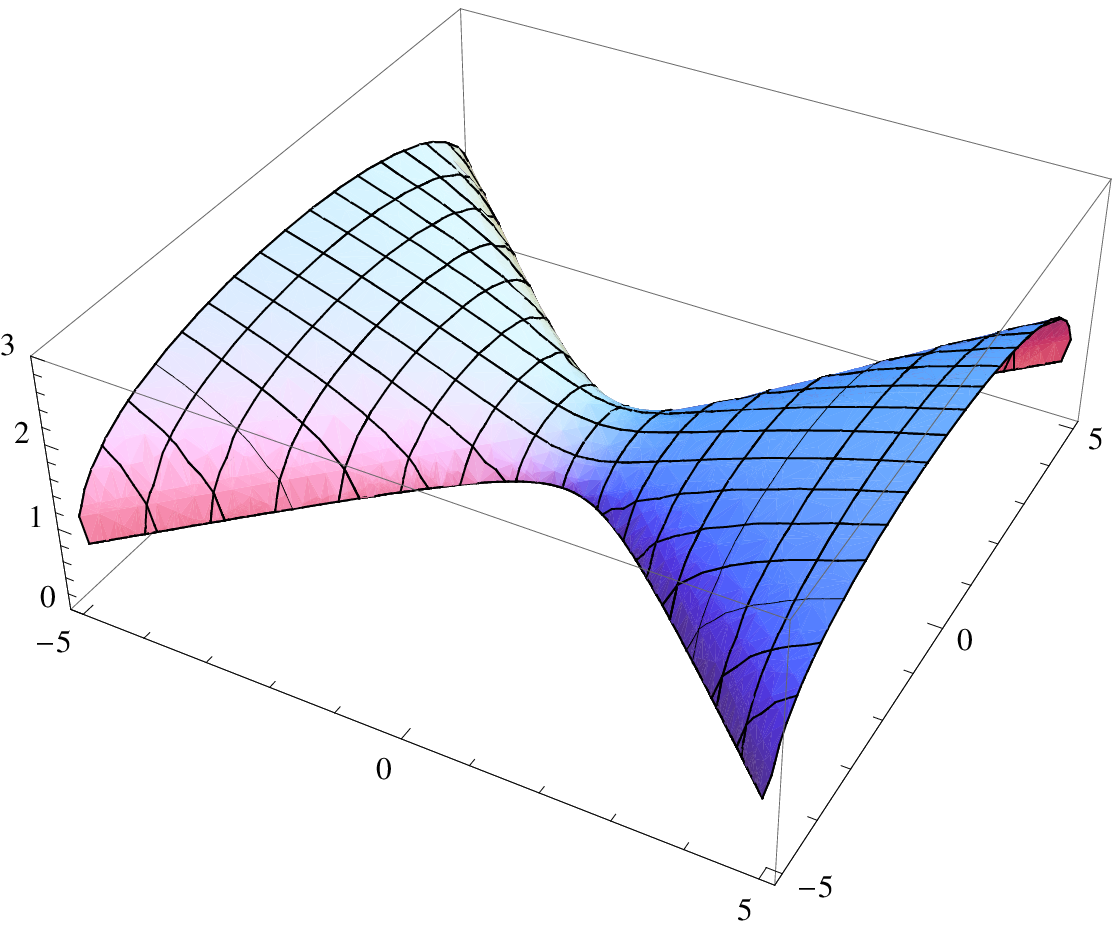}
\put(-195, 80){$d(t, y)$}
\put(-105, 0){$t$}
\put(-5,40){$y$}\\
(a) \hskip 7cm (b) ~~~~~~
  \caption{\baselineskip 14pt 
 The time evolution of the proper distance between two dynamical 1-branes 
 for (a) $M_1=10$, $M_2=1$ and (b) $M_1=2$\,, $M_2=1$ in the six-dimensional 
 Nishino-Salam-Sezgin gauged supergravity. 
 We fix $c_3=0$\,, $d_s=3$\,, $z_0=1$, and $\Lambda=2$. 
For $M_1\ne M_2$, a singularity appears at $t=t_{\rm s}<0$ 
when the distance is
still finite. Then, the solution does not describe the collision 
of two 0-branes.}
  \label{fig:ns2}
 \end{center}
\end{figure}

On the other hand, we can discuss a brane collision for $d_s=3$ and $t<0$.
If $M_1\ne M_2$, a singularity appears at $t=t_{\rm s}<0$ when 
the distance is still finite (see Fig.~\ref{fig:ns2}). 
This is just the same as the case in 
Sec.~\ref{sec:ns-c}. However, if $M_1=M_2=M$, the result completely changes 
(Fig.~\ref{fig:ns}).
Since the distance eventually vanishes at $t=t_{\rm c}$, 
two 1-branes collide with each other. 
The proper length for fixed $y$ decreases as time increases from $t=-\infty$, 
and it eventually vanishes at $t=t_{\rm c}$. Hence, one 1-brane approaches 
the other as time evolves, causing the complete collision at $t=t_{\rm c}$. 
If we fix the 1-brane charges such that $M_1=M_2=M$, the branes first 
collide at larger $|y|$, and as time progresses, the subsequent collisions 
occur at the smaller $|y|$. 
We show $d(t, y)$ integrated numerically in 
Figs.~\ref{fig:ns} $-$ \ref{fig:ns2}. 

\begin{figure}[h]
 \begin{center}
\includegraphics[keepaspectratio, scale=0.55, angle=0]{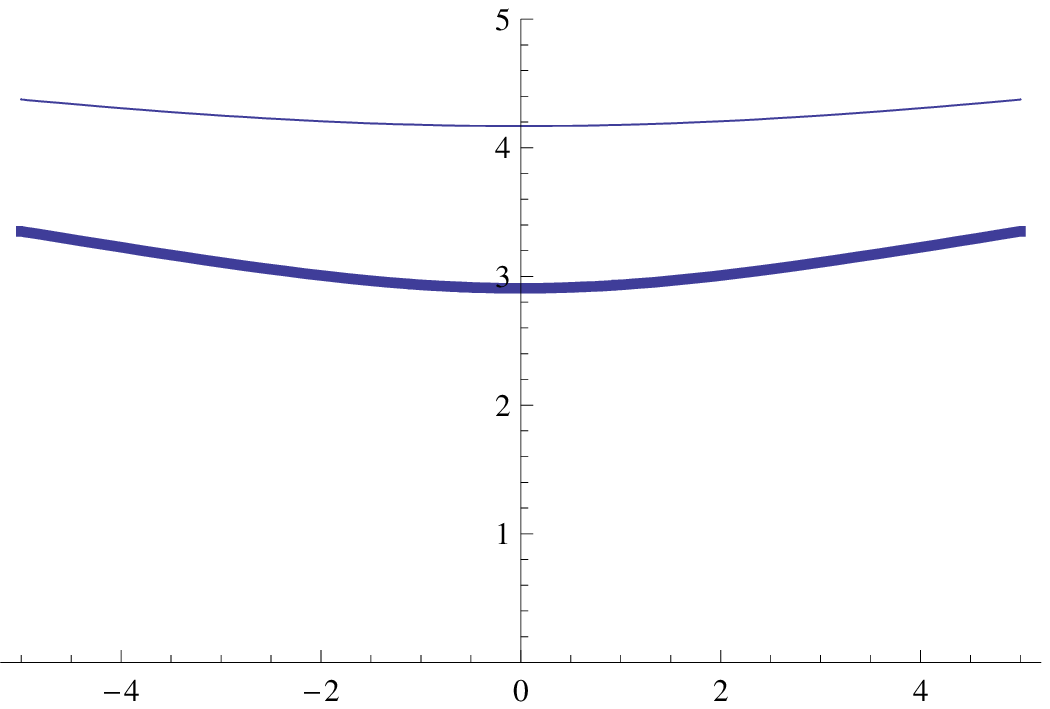}
\put(-95,120){$d(t)$}
\put(10,10){$t$}
\hskip 2.0cm
\includegraphics[keepaspectratio, scale=0.55, angle=0]{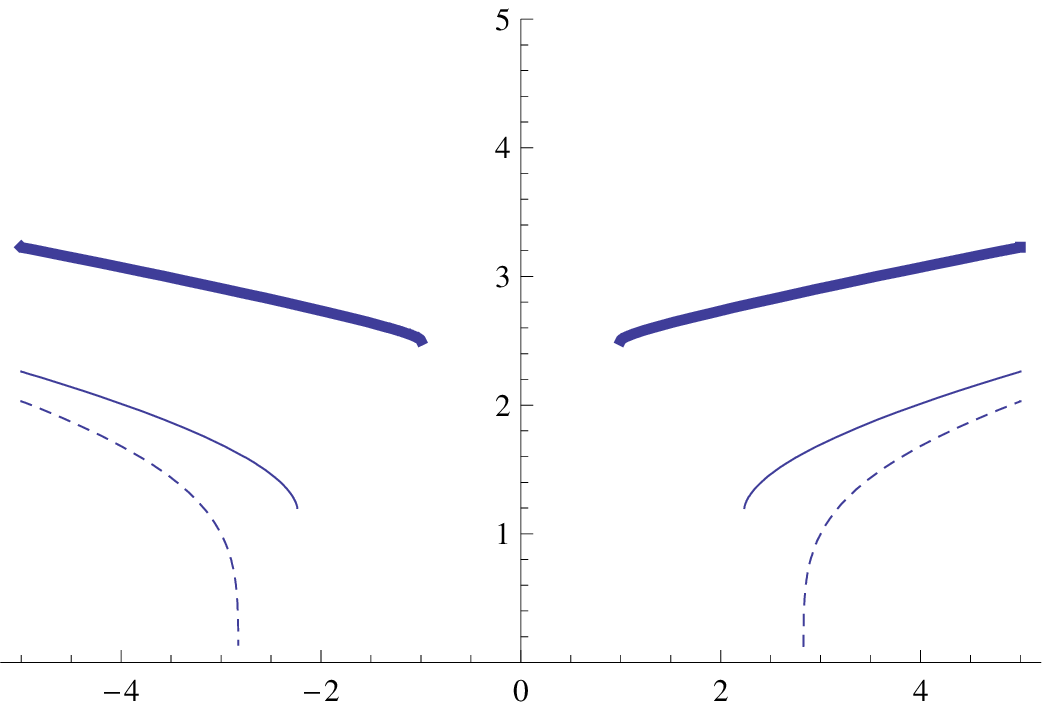}
\put(-95,120){$d(t)$}
\put(10,10){$t$}\\
(a) \hskip 7cm (b) ~~~~~~
  \caption{\baselineskip 14pt 
(a) The proper distance between two dynamical 1-branes at $y=0$ and 
$\bar{z}=0$ for the case of $d_s=0$ in the six-dimensional 
 Nishino-Salam-Sezgin gauged supergravity is depicted. 
 We fix $c_4=0$\,, $z_0=1$\,, and $\Lambda=2$\,. For $t<0\,,$ 
the proper length decreases as time increases. 
The bold line denotes the case of $M_1=M_2=1$, while the solid
one corresponds to the $M_1=10$, $M_2=1$ case. (b) For the case of 
$M_1=M_2$ in the six-dimensional Nishino-Salam-Sezgin gauged supergravity, 
the time evolution of the proper distance between two dynamical 1-branes at 
$y=3$ and $\tilde{z}=0$ given in (\ref{nsc:distance:Eq}) is depicted. 
We fix $c_4=0$\,, $z_0=1$\,, and $\Lambda=2$. We show the lengths 
for $d_s=0$ (bold line), $d_s=1$ (solid line), and $d_s=3$ (dashed line). 
The proper distance rapidly vanishes near where two
1-branes collide in the case of $d_s=3$, 
while in the case of $d_s\ne 3$, 
is still finite when a curvature singularity appears.}
  \label{fig:ns3}
 \end{center}
\end{figure}

We also calculate the distance $d(t, y)$ at $y=0$ and $\bar{z}=0$ 
between two branes before the singularity appears except for the case 
of $d_s=3$ if $M_1=M_2$. 
The proper length is also given by Eq. (\ref{nsc:distance:Eq}). 
In the present case, $d$ is a monotonically decreasing function 
of $t^2$ when $t<0$. 
We show the time evolution of the distance in Fig.~\ref{fig:ns3} 
for the case of $M_1=M_2$. 

On the other hand, for the case of ${M}_1\ne {M}_2$, 
a singularity appears, when the proper distance is still finite. 
For the period of $t>0$, the behavior of six-dimensional 
spacetime is the time reversal of the period of $t<0$. 
We show the proper distance $d(t)$ integrated numerically 
in Fig.~\ref{fig:ns3} for the cases of $d_s=3$ and $d_s\ne 3$.

\subsection{Romans' six-dimensional gauged supergravity}
Similarly, for the six-dimensional Romans' theory \cite{Romans:1985tw},
following the discussion in Ref.~\cite{Nunez:2001pt},
the coupling  
of the 3-form and of the 2-form field strengths to the dilaton are  
given by $\epsilon_rc_r=-1/\sqrt{2}$ and $\epsilon_sc_s=\sqrt{2}$, 
respectively:
\Eqr{
S&=&\frac{1}{2\kappa^2}\int \left[\left(R+2\e^{\phi/\sqrt{2}}
\lambda\right)\ast{\bf 1}
 -\frac{1}{2}\ast d\phi \wedge d\phi\right.\nn\\
 &&\left.\hspace{-0.5cm}
 -\frac{1}{2\cdot 2!}
 \e^{-\phi/\sqrt{2}}\ast F_{(2)}\wedge F_{(2)}
 -\frac{1}{2\cdot 3!}
 \e^{\sqrt{2}\phi}\ast F_{(3)}\wedge F_{(3)} \right],
\label{ro:action:Eq}
}
where $R$ denotes the Ricci scalar constructed from the six-dimensional 
metric $g_{MN}$\,, $\kappa^2$ is the six-dimensional gravitational constant,  
$\ast$ is the Hodge operator in the six-dimensional spacetime, 
$\phi$ denotes the scalar field, $\lambda>0$ is cosmological 
constants, and $F_{\left(3\right)}$ and $F_{\left(2\right)}$ 
are 3-form and 2-form field strengths, respectively.
This has a negative scalar potential. 
In terms of Eq.~(\ref{pl:c:Eq}), Romans' model 
is given by choosing $\Lambda_r=-\lambda$, 
$\Lambda_s=0$, $N_r=2$, and $N_s=4$. 

From the six-dimensional action (\ref{ro:action:Eq}), 
we find the field equations: 
\Eqrsubl{ro:equations:Eq}{
&&R_{MN}=-\frac{1}{2}\e^{\phi/\sqrt{2}}\lambda g_{MN}
+\frac{1}{2}\pd_M\phi \pd_N \phi
+\frac{\e^{-\phi/\sqrt{2}}}{2\cdot 2!}
\left(2F_{MA}{F_N}^A
-\frac{1}{4} g_{MN} F^2_{\left(2\right)}\right)\nn\\
&&~~~~~~~~~+\frac{\e^{\sqrt{2}\phi}}{2\cdot 3!}
\left(3F_{MAB} {F_N}^{AB}
-\frac{1}{2} g_{MN} F_{\left(3\right)}^2\right)\,,
   \label{ro:Einstein:Eq}\\
&&\lap\phi+
\frac{\sqrt{2}}{4\cdot 2!}\,\e^{-\phi/\sqrt{2}}\,F^2_{\left(2\right)}
-\frac{\sqrt{2}}{2\cdot 3!}\,\e^{\sqrt{2}\phi}\,F^2_{\left(3\right)}
+\sqrt{2}\,\e^{\phi/\sqrt{2}}\,\lambda=0\,, 
   \label{ro:scalar:Eq}\\
&&d\left[\e^{-\phi/\sqrt{2}}\ast F_{\left(2\right)}\right]=0\,,
   \label{ro:gauge-r:Eq}\\
&&d\left[\e^{\sqrt{2}\phi}\ast F_{\left(3\right)}\right]=0\,,
   \label{ro:gauge-s:Eq}
}
where $\lap$ denotes the Laplace operator with respect to the six-dimensional 
metric $g_{MN}$\,. 

We assume the six-dimensional metric of the form (\ref{nss:metric:Eq}). 
The scalar field and the gauge field strengths are assumed to be
\Eqrsubl{ro:ansatz:Eq}{
\e^{\phi}&=&h_2^{-\sqrt{2}/2}\,h_3^{\sqrt{2}/2},
  \label{ro:phi:Eq}\\
F_{\left(2\right)}&=&d\left[\sqrt{2}\,h^{-1}_2(t, y, z)\right]\wedge dt,
  \label{ro:fr:Eq}\\
F_{\left(3\right)}&=&d\left[h^{-1}_3(t, y, z)\right]\wedge dt\wedge dy\,.
  \label{ro:fs:Eq}
}
The Einstein equations (\ref{ro:Einstein:Eq}) then reduce to
\Eqrsubl{ro:cEinstein:Eq}{
&&\frac{5}{4}h_2^{-1}\pd_t^2h_2
+\frac{3}{4}h_3^{-1}\pd_t^2h_3+
\frac{1}{4}h_2^{-2}\left(3h_2^{-1}\pd_y^2h_2+h_3^{-1}\pd_y^2h_3\right)
+\frac{1}{4}h_2^{-2}h_3^{-1}
\left(3h_2^{-1}\lap_{\Zsp}h_2+h_3^{-1}\lap_{\Zsp}h_3\right)\nn\\
&&~~~+\frac{1}{2}h_2^{-2}\lambda
+\frac{1}{4}\left(\pd_t\ln h_2\right)^2
+\frac{7}{4}\pd_t\ln h_2\pd_t\ln h_3
+\frac{3}{4}h_2^{-2}\pd_y\ln h_2\pd_y\ln h_3=0,
 \label{ro:cEinstein-tt:Eq}\\
&&h_2^{-1}\pd_t\pd_y h_2+h_3^{-1}\pd_t\pd_y h_3
+\pd_t\ln h_3\pd_y\ln h_2=0\,,
 \label{ro:cEinstein-ty:Eq}\\
&&2h_2^{-1}\pd_t\pd_a h_2+h_3^{-1}\pd_t\pd_a h_3=0\,, 
 \label{ro:cEinstein-ta:Eq}\\
&&\frac{1}{4}h_2^2\left(h_2^{-1}\pd_t^2h_2-h_3^{-1}\pd_t^2h_3\right)
-\frac{1}{4}\left(h_2^{-1}\pd_y^2h_2+3h_3^{-1}\pd_y^2h_3\right)
-\frac{1}{4}h_3^{-1}
\left(h_2^{-1}\lap_{\Zsp}h_2-h_3^{-1}\lap_{\Zsp}h_3\right)\nn\\
&&~~~+\frac{1}{2}\lambda+\frac{1}{4}\left(\pd_th_2\right)^2
-\frac{1}{4}h_2^2\pd_t\ln h_2\pd_t\ln h_3
-\frac{5}{4}\pd_y\ln h_2\pd_y\ln h_3=0\,,
 \label{ro:cEinstein-yy:Eq}\\
&&h_3^{-1}\pd_y\pd_a h_3+2\pd_y\ln h_2\pd_a\ln h_2=0\,,
 \label{ro:cEinstein-ya:Eq}\\
&&R_{ab}(\Zsp)+\frac{1}{4}h_2^2h_3u_{ab}
\left(h_2^{-1}\pd_t^2 h_2+h_3^{-1}\pd_t^2 h_3\right)
-\frac{1}{4}h_3u_{ab}
\left(h_2^{-1}\pd_y^2 h_2+h_3^{-1}\pd_y^2 h_3\right)\nn\\
&&~~~-\frac{1}{4}u_{ab}
\left(h_2^{-1}\lap_{\Zsp}h_2+h_3^{-1}\lap_{\Zsp}h_3\right)
+\frac{1}{4}h_2^2h_3u_{ab}\left[\left(\pd_t\ln h_2\right)^2
+3\pd_t\ln h_2\pd_t\ln h_3\right]\nn\\
&&~~~-\frac{1}{4}h_3u_{ab}\pd_y\ln h_2\pd_y\ln h_3
+\frac{1}{2}u_{ab}h_3\,\lambda=0,
  \label{ro:cEinstein-ab:Eq}
}
where $\triangle_{\Zsp}$ denotes the Laplace operator on Z space and  
$R_{ab}(\Zsp)$ is the Ricci tensor 
constructed from the metric $u_{ab}(\Zsp)$\,.

We next consider the gauge field. 
Under the ansatz (\ref{ro:ansatz:Eq}), 
the Bianchi identity is automatically satisfied. Also the equation of motion
for the gauge field becomes 
\Eqrsubl{ro:gauge2:Eq}{
&&d\left[h_3^{-1}\pd_y h_2\,\Omega(\Zsp)
+\pd_a h_2\,dy\wedge\left(\ast_{\Zsp}dz^a\right)\right]=0,
  \label{ro:gauge2-r:Eq}\\
&&d\left[\pd_a h_3\left(\ast_{\Zsp}dz^a\right)\right]=0,
  \label{ro:gauge2-s:Eq}
 }
where $\ast_{\rm Z}$ denotes the Hodge operator on Z.  

Although the roles of the Bianchi identity and field 
equations are interchanged, the net result is the same.
Finally, we consider the equation of motion for the scalar field. 
Substituting the scalar 
field and the gauge field in (\ref{ro:ansatz:Eq}) into the equation 
of motion for the scalar field (\ref{ro:scalar:Eq}), we have
\Eqr{
&&h_2^2h_3\left(h_2^{-1}\pd_t^2h_2-h_3^{-1}\pd_t^2h_3\right)
+h_3\left(\pd_th_2\right)^2
-h_2\pd_th_2\pd_th_3-h_3\left(h_2^{-1}\pd_y^2h_2
-h_3^{-1}\pd_y^2h_3\right)\nn\\
&&~~~~-h_2^{-1}\pd_yh_2\pd_yh_3
-h_2^{-1}\lap_{\Zsp}h_2+h_3^{-1}\lap_{\Zsp}h_3+2h_3\lambda=0\,.
  \label{ro:scalar-e:Eq}
}
Then, the functions $h_2$ and $h_3$ satisfy the equations
\Eqrsubl{ro:scalar-s:Eq}{
&&
\left(\pd_th_2\right)^2+2\lambda+h_2\pd_t^2h_2-h_2^{-1}\lap_{\rm W}h_2=0,
~~~~{\rm For}~~h_3=1\,,
   \label{ro:scalar-s1:Eq}\\
&&-\pd_t^2h_3+\pd_y^2h_3+2\lambda h_3+h_3^{-1}\lap_{\Zsp}h_3=0\,,
~~~~{\rm For}~~h_2=1\,,
   \label{ro:scalar-s2:Eq}
}
where Laplace operator $\lap_{\rm W}$ is defined in Eq.~(\ref{nss:lapW:Eq}). 
If we set $h_2=1$, the field equations give
\Eqrsubl{ro:equation:Eq}{
&&R_{ab}(\Zsp)=0\,,
   \label{ro:Ricci:Eq}\\
&&h_2=1\,,~~~~\lambda=0,~~~~\pd_t^2h_3=\pd_y^2h_3=0\,,
~~~~\triangle_{\Zsp}h_3=0\,.
   \label{ro:h1:Eq}
 }
Now we will focus upon a case by imposing the conditions
\Eq{
u_{ab}=\delta_{ab}\,,~~~~~~h_2=1\,,~~~~~~\lambda=0\,,
 \label{ro:flat:Eq}
 }
where $\delta_{ab}$ is the four-dimensional Euclidean metric. 
Then, the solution for $h_3$ can be obtained explicitly as
\Eq{
h_3(t, y, z)=c_1t+c_2y+c_3+\sum_{l=1}^N\frac{M_l}{|z^a-z^a_l|^2},
 \label{ro:solution-s:Eq}
}
where $c_i~(i=1,~2,~3)$ are constants\,.  

One can easily get the solution
for $h_3=1$\,, $\lambda\ne 0$ and $\pd_th_2\ne 0$
if the roles of $h_2$ and $h_3$ are exchanged.
The solution of field equations is thus expressed as
\Eqrsubl{ro:solution3:Eq}{
h_2(t, v)&=&\pm\sqrt{2i\lambda}\,t+c_5+\sum_{\alpha=1}^{N'}\frac{L_{\alpha}}
{|v^m-v^m_{\alpha}|^3}\,,
 \label{ro:solution3-r:Eq}\\
h_3&=&1\,,
 \label{ro:solution3-s:Eq}
}
where $c_5$, $v^m_\alpha$, and $L_\alpha$ are constants, and 
the five-dimensional coordinate $v^m$ defined by \eqref{nss:lapW:Eq}. 
Hence, there is no cosmological 0-brane solution in terms of the 
ansatz of fields (\ref{nss:metric:Eq}) and (\ref{ro:ansatz:Eq}) if 
$\lambda\ne 0$. 

\subsection{Romans' five-dimensional gauged supergravity}
Finally, we consider the five-dimensional Romans' theory \cite{Romans}. 
The five-dimensional action is given by
\Eqr{
S&=&\frac{1}{2\kappa^2}\int \left[\left(R+2\e^{2\phi/\sqrt{6}}
\bar{\lambda}\right)\ast{\bf 1}
 -\frac{1}{2}\ast d\phi \wedge d\phi\right.\nn\\
 &&\left.\hspace{-0.5cm}
 -\frac{1}{2\cdot 2!}
 \e^{-2\phi/\sqrt{6}}\ast F_{(2)}\wedge F_{(2)}
 -\frac{1}{2\cdot 2!}
 \e^{4\phi/\sqrt{6}}\ast H_{(2)}\wedge H_{(2)} \right],
\label{ro5:action:Eq}
}
where the expectation values of Yang-Mills potential is assumed to vanish, 
$R$ denotes the Ricci scalar constructed from the five-dimensional metric 
$g_{MN}$\,, 
$\kappa^2$ is the five-dimensional gravitational constant,  
$\ast$ is the Hodge operator in the five-dimensional space-time, 
$\phi$ denotes the scalar field, $\bar{\lambda}>0$ is cosmological 
constant, $F_{\left(2\right)}$ and $H_{\left(2\right)}$ 
are two-form field strengths, 
and the couplings of the 2-form field strengths and cosmological constant 
to the dilaton are given by $\epsilon_rc_r=-2/\sqrt{6}$, 
$\epsilon_sc_s=4/\sqrt{6}$, $\alpha_r=2/\sqrt{6}$\,, 
in the action (\ref{pl:action:Eq}),  respectively.
This has also a negative scalar potential. 
In terms of Eq.~(\ref{pl:c:Eq}), Romans' five-dimensional model 
is given by setting $\Lambda_r=-\bar{\lambda}$, 
$\Lambda_s=0$, $N_r=2$ and $N_s=4$. 

The five-dimensional action (\ref{ro5:action:Eq}), 
gives the field equations: 
\Eqrsubl{ro5:equations:Eq}{
&&R_{MN}=-\frac{1}{2}\e^{2\phi/\sqrt{6}}\,\bar{\lambda}\,g_{MN}
+\frac{1}{2}\pd_M\phi \pd_N \phi
+\frac{\e^{-2\phi/\sqrt{6}}}{2\cdot 2!}
\left(2F_{MA}{F_N}^A
-\frac{1}{3} g_{MN} F^2_{\left(2\right)}\right)\nn\\
&&~~~~~~~~+\frac{\e^{4\phi/\sqrt{6}}}{2\cdot 2!}
\left(2H_{MA} {H_N}^{A}
-\frac{1}{3} g_{MN} H_{\left(2\right)}^2\right),
   \label{ro5:Einstein:Eq}\\
&&\lap\phi+\frac{\sqrt{6}}{6\cdot 2!}\,\e^{-2\phi/\sqrt{6}}\,
F^2_{\left(2\right)}
-\frac{\sqrt{6}}{3\cdot 2!}\,\e^{4\phi/\sqrt{6}}
H^2_{\left(2\right)}+\frac{2\sqrt{6}}{3}\,
\e^{2\phi/\sqrt{6}}\,\bar{\lambda}=0\,,
   \label{ro5:scalar:Eq}\\
&&d\left[\e^{-2\phi/\sqrt{6}}\ast F_{\left(2\right)}\right]=0\,,
   \label{ro5:gauge-r:Eq}\\
&&
d\left[\e^{4\phi/\sqrt{6}}\ast H_{\left(2\right)}\right]=0\,,
   \label{ro5:gauge-s:Eq}
}
where $\lap$ denotes the Laplace operator with respect to the 
five-dimensional metric 
$g_{MN}$\,. 

We assume the five-dimensional metric of the form
\Eqr{
ds^2&=&h^{2/3}_2(t, z)k_2^{1/3}(t, z)\left[-h^{-2}_2(t, z)
k^{-1}_2(t, z)dt^2+u_{ab}(\Zsp)dz^adz^b\right], 
 \label{ro5:metric:Eq}
}
where $u_{ab}(\Zsp)$ is the four-dimensional metric which
depends only on the four-dimensional coordinates $z^a$. 

The scalar field and the gauge field strengths are assumed to be
\Eqrsubl{ro5:ansatz:Eq}{
\e^{\phi}&=&h_2^{-2/\sqrt{6}}\,k_2^{2/\sqrt{6}},
  \label{ro5:phi:Eq}\\
F_{\left(2\right)}&=&d\left[\sqrt{2}\,h^{-1}_2(t, z)\right]\wedge dt,
  \label{ro5:fr:Eq}\\
H_{\left(2\right)}&=&d\left[k^{-1}_2(t, z)\right]\wedge dt\,.
  \label{ro5:fs:Eq}
}
Then, the field equations are reduced to
\Eqrsubl{ro5:equation:Eq}{
&&
R_{ab}(\Zsp)=0\,,
    \label{ro5:eq-ab:Eq}\\
&&
h_2(t, z)=h_0(t)+\bar{h}(z)\,,~~~~~
k_2(t, z)=k_0(t)+\bar{k}(z)\,,
    \label{ro5:eq-h:Eq}\\
&&
\left(\frac{dh_0}{dt}\right)^2+2\bar{\lambda}=0\,,~~~~~~
\frac{dh_0}{dt}\frac{dk_0}{dt}=0\,,~~~~~~\frac{d^2h_0}{dt^2}=0\,,~~~~~~
\frac{d^2k_0}{dt^2}=0\,,\\
&&\lap_{\Zsp}\bar{h}=0\,,~~~~~~\lap_{\Zsp}\bar{k}=0\,,
    \label{ro5:eq-h2:Eq}
}
where $\triangle_{\Zsp}$ is 
the Laplace operator on $\Zsp$ space, and $R_{ab}(\Zsp)$ is the Ricci tensor
with respect to the metric $u_{ab}(\Zsp)$\,. 
Setting $\bar{\lambda}\ne 0$, there is no cosmological solution because of 
Eq.~(\ref{ro5:eq-h2:Eq}). 

Let us consider the case
\Eq{
u_{ab}=\delta_{ab}\,,~~~~~~\bar{\lambda}=0\,,~~~~~~\frac{dh_0}{dt}=0\,,
 \label{ro5:flat:Eq}
 }
where $\delta_{ab}$ the four-dimensional Euclidean metric. 
Then we can construct the solution 
\Eqrsubl{ro5:solution:Eq}{
h_2(z)&=&c_1+\sum_{\alpha=1}^{N'}\frac{L_{\alpha}}{|z^a-z^a_{\alpha}|^2}\,,\\
k_2(t, z)&=&c_2\,t+c_3+\sum_{l=1}^{N}\frac{M_l}{|z^a-z^a_l|^2}\,,
}
where $c_i~(i=1,~2,~3)$, $L_{\alpha}$, and $M_l$ 
 are the constants. 

Now we discuss the cosmological evolution for time-dependent  
solution (\ref{ro5:solution:Eq}).
We define the cosmic time $\tau$, which is given by
\Eq{
\left(\frac{\tau}{\tau_0}\right) \equiv 
\left(c_2\,t\right)^{2/3}\,,~~~~
\tau_0 \equiv \frac{3}{2c_2}\,,~~~~c_3=0\,.
   \label{ro5:ct2:Eq}
}
The five-dimensional metric can be expressed as 
\Eqr{
&&\hspace{-0.8cm}ds^2=h_2^{-4/3}(z)
\left[1+\left(\frac{\tau}{\tau_0}\right)^{-3/2}\bar{k}(z)
\right]^{-\frac{2}{3}}
\left[-d\tau^2
\right.\nn\\
&&\left.+h_2^2(z)\left\{1+\left(\frac{\tau}{\tau_0}\right)^{-3/2}
\bar{k}(z)\right\}
\left(\frac{\tau}{\tau_0}\right)^{1/2}
\delta_{ab}(\Zsp)dz^adz^b\right],
   \label{ro5:metric4:Eq}
}
where the functions $h_2(z)$ and $\bar{k}(z)$ are given by 
\Eq{
h_2(z)=c_1+\sum_{\alpha=1}^{N'}
\frac{L_{\alpha}}{|z^a-z^a_{\alpha}|^{2-d_{s}}}\,,~~~~~
\bar{k}(z) \equiv c_3+\sum_{l=1}^{N}\frac{M_l}{|z^a-z^a_l|^{2-d_{s}}}\,.
   \label{ro5:sme:Eq}
}
Here $d_{s}$ is the number of smeared dimensions and 
should satisfy $0\le d_{s}\le 3$\,. Here we assume that 
one direction of $z^a~(a=1,\ldots,4)$ is not smeared 
in order to fix the location of our universe in the transverse space. 
Our Universe is given by the solutions with the 
five-dimensional coordinates $t,~z^a~(a=1,\ldots,4)$\,. 
The time direction is written by $t$\,. 
Our choice is to take the three-dimensional 
from the overall transverse space with $z^a$. 
The four-dimensional universe is spanned by $t$, $z^2$, $z^3$ and $z^4$\,, 
for instance. The $z^1$ direction is preserved to measure the position 
of our universe in the overall transverse space of 0-branes. 
Since the metric depends on $z^a$ explicitly, we have to smear 
out ${z}^2$\,, ${z}^3$ and ${z}^4$ so as to define our Universe. 
Then the number of the smeared directions 
$d_{s}$ should satisfy the condition $d_{s}=3$\,. 

Unfortunately, the power exponent of four-dimensional 
universe becomes 1/4. 
Hence, we have to conclude that in order to obtain a realistic expansion
of the universe in this type of models, one has to include additional 
fields on the background.

We study the asymptotic behavior of the dynamical 0-brane background. 
The time dependence in the function $h_2$ can be ignored 
in the limit of $z^a\rightarrow \,z^a_l$, because 
the harmonic function $\bar{k}(z)$ dominates near a position of 0-brane. 
In the limit of $z^a\rightarrow \infty$\,, as function $\bar{k}(z)$ 
vanishes, the system becomes static near 0-brane. 
Then, the function $k_2$ depends only on time in the far region from 0-branes. 
The five-dimensional metric in the limit of $z^a\rightarrow \infty$\,, 
is thus given by
\Eqr{
ds^2=-\left(c_2\,t+c_3\right)^{-2/3}
dt^2+\left(c_2\,t+c_3\right)^{1/3}
u_{ab}dz^adz^b\,.
   \label{ro5:surface:Eq}
}

The metric has singularity at $t=-c_3/c_2$\,. 
Then the five-dimensional spacetime does not have any singularity  
if it is restricted inside the domain satisfied by the conditions,  
\Eq{
h_2(z)=c_1+\sum_{\alpha=1}^{N'}
\frac{L_{\alpha}}{|z^a-z^a_{\alpha}|^{2-d_{s}}}>0\,,~~~~~~
k_2(t, z)=c_2\,t+\bar{k}(z)>0\,,
   \label{ro5:do:Eq}
}
where the function $\bar{k}(z)$ is defined in (\ref{ro5:sme:Eq}).
The five-dimensional spacetime cannot be extended beyond this region.  
Since the spacetime evolves into a curvature singularity, 
the regular spacetime with dynamical 0-branes ends up with 
the singularities. 

Although the evolution of dynamical 0-brane with $c_2>0$ has the time 
reversal one of $c_2<0$, the behavior of the background spacetime strongly 
depends on the signature of $c_2$\,. 
In the following, we will focus on the case with $c_2<0$.  
For $t<0$, the function $h_2$ is positive everywhere. Then 
the spacetime is not singular. 
In the limit of $t\rightarrow -{\infty}$, the solution becomes 
a time-dependent uniform spacetime apart from a position of 0-branes. 
The five-dimensional background geometry can be described as 
a cylindrical form of infinite throat near the dynamical 0-branes,   

Let us consider the time evolution the five-dimensional spacetime. 
At $t=0$\,, the five-dimensional spacetime does not have any curvature 
singularity in the background. 
The background geometry has a cylindrical topology near each 0-brane. 
As time slightly increases, a curvature singularity 
appears far from 0-branes $|z^a-z^a_\alpha|\rightarrow\infty$\,. 
After that, the singular hypersurface cuts off more and more of the 
space as time increases further.  
The singular hypersurface splits and surrounds each of 
the 0-brane throats individually after time continues to evolve.  
The spatial surface is finally composed of two isolated throats.
The time evolution of the five-dimensional 
spacetime for $t<0$ is the time reversal of $t>0$.

We find that the overall transverse space tends to expand asymptotically 
like $t^{1/6}$\,, for any values of fixed $z^a$, in the regular domain 
of the five-dimensional metric (\ref{ro5:surface:Eq}), 
while the solutions describe static 
0-branes near the positions of the branes. 
In the far from 0-branes, where $|z^a-z^a_{\alpha}| 
\rightarrow \infty$\,, the background geometry becomes 
FRW universes with the power law expansion $t^{1/6}$\,. 

Next we consider the case of the near-horizon limit that 
the spacetime metric and the functions $h_2$, 
$k_2$ are given by (\ref{ro5:solution:Eq}). 
If we consider the case where all 0-branes are located at the origin of
the Z space, we have
\Eqrsubl{ro5:hk2:Eq}{
h_2(r)&=&c_1+\frac{L}{r^2}\,,\\
k_2(t, r)&=&c_2t+c_3+\frac{M}{r^2}\,,~~~~~~~~
r^2 \equiv \delta_{ab}\, z^a z^b\,,
}
where $L$, $M$ are the total mass of 0-branes  
\Eq{
L \equiv \sum_{\alpha=1}^{N'} L_{\alpha}\,,~~~~~~
M \equiv \sum_{l=1}^{N} M_{l}\,. 
}
Since the dependence on $t$ in (\ref{ro5:hk2:Eq}) is negligible 
in the near-horizon limit $r\rightarrow 0$\,, 
 the five-dimensional metric is reduced to the following form 
\Eqrsubl{ro5:nh3-metric:Eq}{
ds^2&=&ds^2_{\rm AdS_2}+L^{2/3}M^{1/3}d\Omega^2_{(3)}\,, \\
ds^2_{\rm AdS_2}&\equiv&L^{-4/3}M^{-2/3}
\left(-r^4dt^2+\frac{L^2M}{r^2}\,dr^2\right)\,,
}
where $\delta_{ab}\, dz^a dz^b=dr^2+r^2d\Omega^2_{(3)}$ has been used. 
The line elements of a two-dimensional AdS space (AdS$_2$) 
and a three-sphere with the unit radius (${\rm S}^3$) 
are given by $ds^2_{\rm AdS_2}$ and $d\Omega^2_{(3)}$\,, respectively. 
Thus we see that the near-horizon limit of the 0-brane system is a 
AdS$_2$ with a certain internal 3-space.

Before closing this subsection, we discuss the collision of 0-brane. 
There are two kinds of 0-brane in the five-dimensional spacetime. 
One is static 0-brane coming from the function $h_2(z)$\,. The other is 
dynamical 0-brane given by $k_2(t, z)$\,. 
We set the two dynamical 0-branes at $\,\vect{z}_1=(0, 0,\cdots, 0)$ 
and $\,\vect{z}_2=(P, 0,\cdots, 0)$\,, 
where $P$ is a constant. 
On the other hand, we suppose that $N'$ static 
0-branes are sitting at a point, 
\Eq{
\vect{z}_{1} = \cdots = \,\vect{z}_{N'} \equiv \,\vect{z}_0
=(z_0^1\,,\,0\,,\,\cdots\,,\,0)\,.
  \label{ro5:point:Eq}
} 
Now we consider the following quantity
\Eq{
\tilde{z}=\sqrt{\left(z^2\right)^2+\left(z^3\right)^2+\cdots 
+\left(z^{4-d_{s}}\right)^2}\,.
}
Then the proper length at $\tilde{z}=0$ between the two dynamical 0-branes 
is given by  
\Eqr{
d(t)&=&\int^{P}_0 dz^1\left(c_1+\frac{L}{|z^1-z^1_0|^{2-d_{s}}}\right)^{1/3}
\left(c_2\,t+c_3+\frac{M_1}{|z^1|^{2-d_{s}}}
+\frac{M_2}{|z^1-P|^{2-d_{s}}}\right)^{1/6}\,, 
\label{ro5:distance:Eq}
}
where $M_1$ and $M_2$ are the charges of the dynamical 
0-brane and $L$ is defined by
\Eq{
L=\sum_{\alpha=1}^{N'}L_\alpha\,.
  \label{ro5:L:Eq}
} 
For $c_2=-1$, the length $d(t)$ is a monotonically decreasing 
function of time. Since the time evolution of the proper length 
depends on the number of the smeared directions $d_{s}$\,,  
we shall analyze it for each of the values of $d_{s}$ below. 

\begin{figure}[h]
 \begin{center}
\includegraphics[keepaspectratio, scale=0.55, angle=0]{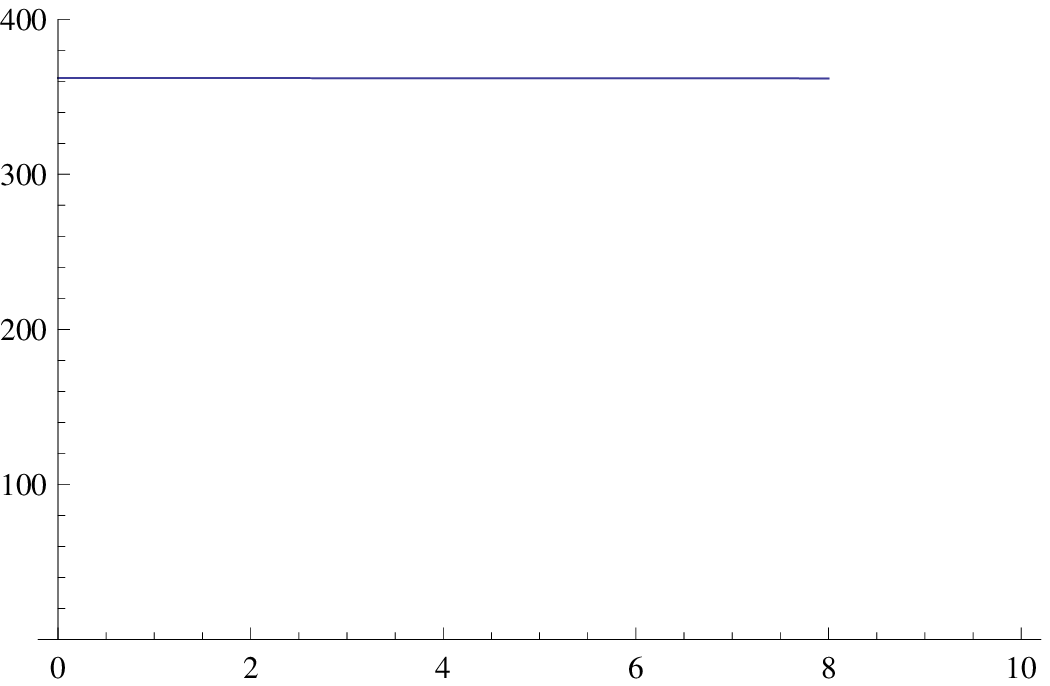}
\put(-175,120){$d(t)$}
\put(10,10){$t$}
\hskip 2.0cm
\includegraphics[keepaspectratio, scale=0.55, angle=0]{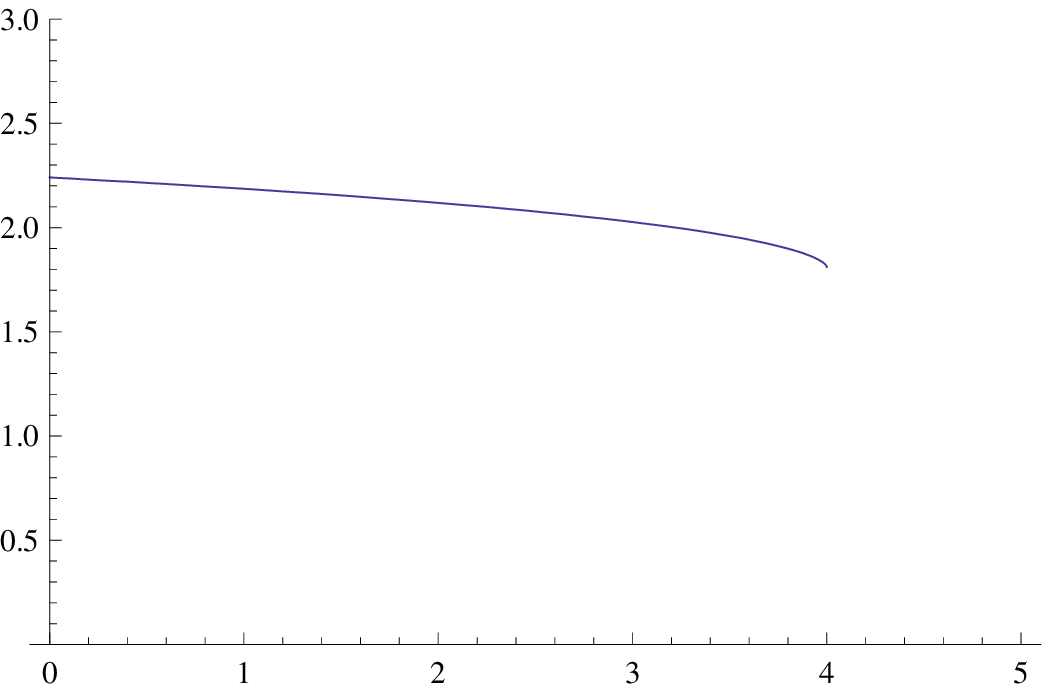}
\put(-175,120){$d(t)$}
\put(10,10){$t$}\\
(a) \hskip 7cm (b) ~~~~~~
  \caption{\baselineskip 14pt 
 The behavior of the proper distance between two dynamical 0-branes 
 for $d_s=0$ (a) and $d_s=1$ (b) in  
 the five-dimensional Romans' theory. For both cases, 
 the two dynamical 0-brane charges 
are identical, $M_1=M_2=1$ and the parameters are taken as $c_1=0$\,, 
$c_2=-1$\,, $c_3=0$\,, $L=1$\,, $z_0^1=0$\,, 
 and $P=1$. 
The result is also the same and a singularity appears before
collision of dynamical 0-branes.}
  \label{fig:ro0}
 \end{center}
\end{figure}

First we consider the case with $d_s \le 2$. 
For $d_2=2$, the harmonic functions $h_2$ and $k_2$ diverges both at 
infinity and near 0-branes. In particular, 
there is no regular spacetime region near 0-branes 
because of $h_2\rightarrow\infty$ and $\bar{k}\rightarrow\infty$\,.  
Then, these are not physically relevant. Hence, we show 
the proper length in Fig.~\ref{fig:ro0} for the cases with 
$d_s=0$ and $d_s=1$. For both cases the singularity between 
two dynamical 0-branes appears before collision because a singularity appears
before the proper distance becomes zero. Although two dynamical 0-branes 
initially approach very slowly, the singular hypersurface suddenly 
appears at a finite distance, and the spacetime finally splits into 
two isolated 0-brane throats. 
Therefore we cannot analyze the collision of the dynamical 0-branes in these
examples. 

However, for the case with $d_{s}=3$\,, 
the function $h_2$ and $\bar{k}$ are written by the linear function of 
$z^a$\,.
If we assume that the $z^a$ directions apart from $z^1$ are smeared,  
the time evolution of the proper length is different from the previous case.  
Hence, the harmonic functions $h_2$ and $\bar{k}$ are expressed as  
\Eq{
h_2\left(z^1\right)=c_1+\sum_{\alpha=1}^{N'}L_{\alpha}
|z^1-z^1_{\alpha}|\,,~~~~~~~
\bar{k}\left(z^1\right)=c_3+\sum_{l=1}^{N}M_l
|z^1-z^1_l|\,.
    \label{ro5:bh2:Eq}
}

We study the dynamics of 0-branes, where one 0-brane 
charge $M_1$ is located at $z^1=0$ and the other $M_2$ at $z^1=P$\,. 
The proper distance between the two dynamical 0-branes is given by
\Eqr{
d(t)&=&\int^P_0 dz^1 \left(c_1+L|z^1-z_0^1|\right)^{1/3}
   \left[c_2\,t+c_3+\left(M_1|z^1|+M_2|z^1-P|\right)\right]^{1/6}\,,
  \label{ro5:length:Eq}
}
where we assume again that $N'$ static 
0-branes are sitting at a point $z^1=z_0^1$\,, and $L$ is defined by 
(\ref{ro5:L:Eq}). 
For $c_2<0$, the proper distance decreases with time. 
Setting $M_1\ne M_2$\,, a curvature singularity appears again at 
a certain finite time $t=t_{\rm s}$\, before the dynamical 0-branes 
collide. Then, $t_{\rm s}$ is written by 
\Eq{
t_{\rm s}\equiv-\frac{c_3+M_1|z^1|+M_2|z^1-P|}{c_2}\,. 
}
This is the same result as the case with $d_{s}\leq 2$\,.

On the other hand, two 0-branes have the same brane charge $M_1=M_2=M$\,, 
the proper distance vanishes at a certain finite time $t=t_{\rm c}$\,, 
where $t_{\rm c}$ is defined by   
\Eq{
t_{\rm c} \equiv -\frac{c_3+MP}{c_2}\,.
}
Then two dynamical 0-branes collide completely. 

\begin{figure}[h]
 \begin{center}
\includegraphics[keepaspectratio, scale=0.55, angle=0]{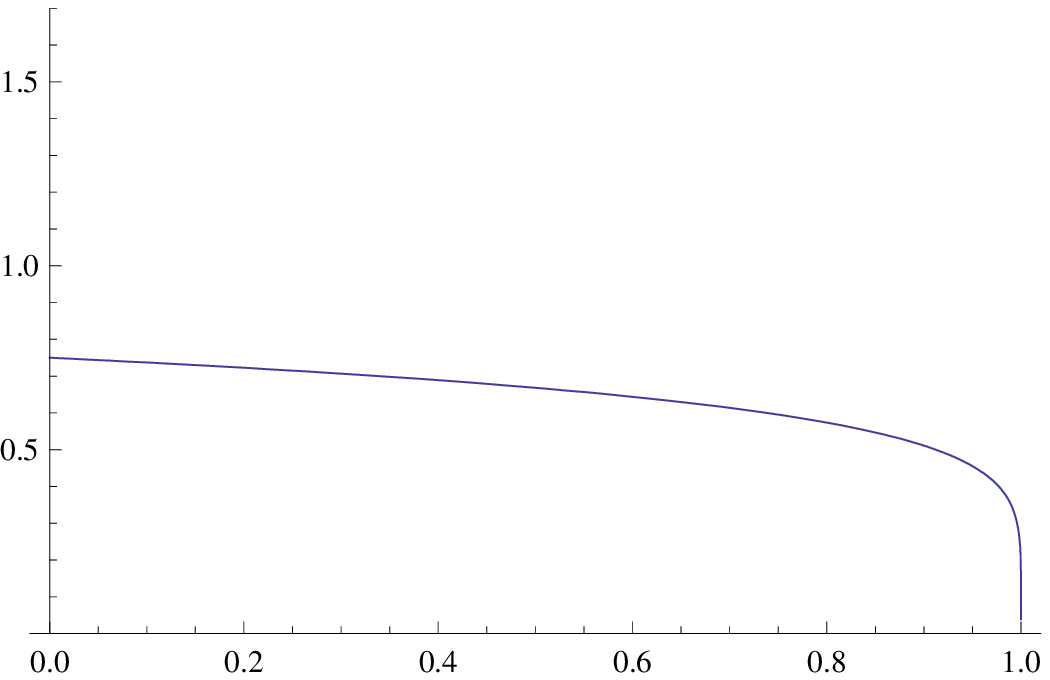}
\put(-175,115){$d(t)$}
\put(10,10){$t$}
\hskip 2.0cm
\includegraphics[keepaspectratio, scale=0.55, angle=0]{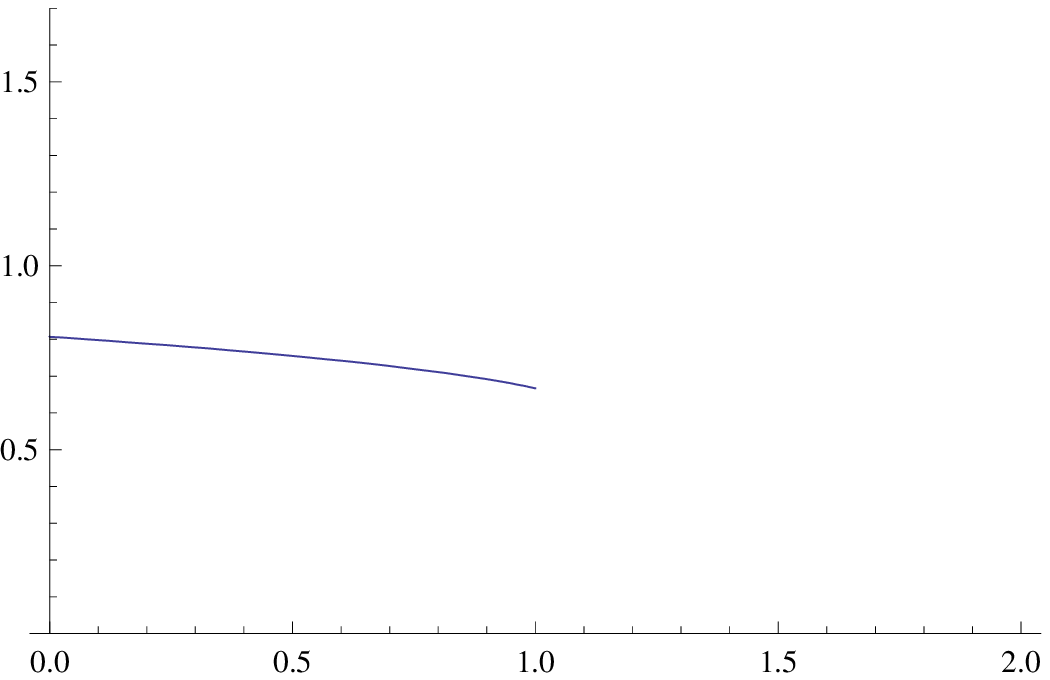}
\put(-175,115){$d(t)$}
\put(10,10){$t$}
\\
(a) \hskip 7cm (b) ~~~~~~
  \caption{\baselineskip 14pt 
 The behavior of the proper distance between two dynamical 0-branes 
 for $M_1=M_2=1$ (a) and $M_1=2$, $M_2=1$ (b) in 
 the five-dimensional Romans' theory. We fix $d_s=3$\,, $c_1=0$\,, 
 $c_2=-1$\,, $c_3=0$\,, $z^1_0=0$\,, 
$L=1$\,, and $P=1$. 
The proper length rapidly vanishes near two
0-branes collide for the case of $M_1=M_2=1$. 
For the case of $M_1=2$, $M_2=1$, it 
is still finite when a curvature singularity appears.}
  \label{fig:ro}
 \end{center}
\end{figure}

If we set $z^1_0=0$\,, 
for simplicity, the proper length between two dynamical 0-branes 
can be written by 
\Eqr{
d(t)=\frac{3}{4L}\left[-c_1^{4/3}+\left(c_1+LP\right)^{4/3}\right]
\left[c_2(t-t_{\rm c})\right]^{1/6}\,.
} 
If we choose the physical parameters as $c_1=0$\,, 
$c_2=-1$\,, $c_3=0$\,,  $P=1$\,, $z^1_0=0$\,, and 
$L=1$\,, the proper distance $d(t)$ is depicted in 
Fig.~\ref{fig:ro} for the two cases (a) the same 0-brane charges 
$M_1=M_2=1$ and (b) different 
charges $M_1=2$, $M_2=1$\,. 
For the case (a), the two dynamical 0-branes can collide completely. 
On the other hand, in the case (b) a singularity appears before the 
collision of dynamical 0-branes, as we have already discussed in 
Sec.\,\ref{sec:ns-c}.

\section{The Instability of the dynamical brane background}
\label{sec:in}
In this section, we briefly discuss the nature of the singularities 
appearing in the time-dependent solutions and present the stability 
analysis for the dynamical brane background. We follow the method 
used by \cite{oai1, Quevedo:2002tm, oai2, Cornalba:2003ze} (See also 
\cite{Grojean:2001pv, Cornalba:2002fi, Cornalba:2002nv}) and 
present the preliminary analysis performed, 
where the Klein-Gordon modes are analyzed. 
An analysis of such possibility will definitely make the property of 
singularity more clear, even if it is just a simple preliminary 
study to asses this issue. 

\subsection{The dynamical 0-brane background in 
Nishino-Salam-Sezgin gauged supergravity}
Let us first consider the stability for the 0-brane solution in 
Nishino-Salam-Sezgin gauged supergravity. 
The six-dimensional metric becomes static space near 0-brane 
while the background depends only on the times far 
from 0-brane. We will study the stability in the 
0-brane solution far from the branes. 

For the limit $r\rightarrow\infty$ in the solution (\ref{nh:h2:Eq}), 
the six-dimensional metric is expressed as 
\Eqrsubl{st:metric:Eq}{
ds^2&=&-\left(\epsilon\sqrt{2\Lambda}\,t+c_4\right)^{-3/2}
dt^2+\left(\epsilon\sqrt{2\Lambda}\,t+c_4\right)^{1/2}
\delta_{mn}(\Wsp)dv^mdv^n\,,\\
\delta_{mn}(\Wsp)dv^mdv^n&\equiv&dr^2+r^2\omega_{ij}(\Ssp^4)d\xi^id\xi^j\,, 
}
where $\omega_{ij}(\Ssp^4)$ denotes the metric of four-dimensional sphere. 
The six-dimensional metric has curvature singularity at 
$t=-c_4/\epsilon\sqrt{2\Lambda}$\,. 
In order to study the stability, we consider the Klein-Gordon equation for 
a massive scalar field propagating in the background (\ref{st:metric:Eq}):
\Eq{
-\frac{1}{\sqrt{-g}}\pd_M\left(\sqrt{-g}g^{MN}\pd_N\varphi\right)
+m^2\varphi=0\,,
   \label{st:KG:Eq}
}
where $g$ denotes the determinant of the six-dimensional metric 
(\ref{st:metric:Eq}). 
In terms of the metric (\ref{st:metric:Eq}), 
the Klein-Gordon equation can be written by
\Eq{
\pd_t\left[\left(\epsilon\sqrt{2\Lambda}\,t+c_4\right)^2\pd_t\varphi\right]
-r^{-4}\pd_r\left(r^4\pd_r\varphi\right)-\frac{1}{r^2}\lap_{\Ssp^4}\varphi
+\left(\epsilon\sqrt{2\Lambda}\,t+c_4\right)^{1/2}m^2\varphi=0\,,
}
where $\lap_{\Ssp^4}$ denotes the Laplace operator on the $\Ssp^4$. 
The six-dimensional metric involved permits separation of variables, 
so we take
\Eq{
\varphi=\varphi_0(t)\varphi_1(r)\varphi_2(\xi)\,,
   \label{st:ansatz:Eq}
}
where the functions $\varphi_1(r)$, $\varphi_2(\xi)$ obeys the 
eigenvalue equations
\Eq{
\lap_\Wsp\varphi_1(r)\varphi_2(\xi)
=-\lambda_{\Wsp}^2\,\varphi_1(r)\varphi_2(\xi)\,.
   \label{st:e-y:Eq}
}
Here $\lap_\Wsp$ is the Laplace operator on the W space, 
$\lambda_{\Wsp}$ is eigenvalue and 
$\varphi_1(r)$, $\varphi_2(\xi)$ satisfy
\cite{Dowker:1988pp}
\Eq{
\varphi_1(r)=\frac{1}{r}\left[b_1J_{\nu}(\lambda_{\Wsp} r)
+b_2Y_{\nu}(\lambda_{\Wsp} r)
\right],~~~~~
\lap_{\Ssp^4}\varphi_2(\xi)=-\lambda^2_{\Ssp^4}\varphi_2(\xi)\,,
}
where $b_1$, $b_2$ are constants, and 
 $J_{\nu}$, $Y_{\nu}$ denote the Bessel functions and $\nu$ is 
related to the eigenvalue $\lambda^2_{\Ssp^4}$ as
\Eq{
\nu^2=\lambda^2_{\Ssp^4}+\frac{9}{4}\,.
}
The Klein-Gordon equation thus is rewritten by 
\Eq{
\frac{d^2\varphi_0}{dt^2}
+\frac{2\epsilon\sqrt{2\Lambda}}{\left(\epsilon\sqrt{2\Lambda}t+c_4\right)}
\frac{d\varphi_0}{dt}
+\frac{1}{\left(\epsilon\sqrt{2\Lambda}t+c_4\right)^2}\left[\lambda_{\Wsp}^2
+\left(\epsilon\sqrt{2\Lambda}t+c_4\right)^{1/2}m^2\right]\varphi_0=0\,.
}

Then the solution for $\varphi_0$ is oscillatory, having the form 
\Eqr{
\varphi_0(t)&=&\frac{\Lambda}
{2m^2\left(\epsilon\sqrt{2\Lambda}t+c_4\right)^{1/2}}\,\left[\eta_1\,
\Gamma\left(1-\gamma\right)J_{-\gamma}
\left\{\frac{4m\left(\epsilon\sqrt{2\Lambda}t+c_4\right)^{1/4}}
{\epsilon\sqrt{2\Lambda}}\right\}\right.\nn\\
&&\left.+\eta_2\,\Gamma\left(1+\gamma\right)J_{\gamma}
\left\{\frac{4m\left(\epsilon\sqrt{2\Lambda}t+c_4\right)^{1/4}}
{\epsilon\sqrt{2\Lambda}}\right\}\right]\,, 
  \label{st:solutions:Eq}
}
where $\eta_1$, $\eta_2$ are constants, $J_{-\gamma}$, $J_{\gamma}$ denote 
the Bessel functions and $\gamma$ is given by 
\Eq{
\gamma=2\sqrt{1-\frac{2\lambda_{\Wsp}^2}{\Lambda}}\,.
  \label{st:gamma:Eq}
}

Let us consider the energy of the Klein-Gordon modes to study whether 
the instability occurs or not. 
Using the asymptotic solution 
(\ref{st:solutions:Eq}),
we will see that $E\rightarrow\infty$ as the singularity is approached, where 
there is a curvature singularity at $t=-c_4/\epsilon\sqrt{2\Lambda}$ 
in the six-dimensional background (\ref{st:metric:Eq}). 
Since the velocity is well-behaved besides the singularity,  
the energy of the Klein-Gordon modes can estimate as 
\Eq{
E=-u^M\pd_M\varphi\,,~~~~~u=\alpha\pd_t+\beta\pd_r\,,
   \label{st:e:Eq}
}
where $u$ is velocity. 
In terms of the normalization condition $u^2=-1$, the behavior of the 
$\alpha$ and $\beta$ are determined in order to remain
non-singular. 
Then, we find 
\Eq{
-\alpha^2\left(\epsilon\sqrt{2\Lambda}\,t+c_4\right)^{-3/2}
+\beta^2\left(\epsilon\sqrt{2\Lambda}\,t+c_4\right)^{1/2}=-1\,.
}
As $t\rightarrow 0$, $\alpha$ and $\beta$ have to behave 
\Eq{
\alpha\sim \left(\epsilon\sqrt{2\Lambda}\,t+c_4\right)^{3/4}\,,~~~~~
\beta\sim \left(\epsilon\sqrt{2\Lambda}\,t+c_4\right)^{-1/4}\,,
   \label{st:coe:Eq}
} 
in the limit $r\rightarrow\infty$ for the dynamical 0-brane background. 
Upon setting (\ref{st:coe:Eq}), one then finds
\Eq{
-E=\alpha\pd_t\varphi+\beta\pd_r\varphi\,.
}

In terms of the asymptotic solution (\ref{st:solutions:Eq}) 
with $\gamma$ (\ref{st:gamma:Eq}), we find that $E\rightarrow\infty$ 
as the singularity is approached if we set $\epsilon=1$ and 
$\Lambda>0$. 
Hence the 0-brane solution implies 
that the energy momentum tensor of the scalar field mode diverges 
far from 0-brane. 
However, it is necessary to study a full analysis of the metric perturbations 
whether the mode of Klein Gordon field is not likely to 
destabilize the metric modes near the singularity or not.

\subsection{The dynamical 1-brane background in 
Nishino-Salam-Sezgin gauged supergravity}
Next we consider the stability for the 1-brane solution in 
Nishino-Salam-Sezgin gauged supergravity. 
For the metric (\ref{nsc:h:Eq}), the harmonic function 
$\tilde{h}(z)$ dominates in the limit of $z^a\rightarrow z^a_l$ 
(near a position of 1-branes) and the time dependence can be ignored. 
Thus the background becomes static. 
On the other hand, in the limit of 
$|z^a|\rightarrow \infty$, $\tilde{h}(z)$ vanishes. 
Then $h_3$ depends on 
$t$ and $y$ in the far region from 1-branes and the
resulting metric is given by 
\Eqrsubl{st1:metric:Eq}{
ds^2&=&\left[\frac{\Lambda}{2}\left(t^2-y^2\right)+c_1 t+
c_2y+c_3\right]^{-1/2}\left(-dt^2+dy^2\right)\nn\\
&&+\left[\frac{\Lambda}{2}\left(t^2-y^2\right)+c_1 t+
c_2y+c_3\right]^{1/2}\delta_{ab}(\Zsp)dz^adz^b\,,\\
\delta_{ab}(\Zsp)dz^adz^b&\equiv&dr^2
+r^2\tilde{\omega}_{ij}(\Ssp^3)d\xi^id\xi^j\,,
}
where $\tilde{\omega}_{ij}(\Ssp^3)$ is 
the metric of three-dimensional sphere. 
For the six-dimensional metric, a curvature singularity may appear at
\Eq{
\frac{\Lambda}{2}\left(t^2-y^2\right)+c_1 t+
c_2y+c_3=0\,.
}
In the following, we will again discuss the stability in the 
1-brane solution far from the branes. 
Let us consider the Klein-Gordon equation for 
a massive scalar field in the six-dimensional background 
(\ref{st1:metric:Eq}):
\Eq{
-\frac{1}{\sqrt{-g}}\pd_M\left(\sqrt{-g}g^{MN}\pd_N\psi\right)
+m^2\psi=0\,,
   \label{st1:KG:Eq}
}
where $g$ denotes the determinant of the six-dimensional metric 
(\ref{st1:metric:Eq}). 
Substituting the six-dimensional metric (\ref{st1:metric:Eq}) 
into the Klein-Gordon equation for 
a massive scalar field (\ref{st1:KG:Eq}), we find
\Eqr{
&&\pd_t\left[\left\{\frac{\Lambda}{2}\left(t^2-y^2\right)+c_1 t\right\}
\pd_t\psi\right]
-\pd_y\left[\left\{\frac{\Lambda}{2}\left(t^2-y^2\right)+c_1 t\right\}
\pd_y\psi\right]
-r^{-3}\pd_r\left(r^3\pd_r\psi\right)\nn\\
&&~~~~~-\frac{1}{r^2}\lap_{\Ssp^3}\psi
+\left[\left\{\frac{\Lambda}{2}\left(t^2-y^2\right)+c_1 t\right\}\right]^{1/2}
m^2\psi=0\,,
   \label{st1:field:Eq}
}
where we set $c_2=c_3=0$\,, and 
$\lap_{\Ssp^3}$ denotes the Laplace operator on the $\Ssp^3$. 
The six-dimensional metric involved permits separation of variables, 
so we take
\Eq{
\psi=\psi_0(t)\psi_1(y)\psi_2(r)\psi_3(\xi)\,,
   \label{st1:ansatz:Eq}
}
where the functions $\psi_2(r)$, $\psi_3(\xi)$ obeys the 
eigenvalue equations
\Eq{
\lap_\Zsp\psi_2(r)\psi_3(\xi)=
-\lambda_{\Zsp}^2\,\psi_2(r)\psi_3(\xi)\,.
  \label{st1:eigen:Eq}
}
Here $\triangle_{\rm Z}$ is the Laplace operator on the Z space, 
$\lambda_{\Zsp}$ is eigenvalue for the equation (\ref{st1:eigen:Eq})\,, 
and $\psi_2(r)$, $\psi_3(\xi)$ are satisfy
\Eq{
\psi_2(r)=\frac{1}{r}\left[b_1J_{\nu}(\lambda_{\Zsp} r)
+b_2Y_{\nu}(\lambda_{\Zsp} r)
\right],~~~~~
\lap_{\Ssp^3}\psi_3(\xi)=-\lambda^2_{\Ssp^3}\psi_3(\xi)\,,
    \label{st1:eigen2:Eq}
}
where $b_1$, $b_2$ are constants, 
$J_{\nu}$, $Y_{\nu}$ denote the Bessel functions and $\nu$ is 
related to $\lambda^2_{\Ssp^3}$\,: 
\Eq{
\nu^2=\lambda^2_{\Ssp^3}+1\,.
    \label{st1:value:Eq}
}

Hence, the Klein-Gordon equation is reduced to 
\Eqr{
&&\psi_1\pd_t\left[\left\{\frac{\Lambda}{2}\left(t^2-y^2\right)+c_1 t\right\}
\frac{d\psi_0}{dt}\right]
-\psi_0\pd_y\left[\left\{\frac{\Lambda}{2}\left(t^2-y^2\right)+c_1 t\right\}
\frac{d\psi_1}{dy}\right]\nn\\
&&~~~~~+\left[\lambda_{\Zsp}^2
+\left\{\frac{\Lambda}{2}\left(t^2-y^2\right)+c_1t\right\}^{1/2}
m^2\right]\psi_0\psi_1=0\,.
   \label{st1:field2:Eq}
}

We shall discuss the massless cases in the following. In terms of 
$c_1=0$ and $m=0$\,, 
the particular solutions of $\psi_0$ and $\psi_1$ are given, respectively, by 
\Eq{
\psi_0(t)=\zeta_1\,t^{-\frac{1}{2}+\rho}+\zeta_2\,t^{-\frac{1}{2}-\rho}\,,~~~~~
\psi_1(y)=\sigma_1\,y^{-\frac{1}{2}+\rho}+\sigma_2\,y^{-\frac{1}{2}-\rho}\,,
  \label{st1:solutions:Eq}
}
where $\zeta_i~(i=1,\,2)$ and $\sigma_i~(i=1,\,2)$ are constants and 
$\rho$ is defined by
\Eq{
\rho=\frac{1}{2}\sqrt{1-\frac{4\lambda_{\Zsp}^2}{\Lambda}}\,.
  \label{st1:rho:Eq}
}

We study the stability in terms of the energy of the Klein-Gordon modes. 
Using the asymptotic solution given by Eq.~(\ref{st1:solutions:Eq}), 
we can estimate the energy 
near the singularity, 
where there is a curvature singularity at $t=\pm y$ 
in the six-dimensional background (\ref{st1:metric:Eq}). 
Since the velocity is well defined besides the singularity,  
the energy of the Klein-Gordon modes can be written as 
\Eq{
E=-v^M\pd_M\psi\,,~~~~~v=\alpha_t\pd_t+\alpha_y\pd_y+\alpha_r\pd_r\,,
   \label{st1:e:Eq}
}
where $v$ is velocity. 
By using the normalization condition $v^2=-1$, the behavior of the 
$\alpha_t$\,, $\alpha_y$\,, and $\alpha_r$ are determined 
in order to remain nonsingular. Then, we find 
\Eq{
\left(-\alpha_t^2+\alpha_y^2\right)
\left[\frac{\Lambda}{2}\left(t^2-y^2\right)\right]^{-1/2}
+\alpha_r^2\left[\frac{\Lambda}{2}\left(t^2-y^2\right)\right]^{1/2}=-1\,.
}
In the limit $r\rightarrow\infty$ and $t\rightarrow 0$, 
for the dynamical 1-brane background (\ref{st1:metric:Eq}), 
$\alpha_t$, $\alpha_y$, and $\alpha_r$ provided  
\Eq{
\left(-\alpha_t^2+\alpha_y^2\right)^{1/2}
\sim \left[\frac{\Lambda}{2}\left(t^2-y^2\right)\right]^{1/4}\,,~~~~~
\alpha_r\sim \left[\frac{\Lambda}{2}\left(t^2-y^2\right)\right]^{-1/4}\,.
   \label{st1:coe:Eq}
} 
If we set the parameters (\ref{st1:coe:Eq}), one then finds
\Eq{
-E=\alpha_t\pd_t\psi+\alpha_y\pd_y\psi+\alpha_r\pd_r\psi\,.
}
In terms of the asymptotic solution (\ref{st1:solutions:Eq}),  
we find that $E\rightarrow\infty$ 
as the singularity is approached at $t=\pm y$\,.

Let us next consider the case $m=0$ and $\Lambda=0$\,. 
In the limit $r\rightarrow\infty$ for the solution (\ref{nh:h3:Eq}), 
the six-dimensional metric becomes 
\Eqrsubl{st1:metric2:Eq}{
ds^2&=&\left(c_1\,t+c_4\right)^{-1/2}
\left(-dt^2+dy^2\right)
+\left(c_1\,t+c_4\right)^{1/2}
\delta_{ab}(\Zsp)dz^adz^b\,,\\
\delta_{ab}(\Zsp)dz^adz^b&\equiv&dr^2+r^2\omega_{ij}(\Ssp^3)d\xi^id\xi^j\,, 
}
where we set $c_2=c_3=0$ and $\omega_{ij}(\Ssp^3)$ denotes 
the metric of the three-dimensional sphere. 
There is a curvature singularity at $t=-c_4/c_1$\,. 

Now we study the behavior of the Klein-Gordon field. 
The scalar field equation (\ref{st1:KG:Eq}) in the six-dimensional background 
(\ref{st1:metric2:Eq}) reads 
\Eq{
\pd_t\left(c_1 t\,\pd_t\psi\right)
-\pd_y\left(c_1 t\pd_y\psi\right)
-r^{-3}\pd_r\left(r^3\pd_r\psi\right)
-\frac{1}{r^2}\lap_{\Ssp^3}\psi+\left(c_1 t\right)^{1/2}
m^2\psi=0\,.
   \label{st1:field3:Eq}
}
If we assume that the scalar field $\psi$ is given by 
(\ref{st1:ansatz:Eq}), where $\psi_2(r)$ and $\psi_3(\xi)$ can be   
written by (\ref{st1:eigen2:Eq}), the function $\psi_1(y)$ is 
determined by the eigenvalue equation:
\Eq{
\frac{d^2\psi_1}{dy^2}=-\lambda_y^2\,\psi_1\,.
} 
Here, $\lambda_y$ is constant. Then, the equation for $\psi_0$ becomes  
\Eq{
\frac{d}{dt}\left(c_1 t\,\frac{d\psi_0}{dt}\right)
+\left[\lambda_y^2\,c_1 t+\lambda_{\Zsp}^2
+\left(c_1 t\right)^{1/2}m^2\right]\psi_0=0\,.
}
For the massless case, the solution of $\psi_0$ is given by the 
oscillatory form 
\Eq{
\psi_0(t)=\e^{-i\lambda_y t}\left[
f_1U\left(\vartheta\,, 1\,, 2i\lambda_y t\right)
+f_2L_{-\vartheta}\left(2i\lambda_y t\right)\right],
  \label{st1:solution:Eq}
}
where $U$ denotes the hypergeometric function, $L_{-\vartheta}$ is the 
Laguerre polynomial, $f_1$ and $f_2$ are constants, 
and $\vartheta$ is defined by 
\Eq{
\vartheta=\frac{1}{2}+\frac{i \lambda^2_\Zsp}{2c_1\lambda_y}\,.
} 
We estimate the energy of the Klein-Gordon modes whether 
the instability exists or not. 
In terms of the asymptotic solution (\ref{st1:solution:Eq}),
we can present the behavior of the energy 
as the singularity is approached. 
Since the velocity is well behaved except for the singularity,  
the energy of the Klein-Gordon modes is given by (\ref{st1:e:Eq})\,. 
By using the normalization condition $v^2=-1$, 
$\alpha_t$, $\alpha_y$, and $\alpha_r$ are determined by 
\Eq{
\left(-\alpha_t^2+\alpha_y^2\right)\left(c_1\,t\right)^{-1/2}
+\alpha_r^2\left(c_1\,t\right)^{1/2}=-1\,.
}
As $t\rightarrow 0$ and $r\rightarrow\infty$, the functions 
$\alpha_t$, $\alpha_y$, and $\alpha_r$ are set to be 
\Eq{
\left(-\alpha_t^2+\alpha_y^2\right)^{1/2}\sim \left(c_1\,t\right)^{1/4}\,,
~~~~~\alpha_r\sim \left(c_1\,t\right)^{-1/4}\,.
   \label{st1:coe2:Eq}
} 
If we use Eq.~(\ref{st1:coe2:Eq}), the energy can be expressed as
\Eq{
-E=\alpha_t\pd_t\psi+\alpha_y\pd_y\psi+\alpha_r\pd_r\psi\,.
   \label{st1:energy2:Eq}
}

Then, for the asymptotic solution (\ref{st1:solution:Eq}), 
the energy becomes $E\rightarrow\infty$ 
as the singularity is approached, that is, $t\rightarrow 0$. 
Since the 1-brane solution gives  
that the energy-momentum tensor of the Klein-Gordon field mode diverges 
in this limit, the mode of the scalar field cannot  
stabilize the metric modes near the singularity.

\subsection{The dynamical 0-brane background in five-dimensional 
Romans' gauged supergravity}
In this subsection, we analyze the stability of the 
dynamical 0-brane background 
in the five-dimensional Romans' gauged supergravity. We will study the 
stability of the scalar field far from 0-branes.
For $r\rightarrow\infty$ in the dynamical 0-brane background 
(\ref{ro5:hk2:Eq}), the five-dimensional metric becomes 
\Eqrsubl{r5:metric:Eq}{
ds^2&=&-\left(c_2\,t+c_3\right)^{-2/3}dt^2
+\left(c_2\,t+c_3\right)^{1/3}
\delta_{ab}(\Zsp)dz^adz^b\,,\\
\delta_{ab}(\Zsp)dz^adz^b&\equiv&dr^2+r^2\omega_{ij}(\Ssp^3)d\xi^id\xi^j\,,
}
where we set $c_1=1$ and 
$\omega_{ij}(\Ssp^3)$ denotes the metric of the three-dimensional sphere. 
There is a curvature singularity at $t=-c_3/c_2$\,. 

Let us consider the behavior of the Klein-Gordon field:
\Eq{
-\frac{1}{\sqrt{-g}}\pd_M\left(\sqrt{-g}g^{MN}\pd_N\varphi\right)
+m^2\varphi=0\,,
   \label{r5:KG:Eq}
}
where $g$ denotes the determinant of the five-dimensional metric 
(\ref{r5:metric:Eq}). 
The scalar field equation (\ref{r5:KG:Eq}) in the five-dimensional background 
(\ref{r5:metric:Eq}) reads 
\Eq{
\pd_t\left[\left(c_2\,t+c_3\right)\,\pd_t\varphi\right]
-r^{-3}\pd_r\left(r^3\pd_r\varphi\right)
-\frac{1}{r^2}\lap_{\Ssp^3}\varphi+\left(c_2\,t+c_3\right)^{1/3}
m^2\varphi=0\,.
   \label{r5:field3:Eq}
}
Here, $\lap_{\Ssp^3}$ denotes the Laplace operator on the $\Ssp^3$\,, 
and the scalar field $\varphi$ is assumed to be
\Eq{
\varphi=\varphi_0(t)\varphi_1(r)\varphi_2(\xi)\,,
\label{r5:ansatz:Eq}
}
where $\varphi_1(r)$ and $\varphi_2(\xi)$ are 
determined by the eigenvalue equation:
\Eq{
\lap_\Zsp\varphi_1(r)\varphi_2(\xi)
=-\lambda_{\Zsp}^2\,\varphi_1(r)\varphi_2(\xi)\,.
   \label{r5:e-z:Eq}
}
Here $\triangle_{\rm Z}$ is the Laplace operator on the Z space, 
$\lambda_{\Zsp}$ is the eigenvalue, and functions
$\varphi_1(r)$ and $\varphi_2(\xi)$ obey
\cite{Dowker:1988pp}
\Eq{
\varphi_1(r)=\frac{1}{r}\left[\bar{b}_1J_{\nu}(\lambda_{\Zsp} r)
+\bar{b}_2Y_{\nu}(\lambda_{\Zsp} r)
\right],~~~~~
\lap_{\Ssp^3}\varphi_2(\xi)=-\lambda^2_{\Ssp^3}\varphi_2(\xi)\,,
}
where $\bar{b}_1$ and $\bar{b}_2$ are constants, 
 $J_{\nu}$ and $Y_{\nu}$ denote the Bessel functions, and $\nu$ is 
related to the eigenvalue $\lambda^2_{\Ssp^3}$ as
\Eq{
\nu^2=\lambda^2_{\Ssp^3}+1\,.
}
By using Eq.~(\ref{r5:e-z:Eq}), the equation for $\varphi_0$ becomes  
\Eq{
\frac{d}{dt}\left[\left(c_2\,t+c_3\right)\,\frac{d\varphi_0}{dt}\right]
+\left[\lambda_{\Zsp}^2
+\left(c_2\,t+c_3\right)^{1/3}m^2\right]\varphi_0=0\,.
}
For $m=0$, the solution of $\varphi_0$ 
is given by the oscillatory form 
\Eq{
\varphi_0(t)=\bar{f}_1J_0\left(\frac{2\lambda_{\Zsp}}{c_2}
\sqrt{c_2\,t+c_3}\right)
+\bar{f}_2Y_0\left(\frac{2\lambda_{\Zsp}}{c_2}\sqrt{c_2\,t+c_3}\right)\,, 
  \label{r5:solution:Eq}
}
where $\bar{f}_1$ and $\bar{f}_2$ are constants. 
We calculate the energy of the Klein-Gordon modes to study the 
stability of the dynamical 0-brane background. 
By using the asymptotic solution (\ref{r5:solution:Eq}),
we can present the behavior of the energy 
as the singularity is approached. 
Since it is possible to calculate the velocity except for the singularity,  
the energy of the Klein-Gordon modes is given by 
\Eq{
E=-u^M\pd_M\varphi\,,~~~~~u=\alpha_t\pd_t+\alpha_r\pd_r\,,
   \label{r5:e:Eq}
}
where $u$ denotes the velocity in the five-dimensional spacetime. 
In terms of the normalization condition $u^2=-1$, 
$\alpha_t$ and $\alpha_r$ are given by 
\Eq{
-\alpha_t^2\left(c_2\,t+c_3\right)^{-2/3}
+\alpha_r^2\left(c_2\,t+c_3\right)^{1/3}=-1\,.
}
As $t\rightarrow -c_3/c_2$ and $r\rightarrow\infty$, the functions 
$\alpha_t$ and $\alpha_r$ are described as  
\Eq{
\alpha_t\sim \left(c_2\,t+c_3\right)^{1/3}\,,
~~~~~\alpha_r\sim \left(c_2\,t+c_3\right)^{-1/6}\,.
   \label{r5:coe2:Eq}
} 
By using Eq.~(\ref{r5:coe2:Eq}), the energy of the scalar field 
can be expressed as
\Eq{
-E=\alpha_t\pd_t\varphi+\alpha_r\pd_r\varphi\,.
   \label{r5:energy2:Eq}
}

For the asymptotic solution (\ref{r5:solution:Eq}), one can note that
the energy becomes $E\rightarrow\infty$ 
as the singularity is approached. 
The dynamical 0-brane solution gives  
that the energy-momentum tensor of the scalar field mode diverges 
in the limit $t\rightarrow-c_3/c_2$\,. 
Hence, the mode of the scalar field cannot  
stabilize the metric modes near the singularity.

\subsection{Intersection involving the $0-p_I$-brane background in the 
$D$-dimensional asymptotically power-law expanding universe}
Now we investigate the stability 
analysis for the dynamical $0-p_I$-brane background. 
The geometry of the $0-p_{I'}$-brane system becomes a static structure 
near branes, while the background geometry depends only on the time 
in the far region from branes. By setting $B=0$ in the $D$-dimensional 
background (\ref{sa:exact:Eq}), the metric in the limit 
$z^a\rightarrow\infty$ is thus given by
\Eqrsubl{sp:metric:Eq}{
ds^2&=&-\left(At\right)^{a_0}dt^2
+\left(At\right)^{b_0}\delta_{ab}(\Zsp)dz^adz^b\,,\\
\delta_{ab}(\Zsp)dz^adz^b&=&dr^2
+r^2\bar{\omega}_{ij}(\Ssp^{D-2})d\xi^id\xi^j\,,
}
where $\bar{\omega}_{ij}(\Ssp^{D-2})$ denotes the metric of the 
$(D-2)$-dimensional sphere and $a_0$ and $b_0$ are defined, respectively, by 
\Eq{
a_0=-\frac{D-3}{D-2}\,,~~~~b_0=\frac{1}{D-2}\,.
 \label{sp:para:Eq}
}
The $D$-dimensional spacetime has singularities at $t=0$\,. 

Let us consider the Klein-Gordon equation 
to discuss 
the stability analysis 
\Eq{
-\frac{1}{\sqrt{-g}}\pd_M\left(\sqrt{-g}g^{MN}\pd_N\varphi\right)
+m^2\varphi=0\,,
   \label{sp:KG:Eq}
}
where $g$ denotes the determinant of the $D$-dimensional metric 
(\ref{sp:metric:Eq}). Equation (\ref{sp:KG:Eq}) on the 
$D$-dimensional background (\ref{sp:metric:Eq}) becomes 
\Eq{
\pd_t\left(At\pd_t\varphi\right)
-r^{-(D-2)}\pd_r\left(r^{D-2}\pd_r\varphi\right)
-\frac{1}{r^2}\lap_{\Ssp^{D-2}}\varphi+\left(At\right)^{b_0}
m^2\varphi=0\,,
   \label{sp:KG2:Eq}
}
where $\lap_{\Ssp^{D-2}}$ denotes the Laplace operator on the $\Ssp^{D-2}$\,. 
We assume that the scalar field $\varphi$ is 
given by 
\Eq{
\varphi=\varphi_0(t)\varphi_1(r)\varphi_2(\xi)\,,
   \label{sp:ansatz:Eq}
}
where the functions $\varphi_1(r)$ and $\varphi_2(\xi)$ obey the 
eigenvalue equations
\Eq{
\lap_\Zsp\varphi_1(r)\varphi_2(\xi)
=-{\lambda}_{\Zsp}^2\,\varphi_1(r)\varphi_2(\xi)\,.
   \label{sp:eigen:Eq}
} 
Here $\lap_{\Zsp}$ denotes the Laplace operator on the $\Zsp$ space, and 
$\lambda_{\Ysp}$ is the eigenvalue for the equation. 

The functions $\varphi_1(r)$ and $\varphi_2(\xi)$ also satisfy the 
equations \cite{Dowker:1988pp}
\Eq{
\varphi_1(r)=\frac{1}{r}\left[b_3J_{\bar{\nu}}(\lambda_{\Zsp} r)
+b_4Y_{\bar{\nu}}(\lambda_{\Zsp} r)
\right],~~~~~
{\lap}_{\Ssp^{D-2}}\varphi_2(\xi)
=-{\lambda}_{\Ssp^{D-2}}^2\,\varphi_2(\xi)\,,
   \label{sp:eigen2:Eq}
} 
where $b_3$ and $b_4$ are constants, 
 $J_{\bar{\nu}}$ and $Y_{\bar{\nu}}$ denote the Bessel functions, and 
 $\bar{\nu}$ is related to the eigenvalue $\lambda^2_{\Ssp^{D-2}}$ as
\Eq{
\bar{\nu}^2=\lambda^2_{\Ssp^{D-2}}+\frac{\left(D-3\right)^2}{4}\,.
}
By using Eqs.~(\ref{sp:metric:Eq}), (\ref{sp:ansatz:Eq}), 
and (\ref{sp:eigen:Eq}), the field equation for $\varphi_0$ becomes 
\Eq{
\frac{d}{dt}\left(A t\,\frac{d\psi_0}{dt}\right)
+\left[\lambda_{\Zsp}^2
+\left(At\right)^{b_0}m^2\right]\varphi_0=0\,.
   \label{sp:field2:Eq}
}
Let us consider the case of $m=0$\,. 
The solution of $\varphi_0$ is given by the oscillating form  
\Eq{
\varphi_0(t)=f_3J_0\left(2\lambda_{\Zsp}\sqrt{A^{-1}\,t}\right)
+f_4Y_0\left(2\lambda_{\Zsp}\sqrt{A^{-1}\,t}\right)\,, 
  \label{sp:solution:Eq}
}
where $f_3$ and $f_4$ are constants and $J_0$ and $Y_0$ 
are the Bessel functions. 
The energy of the Klein-Gordon modes can be calculated by   
\Eq{
E=-u^M\pd_M\varphi\,,~~~~~u=\alpha\pd_t+\beta\pd_r\,,
   \label{sp:e:Eq}
} 
where $u$ is velocity. Then, $\alpha$ and $\beta$ are determined by 
\Eq{
-\alpha^2\left(A\,t\right)^{a_0}+\beta^2\left(A\,t\right)^{b_0}=-1\,,
  \label{sp:ab:Eq}
}
where we used the normalization condition $u^2=-1$\,. 
In the case of $t\rightarrow 0$ and $r\rightarrow\infty$, 
$\alpha$ and $\beta$ must behave as
\Eq{
\alpha\sim \left(A\,t\right)^{-a_0/2}\,,~~~~~
\beta\sim \left(A\,t\right)^{-b_0/2}\,,
   \label{sp:coe:Eq}
}
in order to remain nonsingular. 

If we use the expression (\ref{sp:coe:Eq}), the energy of the scalar field 
is given by 
\Eq{
-E=\alpha\,\pd_t\varphi+\beta\,\pd_r\varphi\,.
   \label{sp:e2:Eq}
} 
For the asymptotic solution (\ref{sp:solution:Eq}), 
one can note that the energy becomes $E\rightarrow\infty$ 
as the singularity is approached. Hence, 
the energy-momentum tensor of the Klein-Gordon field mode diverges. 
The classical solution gives the mode of the scalar field which cannot  
stabilize the metric modes near the singularity.

\subsection{Intersection involving the $0-p_{I'}$-brane background 
in the $D$-dimensional asymptotically de Sitter universe}
Finally, we discuss the stability analysis for the $0-p_{I'}$-brane 
in the asymptotically de Sitter universe. If we set $\tilde{c}=0$ 
and take $z^a\rightarrow\infty$ in the 
background (\ref{ds2:exact:Eq}), the $D$-dimensional metric becomes 
\Eqrsubl{sd:metric:Eq}{
ds^2&=&-d\tau^2
+\left(c_0\,\e^{c_0\tau}\right)^{2/(D-3)}\delta_{ab}(\Zsp)dz^adz^b\,,\\
\delta_{ab}(\Zsp)dz^adz^b
&=&dr^2+r^2\bar{\omega}_{ij}(\Ssp^{D-2})d\xi^id\xi^j\,,
}
where $\bar{\omega}_{ij}(\Ssp^{D-2})$ 
denotes the metric of the $(D-2)$-dimensional sphere, 
$c_0$ is given by (\ref{ds2:c0:Eq}), and the cosmic time $\tau$ is 
defined by (\ref{ds2:cosmic:Eq}). 
There is a curvature singularity at $\tau\rightarrow-\infty$ 
in the $D$-dimensional spacetime. In the following, we set $c_0>0$\,. 
Otherwise, the scale factor of $D$-dimensional spacetime becomes 
complex or negative. 

We consider the Klein-Gordon field 
to analyze the stability 
\Eq{
-\frac{1}{\sqrt{-g}}\pd_M\left(\sqrt{-g}g^{MN}\pd_N\varphi\right)
+m^2\varphi=0\,,
   \label{sd:KG:Eq}
}
where $g$ is the determinant of the six-dimensional metric 
(\ref{sd:metric:Eq}).  
Substituting the $D$-dimensional metric (\ref{sd:metric:Eq}) into 
Eq.~(\ref{sd:KG:Eq}), we obtain 
\Eq{
\left(c_0\,\e^{c_0\tau}\right)^{-1}
\pd_\tau\left[\left(c_0\,\e^{c_0\tau}\right)^{\frac{D-1}{D-3}}
\pd_\tau\varphi\right]
-r^{-(D-2)}\pd_r\left(r^{D-2}\pd_r\varphi\right)
-\frac{1}{r^2}\lap_{\Ssp^{D-2}}\varphi
+\left(c_0\,\e^{c_0\tau}\right)^{\frac{2}{D-3}}
m^2\varphi=0\,,
   \label{sd:KG2:Eq}
}
where $\lap_{\Ssp^{D-2}}$ denotes the Laplace operator on the $\Ssp^{D-2}$\,. 

We assume an ansatz for the scalar field $\varphi$\,: 
\Eq{
\varphi=\varphi_0(\tau)\varphi_1(r)\varphi_2(\xi)\,,
   \label{sd:ansatz:Eq}
}
where the functions $\varphi_1(r)$ and $\varphi_2(\xi)$ satisfy the 
eigenvalue equation 
\Eq{
\lap_\Zsp\varphi_1(r)\varphi_2(\xi)
=-{\lambda}_{\Zsp}^2\,\varphi_1(r)\varphi_2(\xi)\,,
   \label{sd:eigen:Eq}
} 
and obey the equations 
\Eq{
\varphi_1(r)=\frac{1}{r}\left[b_5J_{\tilde{\nu}}(\lambda_{\Zsp} r)
+b_6Y_{\tilde{\nu}}(\lambda_{\Zsp} r)
\right],~~~~~
{\lap}_{\Ssp^{D-2}}\varphi_2(\xi)
=-{\lambda}_{\Ssp^{D-2}}^2\,\varphi_2(\xi)\,.
   \label{sd:eigen2:Eq}
}
Here $\triangle_{\rm Z}$ is the Laplace operator on the Z space, and 
$b_5$ and $b_6$ denote constants, 
 $J_{\tilde{\nu}}$ and $Y_{\tilde{\nu}}$ are the Bessel functions, and 
 $\tilde{\nu}$ is written by the eigenvalue $\lambda^2_{\Ssp^{D-2}}$ as 
\Eq{
\tilde{\nu}^2=\lambda^2_{\Ssp^{D-2}}+\frac{\left(D-3\right)^2}{4}\,.
}
In terms of Eqs.~(\ref{sp:eigen:Eq}), (\ref{sd:metric:Eq}),  
and (\ref{sd:ansatz:Eq}), the field equation for $\varphi_0$ becomes 
\Eq{
\left(c_0\,\e^{c_0\tau}\right)^{-1}
\frac{d}{d\tau}\left[\left(c_0\,\e^{c_0\tau}\right)^{(D-1)/(D-3)}\,
\frac{d\varphi_0}{d\tau}\right]+\left[\lambda_{\Zsp}^2
+\left(c_0\,\e^{c_0\tau}\right)^{2/(D-3)}m^2\right]\varphi_0=0\,.
   \label{sd:field:Eq}
}
Let us first consider the solution of $\varphi_0$ for $D=5$ and $D=6$\,.   
In the case of $D=5$, the solution of $\varphi_0$ can be expressed as 
\Eqr{
\varphi_0(\tau)&=&
\lambda_{\Zsp}^2\,c_0^{-3}\,\e^{-c_0\tau}\left[
f_5\,\Gamma\left(1-\ell_1\right)\,
J_{-{\ell_1}}
\left(2\lambda_{\Zsp}\,c_0^{-3/2}\,\e^{-c_0\tau/2}\right)\right.\nn\\
&&\left.+f_6\,\Gamma\left(1+\ell_1\right)\,
J_{\ell_1}\left(2\lambda_{\Zsp}\,c_0^{-3/2}\,\e^{-c_0\tau/2}\right)\right], 
  \label{sd:solution:Eq}
}
where $f_5$ and $f_6$ are constants, $J_{\ell_1}$ and $J_{-{\ell_1}}$ 
are the Bessel functions, and $\ell_1$ is defined by 
\Eq{
\ell_1=2\sqrt{1-\left(\frac{m}{c_0}\right)^2}\,.
}
On the other hand, setting $D=6$\,, 
we can also find the solution of $\varphi_0$\,:
\Eqr{
\varphi_0(\tau)&=&\frac{9}{4}\,\sqrt{\frac{3}{2}}\,
\lambda_{\Zsp}^{5/2}\,c_0^{-10/3}\,\e^{-5c_0\tau/6}\left[
f_7\,\Gamma\left(1-\ell_2\right)\,
J_{-\ell_2}\left(3\lambda_{\Zsp}\,c_0^{-4/3}\,
\e^{-c_0\tau/3}\right)\right.\nn\\
&&\left.+f_8\,\Gamma\left(1+\ell_2\right)\,
J_{\ell_2}\left(3\lambda_{\Zsp}\,c_0^{-4/3}\,\e^{-c_0\tau/3}\right)\right]. 
  \label{sd:solution2:Eq}
}
Here $f_7$ and $f_8$ denote constants, $J_{\ell_2}$ and $J_{-{\ell_2}}$ 
are the Bessel functions, and the constant $\ell_2$ is given by 
\Eq{
\ell_2=\frac{1}{2}\,\sqrt{25-36\left(\frac{m}{c_0}\right)^2}\,.
}

For $D\ge 4$, the solution of $\varphi_0$ 
can be written in the following form:
\Eqr{
\varphi_0(\tau)&=&\left(\frac{D-3}{2}\right)^{\frac{D-1}{2}}\,
\lambda_{\Zsp}^{\frac{D-1}{2}}\,c_0^{-\frac{(D-1)(D-2)}{2(D-3)}}\,
\e^{-\frac{D-1}{2(D-3)}c_0\tau}\nn\\
&&\times\left[
\bar{f}\,\Gamma\left(1-\ell\right)\,
J_{-\ell}\left((D-3)\,\lambda_{\Zsp}\,c_0^{-\frac{D-2}{D-3}}\,
\e^{-\frac{1}{D-3}c_0\tau}\right)\right.\nn\\
&&\left.+\tilde{f}\,\Gamma\left(1+\ell\right)\,
J_{\ell}\left((D-3)\,\lambda_{\Zsp}\,c_0^{-\frac{D-2}{D-3}}\,
\e^{-\frac{1}{D-3}c_0\tau}\right)\right], 
  \label{sd:solution-g:Eq}
}
where $\bar{f}$ and $\tilde{f}$ are constants, $J_{\ell}$ and $J_{-{\ell}}$ 
denote the Bessel functions, and $\ell$ is defined by 
\Eq{
\ell=\frac{1}{2}\,\sqrt{(D-1)^2-4(D-3)^2\left(\frac{m}{c_0}\right)^2}\,.
}

Since the energy of the scalar field 
can be written as (\ref{sp:e:Eq}), the energy of the Klein-Gordon modes can 
be given by the expression (\ref{sp:e2:Eq})\,. Then, we can find 
$\alpha$ and $\beta$ in this way:  
\Eq{
-\alpha^2+\beta^2\left(c_0\,\e^{c_0\tau}\right)^{2/(D-3)}=-1\,,
  \label{sd:ab:Eq}
}
where we used the normalization condition $u^2=-1$\,. 
In the case of $\tau\rightarrow -\infty$ and $r\rightarrow\infty$, 
$\alpha$ and $\beta$ have to behave as
\Eq{
\alpha\sim {\rm constant}\,,~~~~~
\beta\sim \left(c_0\,\e^{c_0\tau}\right)^{-1/(D-3)}\,.
   \label{sd:coe:Eq}
}
For the solution (\ref{sd:solution-g:Eq})\,, 
the energy becomes $E\rightarrow\infty$ in the limit 
$\tau\rightarrow -\infty$\,. 
Since the energy is not convergent with the asymptotic solution, 
the mode of the scalar field does not stabilize the metric modes 
near the singularity.


\section{Conclusion and Discussion}
  \label{sec:cd}
In this paper, we have discussed the 
time-dependent intersecting branes 
with cosmological constants for not only the delocalized case but also the 
partially localized one in $D$-dimensional gravitational theory. 
We are everywhere brief, and on some points, we simply call attention 
to questions that might be investigated in the future.
The function $h_{I}$ depends on time as well as the coordinate 
of the relative and overall transverse spaces. 
The coupling constants between the field strengths 
and the dilaton are given by the assumptions (\ref{pl:ansatz:Eq}) or  
(\ref{cn:fields:Eq}) and depend on the parameter $N$.
In the case of the eleven- or ten-dimensional supergravity
theory, the dilaton coupling requires $N=4$. 
The power of the time dependence depends
on the number of the brane and total dimensions with
the parameter $N$ of the dilaton coupling constant. 

An exceptional case arises if the parameter $N$ in the dilaton coupling 
takes another value than 4. There are static solutions with $N\ne 4$  
in the lower-dimensional supergravity theories as well as 
Einstein-Maxwell theory \cite{Lu:1995cs, Cvetic:2000dm}. 
In this case one gets asymptotically power-law expanding solutions 
if the dilaton is nontrivial. For the trivial dilaton, 
the Einstein equations give an asymptotically de Sitter solution 
for a single 2-form field strength. 
Since the cosmological constant is related to 
the field strength, 
the time derivative of the warp factor arises only 
from the Ricci tensor and can be compensated
by the cosmological constant in the Einstein equations. 
This is the same structure as in Refs.
\cite{Kastor:1992nn, Maeda:2010aj, Minamitsuji:2010kb} and 
the generalization of the solutions 
\cite{Kastor:1992nn, Maki:1992tq, Maki:1994wc}.
In $N=4$ case, the equation of motion in the presence of the 
cosmological constant gives the static delocalized or partially localized 
intersecting brane solution because of the ansatz of the fields. 
Thus, one expects that the recipe for picking an accelerating expansion
from the dynamical intersecting brane 
solutions depends on the dilaton coupling 
constant, and this is the case for the proposal in Ref.~
\cite{Minamitsuji:2010kb}. 
Once the de Sitter solution is obtained in the single $p$-brane solutions, 
it is possible to apply it to the intersecting brane systems.

An immediate point is that the time-dependent solutions make dynamical
compactification more or less obvious, 
since cosmological evolution is a general property of the solution 
(with constant parameters) once the function $h_I$ 
is properly endowed with the time dependence. 
The power of the scale factor in some solutions gives an accelerating 
expansion law even in the case that functions $h_I$ depend on 
both the time and coordinates of overall transverse space, while the 
extra dimension will shrink as cosmic time increases. 
However, something is still missing, because the scale factor of our 
Universe diverges at $\tau=\tau_\infty$. At the moment, it is not 
clear how to do this.

We have discussed the dynamics of the brane collisions. 
As the spacetime is contracting in the $D$-dimensional spacetime, 
each 0-brane approaches others as the time evolves for $\tau<0$ 
but separates for $\tau>0$ in the asymptotically 
power-law expanding solutions. Thus 0-branes never collide.
In the case of asymptotically de Sitter solutions, 
all domains between branes are connected at $\tau=0$ ($c_0<0$).  
The domain shrinks as the time decreases, while 
the proper distance becomes constant as $\tau$ increases. 
For the $0-p_{I'}$-brane system ($p\le 7$), 
a singularity appears before 0-branes collide, and eventually the 
topology of the spacetime changes so that 
branes are separated by singular hypersurfaces surrounding each brane 
if branes are not smeared. 
Thus, we cannot describe the collision of two 0-branes 
in terms of these solutions. 
On the other hand, the $0-8$-brane system in ten dimensions or 
the smeared $0-p_{I}$-brane system in $D$-dimensional theory 
can provide examples of colliding branes if they have the 
same brane charges and only one overall transverse space. 
We have analyzed the collision of the brane where 
the $p_0-p_I$-branes are localized at the same position along 
the overall transverse directions, in the case of equal charges. 
The brane collision would not occur 
if the brane charges are different. 
Moreover, if these branes are localized at different positions, 
it raises the possibility that the curvature singularities appear. 

We have also studied the dynamics of the five- or 
six-dimensional supergravity 
models with applications to cosmology and collision of branes. 
First we have discussed the brane solutions 
to study the time evolution in the NSS model. 
In the case of vanishing 3-form field strength in the five-dimensional 
effective theory, the scale factor of our four-dimensional spacetime 
is a linear function of the cosmic time which is the same evolution
as the Milne universe. On the other hand, for the dynamical 
1-brane without 2-form field strength, the solution tells us 
that the function $h$ depends on all the world-volume coordinates of 
the 1-brane. Hence, the contribution of the field strength except for
the 2-form leads to an inhomogeneous universe. 
We have investigated the dynamics of 0-branes and found that, when 
the spacetime is contracting in six dimensions, each 0-brane approaches 
the others as the time evolves. All domains between branes connected 
initially ($t=0$), but it shrinks as $t$ increases.
However, for the 0-brane system without smearing branes, a singularity 
appears before 0-branes collide, and eventually the topology of 
the spacetime changes such that parts of the branes are separated 
by a singular region surrounding each brane. Thus, the
solution cannot describe the collision of two 0-branes. 
In contrast, the smeared 0-brane system with $d_s=4$ can provide
an example of colliding 0-branes and collision of the universes, 
if they have the same brane charges.

We have next constructed the time-dependent 1-brane solution in the NSS  
supergravty model. In the asymptotic far 1-brane region, 
the 1-brane spacetime 
in the NSS model approaches the six-dimensional Milne universe. 
In regions close to the 1-branes, for concreteness, we have studied 
the case of two 1-branes in detail. The 1-brane is approaching the 
other as the time progress for $t<0$. 
We have found that, in the case of $t<0$, 
all of the domains between the 1-branes are initially 
connected, but some region (near small $y$) shrinks as the time 
increases, and eventually the topology of the spacetime changes such 
that parts of the branes are separated by a singular region surrounding each
1-brane. Thus, in the case of $d_s\ne 3$, 1-branes never collide. 
On the other hand, the case of $d_s=3$, for $t<0$, could provide an 
example of colliding 1-branes. We found that the collision time 
depends on both brane charges and the place in the world volume of 
the 1-brane. Since this case has the time-reversal symmetry, the evolution 
for $t>0$ is obtained by the time-reversal transformation. 

We also investigated the time-dependent solution in the five-dimensional 
supergravity model. The power of the scale factor is so small 
that the solutions cannot give a realistic expansion law.
Then, it is necessary to include additional matter on the background 
in order to obtain a realistic expanding universe.

We finally analyzed the classical instability of the dynamical 
brane background towards singularity. In order to present the instability 
of the dynamical brane background, we have estimated whether an 
instability does exist by computing the energy of the Klein-Gordon 
modes. One can find that the energy seen by an observer diverges 
as the curvature singularity is approached. This implies that 
the mode of the 
scalar field is likely to destabilize the background metric modes 
near the singularity. Although this result has been given by preliminary 
analysis, it has made the property of singularity in the dynamical 
brane background more clear. It is also necessary for us to perform a more 
rigorous analysis by considering in detail the metric perturbation 
whether the stability analysis arrives to the same conclusion or not.

A recent study of intersecting 
systems depending on the time coordinate and overall transverse space 
shows that all warp factors in the solutions can depend on time 
\cite{Maeda:2012xb}. 
It is interesting to study if similar more general solutions
can be obtained by relaxing some of our assumptions. 
We hope to report on this subject in the near future elsewhere.

\section*{Acknowledgments}
We thank M. Minamitsuji for numerous valuable 
discussions and careful reading of the manuscript and also 
thank M. Cvetic and G.W. Gibbons for discussions and valuable comments. 



\end{document}